\documentclass[reprint,longbibliography,superscriptaddress,
showpacs,preprintnumbers,
notitlepage,amsmath,amssymb,
article,
]{revtex4-1}
\usepackage{adjustbox}

\usepackage{graphicx}
\usepackage{dcolumn}
\usepackage{bm}
\usepackage{bbm}
\usepackage{diagbox}
\usepackage{braket}

\usepackage{graphicx}
\usepackage{enumerate}
\usepackage{subfigure}
\usepackage{epstopdf}
\usepackage{amsmath,amsthm,amssymb,epsfig,amsfonts,mathrsfs}
\usepackage[colorlinks,
linkcolor=red,
anchorcolor=black,
citecolor=blue]{hyperref}
\usepackage{amssymb}
\usepackage{framed}
\usepackage{tabularx}
\usepackage{booktabs}
\usepackage{float}
\usepackage[table]{xcolor}
\usepackage{array}
\usepackage{physics}
\usepackage{dsfont}
\usepackage{tikz-cd}
\usepackage{MnSymbol}
\hypersetup{colorlinks=true}

\usepackage{tikz}
\usetikzlibrary{arrows}
\usetikzlibrary{arrows.meta}
\usetikzlibrary{shapes.geometric,calc,arrows, positioning,shapes.misc,decorations.markings}
\tikzset{
  big arrow/.style={
    decoration={markings,mark=at position 1 with {\arrow[scale=2,#1]{>}}},
    postaction={decorate},
    shorten >=0.4pt},
  big arrow/.default=black}
\tikzset{every picture/.style={line width=0.75pt}}
\pgfdeclarelayer{edgelayer}
\pgfdeclarelayer{nodelayer}
\pgfsetlayers{edgelayer,nodelayer,main} 
\tikzstyle{none}=[inner sep=0pt] 

\tikzstyle{NodeCross}=[draw, shape=circle, cross out, inner sep=0pt, minimum size=6pt,line width=0.25mm]
\tikzstyle{Circle}=[draw, shape=circle, black,  fill=black, inner sep=0pt, minimum size=6pt]
\tikzstyle{Star}=[draw, shape=star, fill=black, star points=8, inner sep=0pt, minimum size=8pt]

\tikzstyle{DashedLine}=[-, densely dashed, line width=0.25mm]
\tikzstyle{DottedLine}=[-, dotted, line width=0.25mm]
\tikzstyle{ThickLine}=[-, line width=0.25mm]
\tikzstyle{ArrowLineRight}=[-, -{Stealth[scale=1.75]}, line width=0.1mm, scale=5]
\tikzstyle{RedLine}=[-, draw={rgb,255: red,191; green,0; blue,0}, fill=none, line width=0.25mm]
\tikzstyle{DottedRed}=[-, dotted, draw={rgb,255: red,191; green,0; blue,0}, fill=none, line width=0.25mm]
\tikzstyle{DashedLineThin}=[-, densely dashed, line width=0.125mm, fill=none, draw=black]
\tikzstyle{ArrowLineRed}=[-, -{Stealth[scale=1.75]}, draw={rgb,255: red,191; green,0; blue,0}, line width=0.25mm, scale=5]

\newcommand{\1}{\text{\uppercase\expandafter{\romannumeral1}}}
\newcommand{\2}{\text{\uppercase\expandafter{\romannumeral2}}}
\newcommand{\3}{\text{\uppercase\expandafter{\romannumeral3}}}
\newcommand{\4}{\text{\uppercase\expandafter{\romannumeral4}}}
\newcommand{\5}{\text{\uppercase\expandafter{\romannumeral5}}}
\newcommand{\6}{\text{\uppercase\expandafter{\romannumeral6}}}

\def\Z{\mathbb{Z}}

\def\Tr{\mathop{\mathrm{Tr}}\nolimits}

\def\bra#1{{\langle{#1}|}}

\def\ket#1{{|{#1}\rangle}}

\def\unit{{1\kern-.65ex {\rm l}}}
\def\1{{1\kern-.65ex {\rm l}}}

\def\CA{{\cal A}}
\def\CB{{\cal B}}
\def\CC{{\cal C}}

\def\CE{{\cal E}}

\def\CH{{\cal H}}

\def\CL{{\cal L}}
\def\CM{{\cal M}}
\def\CN{{\cal N}}

\def\CS{{\cal S}}
\def\CT{{\cal T}}

\def\CZ{{\cal Z}}

\def\bbZ{{\mathbb{Z}}}

\newcount\hour \newcount\minute
\hour=\time \divide \hour by 60
\minute=\time
\def\now{%
\ifnum \hour<13
  \ifnum \hour=0 \advance \hour by 12 \number\hour:\else \number\hour:\fi%
     \ifnum \minute<10 0\fi%
     \number\minute%
\ A.M.%
\else \advance \hour by -12 \number\hour:%
  \ifnum \minute<10 0\fi%
  \number\minute%
  \ P.M.%
\fi%
}

\makeatother

\def\mb{\mathbb}

\def\mc{\mathcal}
\def\bp{\begin{pmatrix}}
\def\ep{\end{pmatrix}}
\def\be{\begin{equation}}
\def\ee{\end{equation}}
\newcommand{\ba}{\begin{aligned}}
\newcommand{\ea}{\end{aligned}}

\def\ptl{\partial}

\def\dd{{\rm d}}

\newcommand{\kket}[1]{| #1\rrangle }
\newcommand{\bbra}[1]{\llangle #1| }

\begin{document}
\title{Topological Holography for Mixed-State Phases and Phase Transitions}
\author{Ran Luo}
\email{ranluo@pku.edu.cn}
\affiliation{School of Physics, Peking University,
Beijing 100871, China}
\affiliation{Perimeter Institute for Theoretical Physics, Waterloo, Ontario, Canada N2L 2Y5}
\author{Yi-Nan Wang}
\email{ynwang@pku.edu.cn}
\affiliation{School of Physics, Peking University,
Beijing 100871, China}
\affiliation{Center for High Energy Physics, Peking University}
\author{Zhen Bi}
\email{zjb5184@psu.edu}
\affiliation{Department of Physics, The Pennsylvania State University, University Park, Pennsylvania 16802, USA}

\begin{abstract}
We extend the symmetry topological field theory (SymTFT) framework to open quantum systems. Using canonical purification, we embed mixed states into a doubled (2+1)-dimensional topological order and employ the slab construction to study (1+1)-dimensional mixed-state phases through condensable algebras in the doubled SymTFT. Hermiticity and positivity of the density matrix impose additional constraints on allowable anyon condensations, enabling a systematic classification of mixed-state phases --including strong-to-weak symmetry breaking (SWSSB) and average symmetry-protected topological (ASPT) phases. We present examples of mixed-state phase transitions involving SWSSB and show how gauging within the open SymTFT framework reveals connections among different mixed-state phases.

\end{abstract}

\maketitle

\section{Introduction}

Recent advances have transformed our understanding of symmetry, culminating in a unified framework known as symmetry topological field theory (SymTFT). A SymTFT is a \((d+1)\)-dimensional topological field theory \(\mathcal{T}^{\mathrm{Sym}}_{d+1}\) that holographically encodes the global symmetry structure of a \(d\)-dimensional quantum field theory $\CT_d$. In this framework, both a global symmetry and its dual are incorporated in the bulk construction, placing them on equal footing. Gauging a global symmetry can be naturally described by altering boundary conditions in the SymTFT \(\mathcal{T}^{\mathrm{Sym}}_{d+1}\). Moreover, because symmetries are realized as topological operators within the SymTFT, the framework naturally generalizes to include non-invertible (categorical) symmetries. Applications of SymTFT now span a broad range of areas, from string theory to condensed matter physics~\cite{Witten:1998wy,Kong:2020cie,Gaiotto:2020iye,Lichtman2021,Apruzzi:2021nmk,Apruzzi:2022dlm,DelZotto:2022ras,Moradi:2022lqp,Freed:2022qnc,Kaidi:2022cpf,vanBeest:2022fss,Kaidi:2023maf,Chen:2023qnv,Bhardwaj:2023ayw,Apruzzi:2023uma,Cao:2023rrb,Baume:2023kkf,Brennan:2024fgj,Argurio:2024oym,Copetti:2024onh,Bhardwaj:2024igy,Choi:2024tri,Tian:2024dgl,Najjar:2024vmm,Bonetti:2024etn,Kong:2015flk,Ji_2019,Ji:2019jhk,ji2022unifiedview,Chatterjee_2023,Chatterjee2023shadow,chatterjee2024emergent,Chatterjee2024noninvertible,Wen:2024udn,Wen2025igspt,Wen2023igsptbb,huang2025topologicalholography}.

More precisely, a \(d\)-dimensional theory \(\mathcal{T}_d\) can be constructed from its SymTFT \(\mathcal{T}^{\rm Sym}_{d+1}\) via the so-called slab -- or sandwich -- construction. One considers \(\mathcal{T}^{\rm Sym}_{d+1}\) defined on the manifold \(M_d \times [0,1]\), with a \emph{topological boundary} (also called the symmetry boundary) at \(x^{d+1} = 0\), and a \emph{dynamical boundary} (the physical boundary) at \(x^{d+1} = 1\). Upon collapsing the \(x^{d+1}\) direction, this setup yields an effective \(d\)-dimensional theory \(\mathcal{T}_d\), whose global symmetries depend on the choice of boundary conditions at the symmetry boundary. Specifically, topological operators with Dirichlet boundary conditions on the symmetry boundary are interpreted as ``condensed'' and become local operators in \(\mathcal{T}_d\), while those with Neumann boundary conditions generate global symmetries. Within this framework, different phases of \(\mathcal{T}_d\) correspond to different choices of boundary conditions on the dynamical boundary, which are classified by patterns of anyon condensation in \(\mathcal{T}^{\rm Sym}_{d+1}\) -- an approach known as the \emph{categorical Landau paradigm}\cite{Bhardwaj:2023fca,Bhardwaj:2023idu,Bhardwaj:2023bbf,Hai:2023osv,Bhardwaj:2024qrf,Kong:2024ykr,Chen:2024ulc,Bhardwaj:2024wlr,Bhardwaj:2024qiv,Bhardwaj:2025piv}. 

For a \((1+1)\)D theory with global symmetry group \(G\), different phases can be classified via anyon condensation in a \((2+1)\)D topological order described by the Drinfeld center \(\mathcal{Z}(\mathcal{S})\), where the symmetry category is \(\mathcal{S} = \mathrm{Vec}_G\) (or more generally \(\mathcal{S} = \mathrm{Vec}_G^\omega\) in the presence of an anomaly). Anyon condensation in \(\mathcal{Z}(\mathcal{S})\) is classified by condensable algebras within \(\mathcal{Z}(\mathcal{S})\). Gapped phases\cite{KLgappedboundary,Cong2017,Bhardwaj:2023idu,Xu:2022rtj,Hai:2023osv} correspond to the condensation of \emph{Lagrangian} algebras, whereas gapless phases -- either describing critical points between gapped phases or stable gapless phases -- arise from the condensation of \emph{non-Lagrangian} condensable algebras~\cite{Kong2018gapless,Kong2021gapless,Bhardwaj:2024qrf,Chen:2024ulc,huang2025topologicalholography,Wen2025igspt}. This categorical Landau framework has been successfully applied to a wide range of one-dimensional systems, where the associated SymTFT often imposes strong constraints on the conformal field theories that can appear\cite{Ji_2019,Ji:2019jhk,ji2022unifiedview,Chatterjee_2023,Chatterjee2023shadow,chatterjee2024emergent,Chatterjee2024noninvertible}. Extensions of this approach to higher dimensions are actively being explored\cite{Xu2023,Luo2023,Chen2024}. Altogether, SymTFT offers a promising and unifying framework for organizing our understanding of phases and phase transitions in systems with global symmetries, including both gapped and gapless cases.

While most existing literature focuses on global symmetries in closed quantum systems described by pure states, in this work we broaden the scope to include systems in mixed states, where two distinct notions of symmetry -- \emph{strong} and \emph{weak} symmetries -- naturally arise~\cite{Buca2012,Albert2014,Albert2018,Lieu2020,Rakovszky2024}. For a given mixed state \(\rho\), a \emph{strong} (or exact) symmetry is one that acts coherently on each pure-state component of \(\rho\), satisfying $U_g \rho = \phi(g)\, \rho, \quad \phi: G \to U(1)$,  
while a \emph{weak} (or average) symmetry requires invariance only under conjugation: $U_g \rho U_g^\dagger = \rho, \quad \forall g \in G$. Motivated by rapid advances in quantum simulation platforms -- where decoherence often drives systems away from pure states -- there has been growing interest in characterizing phases of matter in open or noisy environments. The behavior of topologically ordered states in open quantum systems is an active area of research\cite{Chen2024separability,Ramanjit2025noisy,Sang2024renormalization,Ellison:2024svg, sang2025mixedstatephases, Sang2025stability, yang2025topologicalmixedstates,Zhang2025swssb1form,lessa2025higherformanomaly,wang2024intrinsic,kikuchi2024anyon}. In addition, recent studies reveal that the landscape of mixed-state phases with symmetries is much richer than its pure-state counterpart. For instance, symmetry-protected topological (SPT) phases have natural extensions to mixed states, giving rise to the notion of \emph{average SPT} (ASPT) phases~\cite{deGroot:2021vdi,Ma:2022pvq, Zhang:2022jul, Lee:2022hog, Ma2025, Zhang2023fracton, Ma:2024kma, xue2024tensornetworkformulationsymmetry, Guo2025lpdo,Sun:2024iwm,Lessa2025multientanglement}. Notably, some ASPT phases have no pure-state analog and are termed \emph{intrinsic ASPTs}\cite{Ma2025}. Likewise, new patterns of spontaneous symmetry breaking emerge, such as \emph{strong-to-weak SSB} (SWSSB)\cite{Lee2023swssb, Lessa:2024gcw,Sala:2024ply,Gu:2024wgc,Shah:2024gzd,zhang2024fluctuationdissipationtheoreminformationgeometry, Guo:2024ecx,kim2024errorthresholdsykcodes,Kuno:2025dqf,Stephen2025, Huang2025,Liu2025wightman,Weinstein2025wightman}, where a strong symmetry is spontaneously broken while its weak counterpart remains preserved. In this work, we aim to develop a SymTFT framework for a systematic understanding of these mixed-state phases, thereby offering a unified topological perspective on symmetry and phase structure in open quantum systems.

In particular, we propose that the SymTFT for an open quantum system -- referred to as the \emph{open SymTFT} -- with global symmetry \(G\), which may include both strong and weak (average) components, is given by the Drinfeld center of a doubled symmetry category, \(\mathcal{Z}(\mathrm{Vec}_{G \times \overline{G}})\), where \(\overline{G}\) denotes a second copy of \(G\).
This doubled structure arises naturally from viewing the density matrix as a state in a doubled Hilbert space via the operator-state correspondence, making it well-suited for encoding the symmetry properties of mixed states\cite{Ma:2024kma,xue2024tensornetworkformulationsymmetry,lu2024bilayer}. Classifying mixed-state phases in this framework amounts to studying the set of condensable algebras \(\mathcal{A}\) within \(\mathcal{Z}(\mathrm{Vec}_{G \times \overline{G}})\). However, not all such algebras correspond to physically realizable phases: the density matrix of a mixed state must satisfy Hermiticity and positivity\cite{Duivenvoorden2017,Ma:2024kma, xue2024tensornetworkformulationsymmetry}, which impose nontrivial constraints on allowable condensation patterns. We will describe how these physical constraints can be interpreted and implemented in the language of anyon condensation. We carry out a detailed analysis of the Lagrangian algebras in \(\mathcal{Z}(\mathrm{Vec}_{G \times \overline{G}})\), and show how they organize into familiar pure-state phases such as symmetry-breaking and symmetry-protected topological phases, as well as more inherent mixed-state phases, including SWSSB and ASPT phases. With the open SymTFT construction, we can also study the phase transition points between mixed-state phases. In particular, transitions between SWSSB phases and either trivial symmetric phases or fully SSB phases are described by non-Lagrangian algebras within the SymTFT framework. These critical points correspond to conformal field theories (CFTs) defined in the doubled Hilbert space. As a result, both order and disorder parameters exhibit power-law correlations at criticality. However, certain one of them must be measured in the doubled space, through Wightman correlation functions\cite{Liu2025wightman,Weinstein2025wightman}, in order to capture the critical behavior.

The rest of the paper is organized as follows. In Sec.~\ref{sec:gf}, we formulate the rules for constructing condensable algebras in open SymTFTs, emphasizing the constraints imposed by Hermiticity and positivity of the density matrix. Combined with the modularity condition, these lead to an algorithm for identifying admissible condensable algebras in mixed-state settings. In Sec.~\ref{sec:example}, we illustrate this framework using several examples, including discrete abelian groups \( G = \mathbb{Z}_2 \), \( \mathbb{Z}_4 \), \( \mathbb{Z}_2 \times \mathbb{Z}_2 \), and non-abelian groups \( G = S_3 \), \( D_4 \). We classify their gapped phases and identify instances of SWSSB and (intrinsic) ASPT phases. In Sec.~\ref{sec:pt}, we analyze phase transitions between mixed-state phases, both from the SymTFT perspective and via explicit lattice models. We highlight transitions involving SWSSB phases and comment on the possibility of intrinsic gapless mixed-state phases. In Sec. \ref{sec:Gaugingopen}, we also discuss the role of gauging in this framework, which provides a unified perspective for relating different mixed-state phases of matter. In Sec.~\ref{sec:eft}, we develop an effective field theory approach motivated by the SymTFT construction and derive ASPT response functions. Within the effective field theory approach, we also discuss the generalization to continuous $U(1)$ symmetry. 

Before proceeding, we offer a clarifying remark. Throughout this work, we adopt the sandwich (or slab) construction, in which the topological boundaries are chosen to be charge-condensed (except in the discussion of gauging). This setup ensures that the resulting effective \((1+1)\)D system possesses a global \( G \) symmetry, with well-defined local order parameter fields. The corresponding dual symmetry \( \hat{G} \), which naturally arises in the SymTFT framework, does not admit local order parameters in this construction. To avoid confusion, we will simply refer to \( G \) as the global symmetry of the system throughout the main text.

\section{General Formulation}\label{sec:gf}
In this section, we present the main result of this work: a set of rules governing anyon condensation in the open SymTFTs. We will focus on the case of a \((1+1)\)D quantum system, where the associated SymTFT lives in a \((2+1)\)D bulk. Throughout this work, we take the global symmetry group \(G\) to act as a strong symmetry. Systems with weak (or average) symmetry will be understood via the mechanism of strong-to-weak symmetry breaking, which will be the result of certain anyon condensation.

To formulate the open SymTFT, we begin by applying the canonical purification procedure, which maps a mixed state described by a density matrix \(\rho\) to a pure state in a doubled Hilbert space. Suppose the density matrix $\rho$ has eigen decomposition as $\rho=\sum_ip_i\ket{\psi_i}\bra{\psi_i}$, then the canonical purification is given by 
\begin{equation}
    |\rho^c\rangle\!\rangle=\sum_i \sqrt{p_i}\ket{\psi_i}\ket{\bar{\psi}_i}
\end{equation}
With canonical purification, a density matrix acting on a Hilbert space \(\mathcal{H}\) is mapped to a pure state in the doubled Hilbert space \(\mathcal{H} \otimes \mathcal{H}^*\). Under this mapping, a symmetry \(G\) of the original system is promoted to a doubled symmetry \((G \times \overline{G}) \rtimes \mathbb{Z}_2\), where \(\overline{G}\) acts on the ancillary copy \(\mathcal{H}^*\). The additional \(\mathbb{Z}_2\) factor, denoted by \(J\), is an anti-unitary symmetry that exchanges \(\mathcal{H}\) and \(\mathcal{H}^*\), and it reflects the Hermitian nature of the density matrix. Another essential property of a density matrix is that it is positive semidefinite -- that is, all eigenvalues \(p_i \geq 0\). This ensures, in particular, that fractional powers of the density matrix are well-defined.

\begin{figure}[!htbp]
    \centering
    \includegraphics[width=0.8\linewidth]{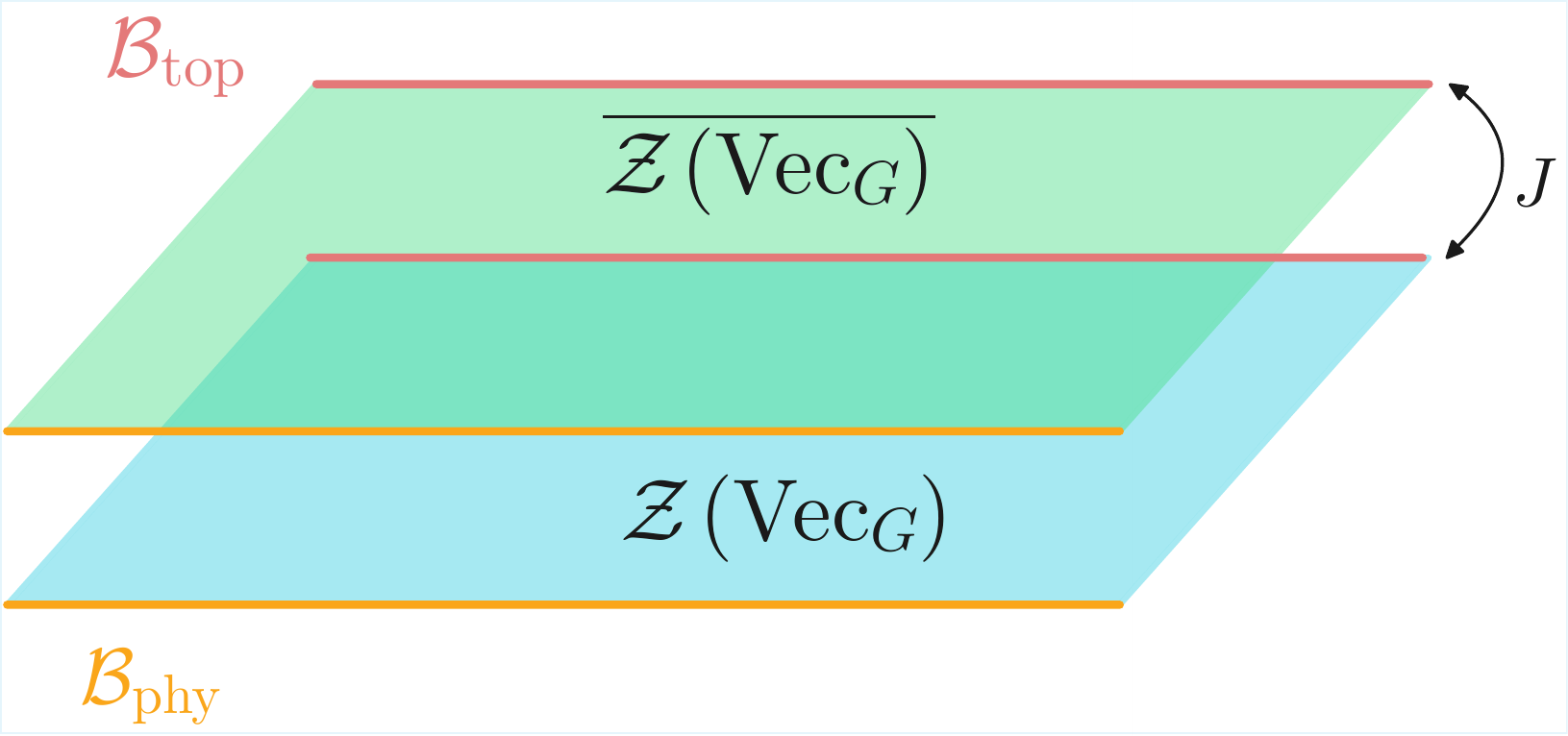}
    \caption{Open SymTFT}\label{fig:opensymTFT}
\end{figure}

We then construct a bulk \((2+1)\)D SymTFT that encodes the doubled symmetry structure via the Drinfeld center $\CZ({\rm Vec}_{G\times \overline{G}})\cong \CZ({\rm Vec}_{G})\boxtimes \overline{\CZ({\rm Vec}_{G})}$, as shown in Fig. \ref{fig:opensymTFT}. In this construction, we do not promote the \(\mathbb{Z}_2\) symmetry \(J\) -- which exchanges the original Hilbert space \(\mathcal{H}\) and its dual \(\mathcal{H}^*\) -- to a full TQFT. There are two key reasons for this. First, from a technical standpoint, it remains unclear how to consistently incorporate anti-unitary symmetries such as \(J\) into a bulk topological field theory. Second, \(J\) reflects a structural constraint of the density matrix: it encodes Hermiticity and, as such, is never spontaneously broken. Since there are no distinct phases associated with \(J\), it does not need to be dynamically realized in the bulk. Instead, \(J\) is treated as an exchange symmetry between the two copies of the theory arising from the doubled Hilbert space.

Since the density matrix is mapped to a pure state in the doubled Hilbert space, its phases can once again be characterized using the SymTFT in the doubled space, via anyon condensation on the topological and physical boundaries. 
(In Appendix~\ref{app:CatLanPar}, we briefly review the basic framework of the categorical Landau paradigm and the classification of phases based on anyon condensation in pure states.) However, due to the Hermitian and positive semidefinite nature of the density matrix, not all condensation patterns in the open SymTFT are physically allowed. These conditions impose additional constraints on the possible anyon condensations, other than the standard requirement of modularity, which we derive in the following section. 

As an aside, tracing out one copy of the doubled Hilbert space yields a reduced density matrix that serves as the open-system analog of the sandwich construction. In this setup, the bulk realizes a mixed-state topological order~\cite{Ramanjit2025noisy,Sang2024renormalization,Ellison:2024svg, sang2025mixedstatephases, Sang2025stability, yang2025topologicalmixedstates,Zhang2025swssb1form,lessa2025higherformanomaly}. This reduced density matrix represents the physical state of the open quantum system.

In this section, we focus on gapped phases. However, we caution the reader about the use of this terminology. In our context, a ``gapped phase" refers specifically to a phase of matter realized as a gapped boundary condition\cite{KLgappedboundary} of the open SymTFT. As we will see through examples, this definition encompasses both symmetric phases and those exhibiting spontaneous symmetry breaking. This usage of ``gapped" differs from more recent definitions of gapped mixed-state phases~\cite{Sang2025stability, Ma2025,yang2025topologicalmixedstates}, which have primarily been used to describe mixed-state generalizations of short-range correlation in pure states. From the examples we have studied, it appears that our notion of gapped phases includes, as a subset, the gapped mixed-state phases defined in earlier works. On the other hand, if a state is gapless according to our definition, it exhibits nontrivial correlations between local operators. By construction, such behavior is excluded by the definition of gapped mixed states used in previous work.

\subsection{J-symmetry and positivity}

The topological order in the doubled Hilbert space is given by \(\mathcal{Z}(\mathrm{Vec}_{G}) \boxtimes \overline{\mathcal{Z}(\mathrm{Vec}_{G})}\). We denote the simple anyons in the first copy by \(\{a\}\), and those in the second copy by \(\{\bar{a}\}\), such that the \(\mathbb{Z}_2\) symmetry \(J\) exchanges \(a \leftrightarrow \bar{a}\). As a result, \(a\) and \(\bar{a}\) carry opposite topological spins, reflecting the anti-unitary nature of the \(J\)-symmetry. To make the discussion more concrete, it is helpful to keep a simple example in mind: the case where \(G = \mathbb{Z}_2\). In this setting, the open SymTFT corresponds to two copies of the toric code model, with anyon content given by \(\{1, e, m, f\} \boxtimes \{1, \bar{e}, \bar{m}, \bar{f}\}\).

The constraint imposed by the \(J\)-symmetry can now be understood more transparently. Suppose we condense a composite anyon of the form \(a\bar{b}\), where \(a\) is an anyon in \(\mathcal{H}\) and \(\bar{b}\) is an anyon in \(\mathcal{H}^*\). For \(a\bar{b}\) to be condensable, it must be a boson, which implies that \(a\) and \(\bar{b}\) must have opposite topological spins. However, invariance under the \(J\)-symmetry requires that the state also condense the mirror anyon \(b\bar{a}\); otherwise, the condensation pattern would explicitly break \(J\). This symmetry condition already eliminates a large class of condensable algebras otherwise allowed in the system. In the case of $G=\bbZ_2$, we are left with only 4 possible Lagrangian algebras:
\begin{align}
    &\mathcal{L}_S=1\oplus e\oplus \bar{e} \oplus e\bar{e} \\ 
    \label{eq:Z2C1}
    &\CA_1=1\oplus m\oplus \bar{m}\oplus m\bar{m} \\
    \label{eq:Z2C2}
    &\CM_1=1\oplus e\bar{e}\oplus m\bar{m}\oplus f\bar{f} \\
    \label{eq:Z2C3}
    &\CM_2=1\oplus e\bar{m}\oplus m\bar{e}\oplus  f\bar{f}
\end{align}

However, among these four choices, only the first three correspond to physical density matrices; the fourth is incompatible with the positivity constraint required of a valid density matrix. In particular, we will argue that the condensation pattern \(\mathcal{M}_2\) is not consistent with positivity, either in the bulk or on the boundary of the SymTFT. 

Consider a situation where the anyon \(e\bar{m}\) condenses in the SymTFT. This implies that the expectation value of the corresponding open string operator remains nonzero at long distances, i.e.,
\begin{equation}
    \Big|\langle\!\langle \rho^c \big| W_e^c(\gamma_{ij}) \otimes W_{\bar{m}}^c(\gamma_{ij})^* \big| \rho^c \rangle\!\rangle\Big|^2 \sim O(1),
    \label{eq:anyoncondensation}
\end{equation}
where \(|\rho^c\rangle\!\rangle\) is the wavefunction of the doubled SymTFT, which can be understood as the canonical purification of the original density matrix. Here, \(\gamma_{ij}\) denotes the path of the open string connecting points \(i\) and \(j\). If the anyon \(e\bar{m}\) condenses in the bulk, then Eq.~\ref{eq:anyoncondensation} should hold for arbitrary positions \(i\) and \(j\) in the bulk. In contrast, if the condensation occurs at the boundary, the endpoints \(i\) and \(j\) lie on the boundary. We now consider the implications of Eq.~\ref{eq:anyoncondensation}, given that \(|\rho^c\rangle\!\rangle\) is the canonical purification of a density matrix. 

\begin{figure}[!htbp]
    \centering
    \includegraphics[width=0.8\linewidth]{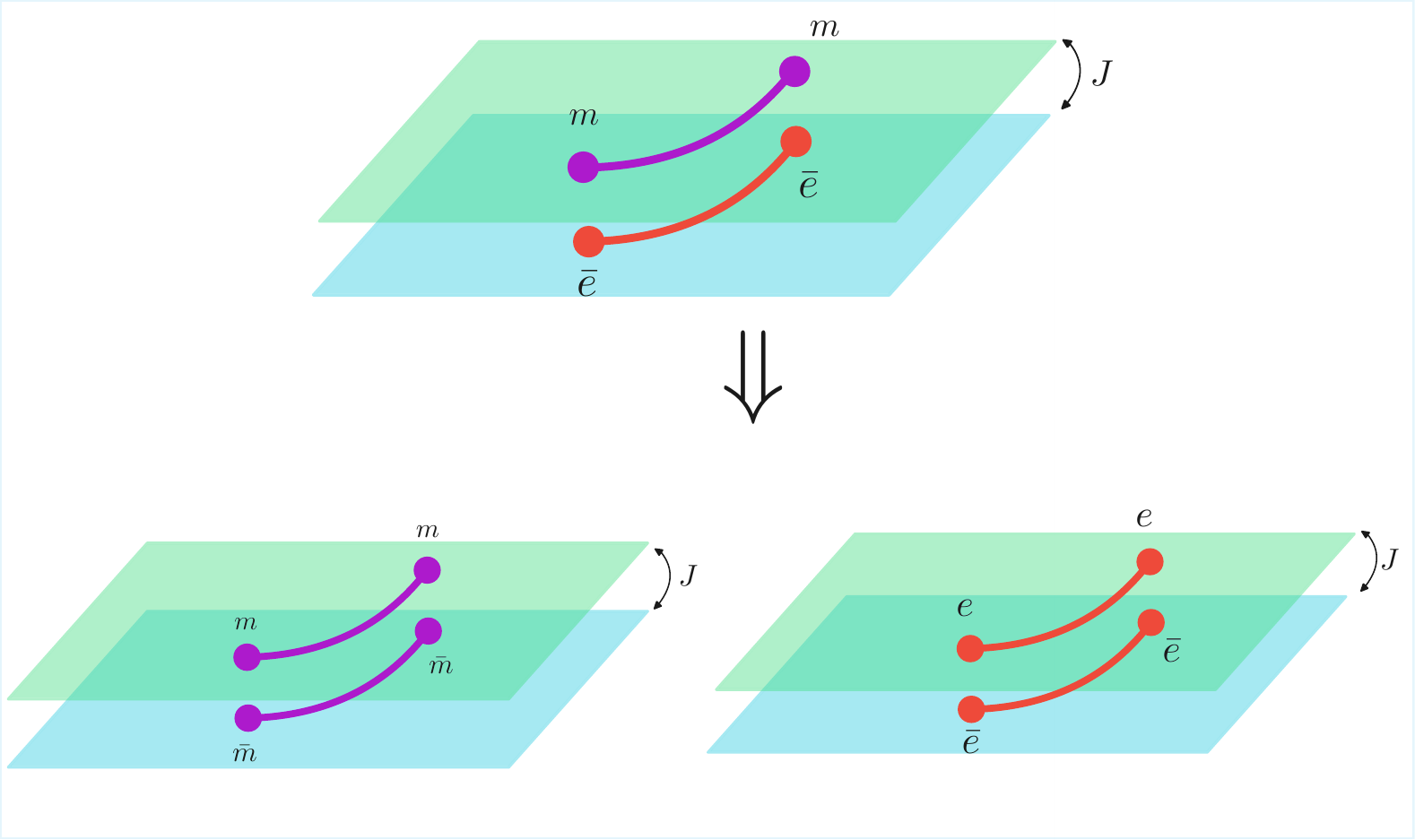}
    \caption{Constraints on condensation patterns due to positivity: if \(m\bar{e}\) condenses, then positivity requires that \(e\bar{e}\) and \(m\bar{m}\) must also be included in the condensation.}\label{fig:positivity}
\end{figure}

Through the positivity of density matrices and the Cauchy–Schwarz inequality, we find the following inequality holds\cite{Duivenvoorden2017,Ma:2024kma, xue2024tensornetworkformulationsymmetry} :
\begin{align}
    O(1) &\sim \Big|\langle\!\langle \rho^c|W_e^c(\gamma_{ij})\otimes W_{\bar{m}}^c(\gamma_{ij})^*|\rho^c\rangle\!\rangle\Big|^2 \\
    &=\Big|\Tr (\sqrt{\rho}W_e\sqrt{\rho}W_m^\dagger)\Big|^2 \\
    &=\Big|\Tr (\rho^{1/4}W_e\rho^{1/4}\rho^{1/4}W_m^\dagger\rho^{1/4})\Big|^2 \\ \nonumber
    &\leq \Tr [(\rho^{1/4}W_e\rho^{1/4}) (\rho^{1/4}W_e\rho^{1/4})^\dagger)] \\ 
    &\times \Tr [(\rho^{1/4}W_m\rho^{1/4})(\rho^{1/4}W_m\rho^{1/4})^\dagger] \\ \nonumber
    &= \langle\!\langle \rho^c|W_e^c(\gamma_{ij})\otimes W_{\bar{e}}^c(\gamma_{ij})^*|\rho^c\rangle\!\rangle \\
    &\times \langle\!\langle{\rho^c}|W_m^c(\gamma_{ij})\otimes W_{\bar{m}}^c(\gamma_{ij})^* |{\rho^c}\rangle\!\rangle
    \label{eq:positivity}
\end{align}
where the states and operators with $^c$ label are in the double space, and the ones without $^c$ label are in the original Hilbert space. This inequality implies that if \(e\bar{m}\) condenses, then \(e\bar{e}\) and \(m\bar{m}\) must also condense in the doubled system due to the positivity requirement of the density matrix, as shown in Fig. \ref{fig:positivity}. However, one can observe that the set \(\{e\bar{m}, m\bar{e}, e\bar{e}, m\bar{m}\}\) cannot be in the same Lagrangian algebra, as there is nontrivial mutual braiding among these anyons. This indicates a tension between the positivity constraint and the modular invariance required for consistent anyon condensation. As a result, \(\mathcal{M}_2\) cannot be realized as a physically admissible condensable algebra in this setting.

\begin{figure}[!htbp]
    \centering
    \includegraphics[width=0.8\linewidth]{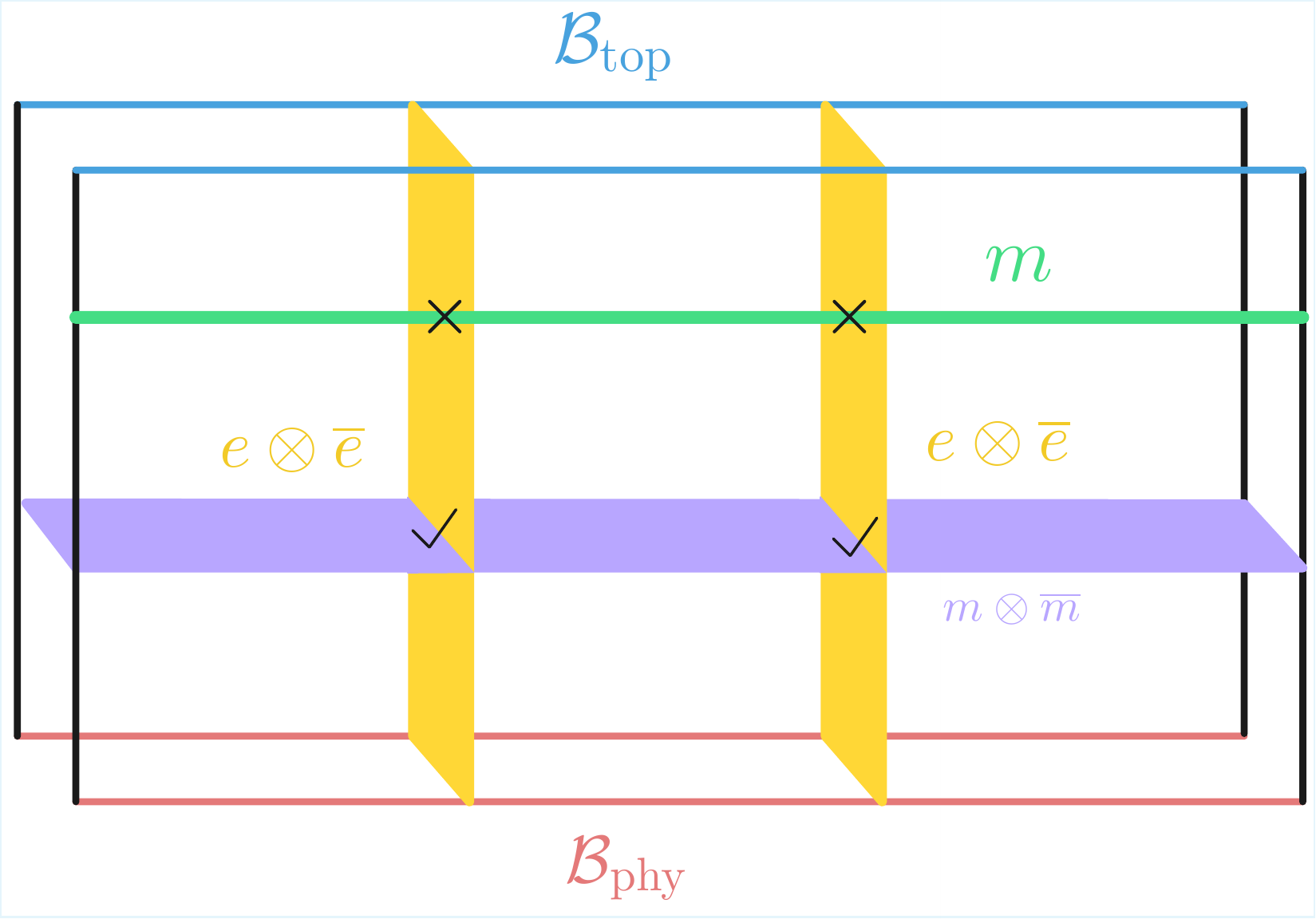}
    \caption{SWSSB and Wightman correlation function of $W_{e\bar{e}}$. $W_{e\bar{e}}$ is a good local order parameter for SWSSB as it is invariant under weak symmetry but charged under strong symmetry}\label{fig:Z2SWSSB}
\end{figure}

Looking back at the allowed condensation patterns in Eqs.~\ref{eq:Z2C1}, \ref{eq:Z2C2}, and \ref{eq:Z2C3}, the first two are straightforward to interpret: they correspond to coherent anyon condensations within the original system. For instance, \(\mathcal{L}_S = (1\oplus  e)\otimes (1\oplus  \bar{e})\) is simply a product of independent condensations in each copy. These phases are smoothly connected to those of pure-state systems: \(\mathcal{L}_S\) corresponds to complete \(\mathbb{Z}_2\) symmetry breaking, while \(\mathcal{A}_1\) represents the symmetric phase. The third case, however, is more intriguing, as it involves the condensation of anyons that couple the two copies of the system. Therefore, it corresponds to a phase that is only realizable in a mixed-state setting, hence the label \(\mathcal{M}\). This pattern corresponds to a spontaneous strong-to-weak symmetry breaking (SWSSB) of the \(\mathbb{Z}_2\) symmetry. We can make this more explicit using the sandwich construction to compute the Wightman correlator, as illustrated in Fig.~\ref{fig:Z2SWSSB}. The condensation pattern implies that the following quantity is nonzero:
\[
\begin{aligned}
&\Big|\langle\!\langle \rho^c \big| W_{e\bar{e}}(x) \otimes W_{e\bar{e}}(y) \big| \rho^c \rangle\!\rangle\Big|^2 \\
&= \mathrm{Tr}\left[\sqrt{\rho} \, W_e(x) W_e(y) \, \sqrt{\rho} \, W_e^\dagger(x) W_e^\dagger(y)\right] \sim O(1),
\end{aligned}
\]
where \(W_{e\bar{e}}\) is a short open string operator corresponding to the anyon \(e\bar{e}\), which terminates on both the topological and physical boundaries. $W_{e\bar{e}}$ serves as a local order parameter for the SWSSB as it is invariant under weak symmetry but charged under strong symmetry. According to Ref.~\cite{Liu2025wightman,Weinstein2025wightman}, Wightman correlators are reliable diagnostics of SWSSB.

Note that the structure of the positivity inequality is completely general. Assuming that an anyon \(a\bar{b}\) is condensed in the doubled space, the inequality implies that both \(a\bar{a}\) and \(b\bar{b}\) must also be condensed in the purified state \(|\rho^c\rangle\!\rangle\). (We also remark that this type of inequality applies more broadly to long-range order in local operators as well.) We also note that the discussion applies to both Abelian and non-Abelian symmetries. In the non-Abelian case, one generally needs to consider open ribbon operators~\cite{KitaevTC2003,Ribbon2008,Ribbon2017}. A subtlety with non-Abelian ribbon operators is that multiple distinct operators can correspond to the same anyon type, which is labeled by a conjugacy class and an irreducible representation of its centralizer. These different operators reflect local degrees of freedom at the endpoints of the open ribbon. However, one can always form a suitable linear combination of ribbon operators within the same anyon class to construct a unit ribbon operator, in which the endpoint degrees of freedom fuse to the trivial sector\cite{Ribbon2008}. 

Next, we describe a systematic procedure for identifying positivity-allowed condensable algebras in open SymTFTs. To do so, we will introduce some necessary notation and outline an algorithm for determining such admissible condensable algebras.
 

\subsection{Mixed-state-allowed Lagrangian algebra}
We begin by reviewing some basic notions of topological orders and condensable algebras. We consider bosonic topological orders described by a modular tensor category (MTC), which is characterized by modular \(S\) and \(T\) matrices. The \(S\) matrix encodes information about anyon braiding, while the \(T\) matrix captures the topological spins of the anyons. For clarity, we denote these matrices as \(S^{\mathrm{top}}\) and \(T^{\mathrm{top}}\) to emphasize their topological nature.

Anyon condensations in an MTC are classified by condensable algebras -- that is, commutative separable algebras in the category~\cite{etingoftencat}. A particularly important class is the Lagrangian algebra, defined as a maximal condensable algebra: adding any additional anyon to it would make it non-condensable. Every Lagrangian algebra takes the form
\be
\mathcal{L} = \bigoplus_{\alpha} N_\alpha \, \alpha\,,
\ee
where \(\alpha\) labels simple anyons and \(N_\alpha\in\mb{N}\) are multiplicities. It satisfies the quantum dimension constraint
\be
\sum_{\alpha} N_\alpha d_\alpha ={\rm Qdim}(\CZ(\CC)) \equiv \sqrt{\sum_{\alpha} d_\alpha^2}\,,
\ee
where \(d_\alpha\equiv S^{\rm top}_{1\alpha}/S^{\rm top}_{11}\) denotes the quantum dimension of \(\alpha\), and ${\rm Qdim}(\CZ(\CC))$ is referred to as the quantum dimension of the MTC.

Gapped phases can be characterized by condensing Lagrangian algebras on the physical boundary. There is an efficient way for finding all possible Lagrangian algebras using the modular \(S\) and \(T\) matrices~\cite{Ji:2019jhk}. Consider putting the SymTFT on a solid torus, on the boundary $T^2$ the TQFT generates a Hilbert space $\CH(T^2)$ spanned by anyon basis, see Appendix.\ref{app:Gauging}. Each Lagrangian algebra \(\mathcal{L} = \bigoplus_{\alpha} N_\alpha \alpha\) corresponds to a partition vector in $\CH(T^2)$:
\be
\mathcal{L} = \bigoplus_{\alpha} N_\alpha \alpha \quad \mapsto \quad \ket{\mathcal{L}} = \sum_\alpha N_\alpha \ket{\alpha} \, .
\ee
Modular invariance of the SymTFT dictates that the partition vector \(\ket{\mathcal{L}}\) should be invariant under the action of both \(S^{\mathrm{top}}\) and \(T^{\mathrm{top}}\),
\be
S^{\mathrm{top}} \ket{\mathcal{L}} = T^{\mathrm{top}} \ket{\mathcal{L}} = \ket{\mathcal{L}} \, .
\ee
The integer solutions with the first entry equal to 1 correspond to all possible Lagrangian algebras in the system.

For a mixed state \(\rho\) with strong symmetry \(G\), we construct its SymTFT as the modular tensor category \(\mathcal{Z}(\mathrm{Vec}_{G \times \overline{G}})\), whose simple anyons are labeled as~\cite{COSTE2000679}
\begin{equation}
    a\bar{b} = (C_a, r_a) \otimes \overline{(C_b, r_b)}\,,
\end{equation}
where \(C_a\) and \(C_b\) are conjugacy classes in \(G\), and \(r_a \in \mathrm{Rep}(N(C_a))\), \(r_b \in \mathrm{Rep}(N(C_b))\) are irreducible representations of the corresponding centralizers in \(G\). The overline indicates that the anyons in the second copy are labeled by the inverses of those in the first copy, reflecting the complex conjugation inherent to the second factor. Using this notation, we can write a condensable algebra \(\mathcal{A}\) as
\be
\mathcal{A} = \bigoplus_{a,b} N_{a,b} \, a\bar{b} \, , \quad N_{a,b} \in \mathbb{Z}.
\ee

The mixed-state-allowed maximal condensable algebra ought to satisfy:
\begin{enumerate}
    \item Modularity. The condensable algebra shall be modular invariant.
    \item $J$-symmetry. The condensable algebra must be symmetric under $\{a\}\leftrightarrow \{\bar{a}\}$ permutation.
    \item Positivity. As previously derived, if $a\bar{b}$ is condensed, then $a\bar{a}$ and $b\bar{b}$ must also condense.
    \item Dimensionality. The dimension of Lagrangian algebra shall be equal to the quantum dimension of $\CZ(\CC)$.
    \item Stability. The condensable algebra shall satisfy a stability condition
    \be N_{a,b}N_{c,d} \le \sum_{e,f} \CN^{e}_{ac} \CN^{f}_{bd} N_{e,f} \ , \ee
    where $\CN$ is the fusion rule of $\CZ(\CC)$ (see \eqref{eq:fusionrule}). This condition ensures that the ground state degeneracy is invariant under perturbation\cite{Lan:2014uaa}.
    
\end{enumerate}
The data of a condensable algebra is conveniently encoded in the integer matrix \(N_{a,b}\), indexed by the anyon labels. The constraints on a valid condensable algebra translate into the following conditions on the matrix \(N\):
\begin{enumerate}
    \item Modularity: \(S N S^\dagger = T N T^\dagger = N\),
    \item J-Symmetry: \(N^T = N\),
    \item Positivity: If \(N_{i,j} \ne 0\), then \(N_{i,i} \ne 0\) and \(N_{j,j} \ne 0\),
    \item Lagrangian condition: \(\sum_{i,j} N_{i,j} \cdot \mathrm{dim}(i)\, \mathrm{dim}(j) = \mathrm{Qdim}(\CZ({\rm Vec}_{G\times\overline{G}}))\),
    \item Stability: \(N_{i,j} \cdot N_{k,l} \leq \sum_{p,q} \mathcal{N}^p_{i,k} \mathcal{N}^q_{j,l} N_{p,q}\),
\end{enumerate}
where \(S\) and \(T\) are the modular \(S\)- and \(T\)-matrices of \(\mathcal{Z}(\mathrm{Vec}_G)\), and \(\mathcal{N}^p_{i,k}\) are the fusion coefficients. We note rule~3 follows directly from our positivity condition, which constrains only the individual matrix elements of \(N\). Remarkably, in every admissible solution we have examined, the resulting matrix \(N\) is automatically positive-semidefinite. We conjecture that this property holds generally for all Lagrangian algebras in the open SymTFT.

While specifying $N_{a,b}$ is the most general way of stating the condensable algebra, for convenience in reading, we also write minimal generating set for condensable algebras when the matrices are too big, i.e. $\CA = \expval{a_1,a_2,\dots}$. 




Once a matrix solution \(N_{a,b}\) is obtained, the first step is to examine its first row, denoted by $N_{[1]}$ (or equivalently, the first column, due to the \(J\)-symmetry which enforces \(N\) to be symmetric). Define the total quantum dimension associated with the first row as
\be
\mathrm{dim}(N_{[1]}) = \sum_i N_{1,i} \, \mathrm{dim}(i).
\ee
If \(\mathrm{dim}(N_{[1]}) = \mathrm{Qdim}(\mathcal{Z}(\mathrm{Vec}_G))\), then the state can be interpreted as a coherent condensation of a Lagrangian algebra within the original system. In this case, the matrix \(N\) factorizes as \(N = N_{[1]} \otimes N_{[1]}^T\), which implies that the condensable algebra takes the form \(\mathcal{A} = \mathcal{A}_{[1]} \otimes \overline{\mathcal{A}_{[1]}}\), corresponding to independent condensations in each copy of the system. Such phases can therefore be understood as pure-state phases. 

On the other hand, if \(\mathrm{dim}(N_{[1]}) < \mathrm{Qdim}(\mathcal{Z}(\mathrm{Vec}_G))\), it indicates that the condensation must involve components that couple the two Hilbert spaces in a nontrivial way and cannot be realized within a pure-state setting. We thus refer to these as \emph{inherent mixed-state phases}. In these cases, we find mixed-state phases such as SWSSB and ASPT phases. In our formalism, ASPT phases always appear in conjunction with SWSSB, as the weak symmetry emerges through the mechanism of strong-to-weak symmetry breaking.

The SWSSB phases are relatively easy to identify, as they involve paired charge condensation of the form \(a\bar{a}\), without individual condensation of either \(a\) or \(\bar{a}\). The ASPT phases, on the other hand, can always be viewed as symmetry-protected topological phases in the doubled Hilbert space. However, not all SPTs in the doubled space correspond to physical ASPT density matrices; those that violate \(J\)-symmetry or positivity are ruled out. A quick diagnostic for ASPT phases is the presence of a decorated domain wall structure in the condensation pattern, which can often be matched to known decorated domain wall constructions of ASPTs. A more rigorous approach is to analyze symmetry localization at the boundary of the ASPT state. In such cases, the global symmetry is localized to the boundary, where it often acts via a projective representation. Finally, one can derive an effective field theory for the slab construction by considering the bulk SymTFT combined with fixed boundary conditions determined by the anyon condensation pattern. The resulting one-dimensional effective theory may exhibit a nontrivial topological response, providing another diagnostic for identifying ASPT phases.

\subsection{Realization of anyon condensations in open SymTFT}

The positivity inequality in Eq.~\ref{eq:positivity} provides a necessary condition for anyon condensation patterns to correspond to a physical, positive semidefinite density matrix. However, it is not immediately obvious that all condensation algebras satisfying Eq.~\ref{eq:positivity} will, in practice, yield positive density matrices. In this section, we argue that all such condensation patterns can be physically realized -- either through coherent anyon condensation, noise quantum channels, or a combination of both. This supports the conclusion that all solutions obtained through our procedure correspond to valid physical density matrices.

First, consider the coherent anyon condensation of a particular anyon \(a\). This can be implemented by modifying the system's Hamiltonian with a term of the form \(-\lambda_a K_a\), where \(K_a\) is a short ribbon operator that creates a pair of anyons \(a\) and \(a^{-1}\). In the limit of large \(\lambda_a\), the kinetic energy associated with the \(a\)-anyon is suppressed, leading to its condensation in the ground state. All pure-state phases can be realized through such coherent condensation processes. Importantly, all coherent condensations of anyons within a Lagrangian algebra do not interfere with one another, as the constituent anyons are mutually bosonic. As a result, their corresponding ribbon operators commute, allowing for simultaneous condensation.

For mixed-state phases, a natural first step is to apply coherent condensation to all anyons in the condensation pattern that can be realized in a pure-state setting. This reduces the problem to identifying the genuinely mixed-state components of the condensation pattern, which we will address next. Inherent mixed-state condensation patterns always fall into two distinct classes:  
1) \emph{Diagonal class}: condensation of pairs \(a\bar{a}\) without individual condensation of \(a\) or \(\bar{a}\), and  
2) \emph{Off-diagonal class}: condensation patterns involving \(\{a\bar{b}, b\bar{a}, a\bar{a}, b\bar{b}\}\), while none of the individual anyons \(a\), \(b\), \(\bar{a}\), or \(\bar{b}\) are condensed.

The diagonal class can be realized through a process we refer to as \emph{incoherent proliferation}, where the system is subjected to a noise channel that simultaneously generates a pair of anyons across the two copies. This process can be modeled by a quantum channel of the form \(\mathcal{E}_a(\rho) \sim K_a \rho K_a^\dagger\), where \(K_a\) is a short ribbon operator that creates an \(a a^{-1}\) pair in the original system. Under the operator-state mapping, this quantum channel corresponds to an imaginary-time evolution of the form \(\sim e^{-\beta K_a \otimes K_a^*}\) \footnote{Strictly speaking, the mapping between a quantum channel and an evolution operator is well-defined in the Choi–Jamiołkowski doubled-space representation. For canonical purification, this correspondence is less straightforward. Nonetheless, for fixed-point wavefunctions of topologically ordered states and in the strong decoherence limit, we expect quantum channels to induce similar effects on the canonically purified state.} acting on the purified state. In the strong-noise limit \(\beta \rightarrow \infty\), this evolution effectively projects onto states in which \(a\bar{a}\) is condensed. Crucially, the projector arising from the incoherent proliferation process commutes with the coherent anyon condensation procedure, as their corresponding short ribbon operators mutually commute.

For Abelian SymTFTs, we can go further and observe that the off-diagonal condensation patterns are automatically accounted for by combining coherent anyon condensation with incoherent proliferation. Specifically, suppose the condensable algebra includes \(\{a\bar{b}, b\bar{a}, a\bar{a}, b\bar{b}\}\). Then, since \(a\bar{b} \cdot b^{-1}\bar{b}^{-1} = ab^{-1}\bar{1}\), the fusion product implies that the anyon \(ab^{-1}\) must also be part of the condensable algebra. This observation shows that if we coherently condense \(ab^{-1}\) and apply incoherent proliferation for \(b\bar{b}\), the off-diagonal condensation \(a\bar{b}\) is automatically generated. Since we have already argued that both \(ab^{-1}\) and \(a\bar{a}\) condensations can be implemented through physically realizable processes -- coherent condensation and noise-induced proliferation -- it follows that all off-diagonal patterns in Abelian SymTFTs can be realized this way. Therefore, we conclude that all Abelian SymTFTs satisfying the positivity-inequality correspond to valid physical density matrices.

There is an alternative method for realizing off-diagonal condensation patterns that also applies to non-Abelian cases. This involves the so-called \emph{coherent noise channel}, defined as \(\mathcal{E}_{a+b}(\rho) \sim (K_a + K_b)\rho(K_a + K_b)^\dagger\), where \(K_a\) and \(K_b\) are short ribbon operators associated with anyons \(a\) and \(b\), respectively. In the non-Abelian case, one must take care to properly account for the boundary local degrees of freedom associated with ribbon endpoints. However, there exists a canonical (and often unique) choice of boundary conditions that resolves this ambiguity. When translated into the purified-state picture, this channel effectively projects onto the off-diagonal condensation pattern involving \(a\bar{b}\) and \(b\bar{a}\). This provides a general method for realizing such condensations beyond the Abelian setting.

With these arguments, we conclude that all condensable algebras passing our positivity screening correspond to physically realizable states in open quantum systems. Indeed, in the examples examined in the next section, we are able to identify all such states as allowed mixed-state phases, for both Abelian and non-Abelian symmetry groups.

\section{Examples of open SymTFTs and mixed-state gapped phases}\label{sec:example}
In this section we present several examples of the classification of mixed-state gapped phases using the open SymTFT. We give explicit examples of $1+1$D systems with $\bbZ_4$, $\bbZ_2\times \bbZ_2$, $S_3$ and $D_4$ symmetries, and analyze the symmetry patterns corresponding to each Lagrangian algebra. Due to length, the complete mixed-state phase classification of $D_4$ symmetry example is shown in Appendix.\ref{app:D4sym}. 

\subsection{Mixed-state gapped phases with $\bbZ_4$ symmetry}
A \(\mathbb{Z}_4\) group can be understood as a central extension of a \(\mathbb{Z}_2\) group by another \(\mathbb{Z}_2\), expressed as the short exact sequence:
\[
1 \rightarrow \mathbb{Z}_2^c \rightarrow \mathbb{Z}_4 \rightarrow \mathbb{Z}_2^q \rightarrow 1.
\]
The open SymTFT associated with \(\mathbb{Z}_4\) symmetry is given by the modular tensor category \(\mathcal{Z}(\mathrm{Vec}_{\mathbb{Z}_4 \times \overline{\mathbb{Z}_4}})\), which contains \(16 \times 16\) simple anyons labeled by
\[
\{ e^a m^b \bar{e}^c \bar{m}^d \} \ , \quad a,b,c,d \in \mathbb{Z}_4.
\]
Among the condensable algebras that satisfy the previously stated constraints, there are seven Lagrangian algebras. Notably, three of these correspond to inherent mixed-state phases.

\begin{figure}[!htbp]
    \centering
    \includegraphics[width=0.8\linewidth]{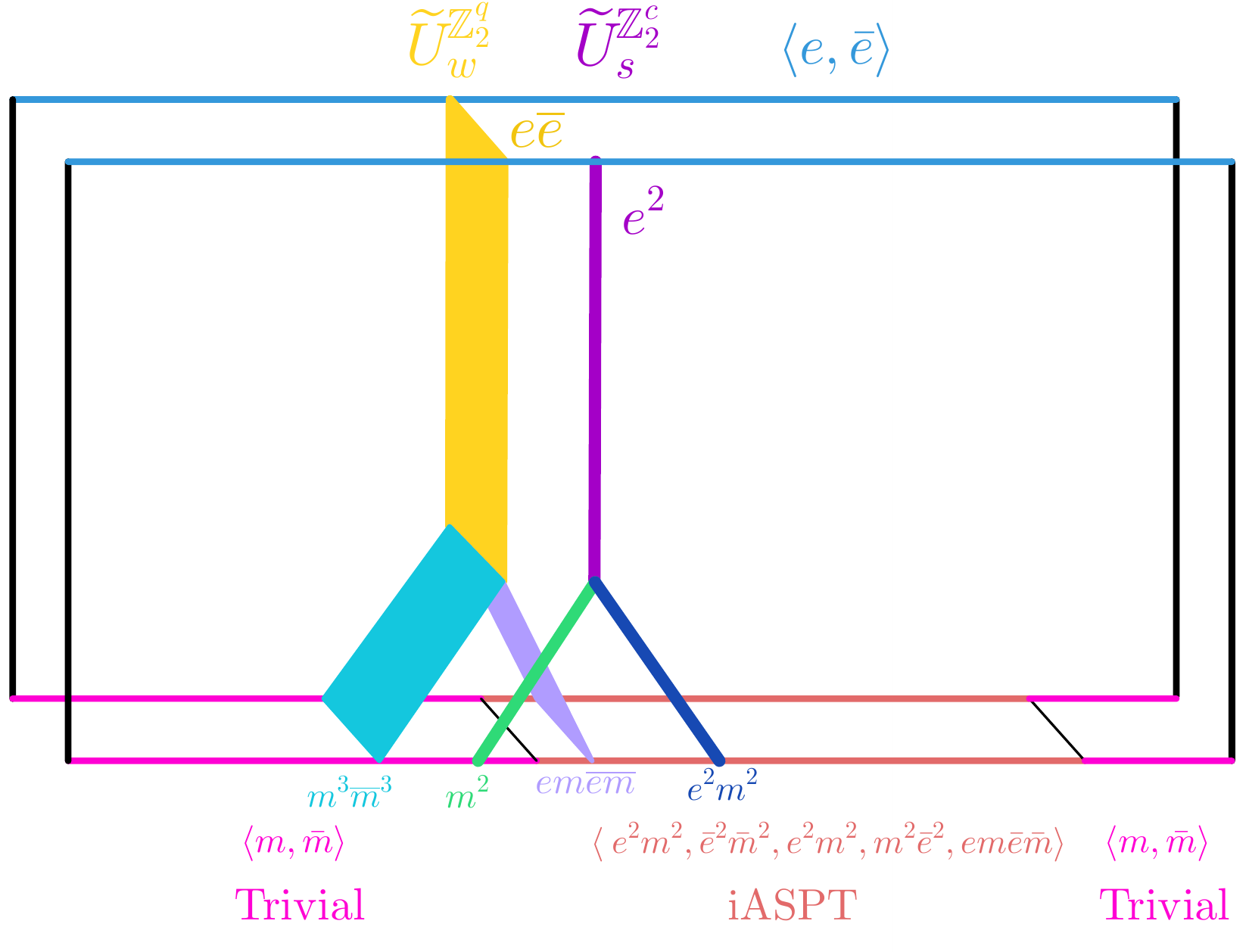}
    \caption{Symmetry localization and projective representation on the boundary of iASPT.}\label{fig:Z4}
\end{figure}

\begin{enumerate}
    \item The Lagrangian algebra generated by all charges condensed
    \[ \CL_S = \expval{e,\bar{e}}  \ .
    \]
    In the following, we assume it to be the topological boundary conditions. Then taking $\CL_S$ also as physical boundary condition would give pure-state phase of complete $\bbZ_4$ SSB;
    \item The pure-state phase of spontaneous symmetry breaking from $\bbZ_4^s\to \bbZ_2^{c,s}$ ($s$ denotes strong symmetries) is realized by choosing the physical boundary condition to be generated by \[ \CA_1 = \expval{e^2,\bar{e}^2 , m^2 ,\bar{m}^2 }\ ; \] 
    \item The pure-state phase of trivial symmetric phase of $\bbZ_4^s$ is realized by choosing the physical boundary condition to condense all the $m$ anyons carrying fluxes
     \[\CA_2 = \expval{m,\bar{m}} \ ; \]
    \item The mixed-state SWSSB phase preserving only $\bbZ_4^w$ ($w$ denotes weak symmetries) is realized by condensing all the diagonal anyons on the physical boundary,
    \[\CM_1 =  \expval{e\bar{e},m\bar{m}} \ ; \]
    \item The mixed-state SWSSB phase with only $\bbZ_2^{c,w}$ weak symmetry remains is realized by the condensable algebra
    \[ \CM_2 = \expval{e^2,e\bar{e},\bar{e}^2,m^2\bar{m}^2} \ ; \]
    \item The mixed-state SWSSB phase with weak $\bbZ_2^{q,w}$ and strong $\bbZ_2^{c,s}$ symmetry is given by choosing physical boundary condition to be
    \[\CM_3 = \expval{ m\bar{m},m^2,\bar{m}^2,e^2\bar{e}^2 } \ ; \]
    \item The mixed-state SWSSB phase, characterized by a weak \(\mathbb{Z}_2^{q,w}\) symmetry and a strong \(\mathbb{Z}_2^{c,s}\) symmetry, and supporting a nontrivial intrinsic ASPT phase, is realized by the following condensation on the physical boundary:
    \[\CM_4 = \expval{e^2\bar{e}^2,e^2\bar{m}^2,e^2m^2, m^2\bar{e}^2,\bar{e}^2\bar{m}^2,em\bar{e}\bar{m} }\ .\]
   In this Lagrangian algebra, we observe that the weak symmetry domain wall \(em\bar{e}\bar{m}\) carries a nontrivial charge under the strong symmetry, and similarly, the strong symmetry domain wall \(e^2m^2\) is charged under the weak symmetry. This mutual charging structure realizes the correct decorated domain wall pattern characteristic of this nontrivial intrinsic ASPT phase. The intrinsic nature of this phase is evident from the fact that no pure-state SPT phase exists in this symmetry class. Thus, realizing this topological phase fundamentally requires a mixed-state setting.

Another way to identify this phase as a nontrivial ASPT is by examining the localized symmetry action at the boundary between the ASPT phase and a trivial symmetric phase, as illustrated in Fig.~\ref{fig:Z4}. In the doubled space, the global symmetry of the system is actually \(\mathbb{Z}_4 \times \mathbb{Z}_2\), where the \(\mathbb{Z}_4\) factor is generated by the weak symmetry (denoted by $\tilde{U}_w^{\bbZ_2^q}$) and the \(\mathbb{Z}_2\) factor (denoted by $\tilde{U}_s^{\bbZ_2^c}$) arises from the strong symmetry. Following the approach of Ref.~\cite{Wen2025igspt}, we analyze the symmetry action along the boundary and find that it forms a projective representation of the \(\mathbb{Z}_4 \times \mathbb{Z}_2\) symmetry. This implies that the state realizes a nontrivial SPT phase in the doubled (purified) system, and hence corresponds to a nontrivial ASPT phase in the mixed-state setting.

\end{enumerate}

\subsection{Mixed-state gapped phases with $\bbZ_2^{(1)}\times \bbZ_2^{(2)}$ symmetry}
In this section, we present the mixed-state phases with \(\mathbb{Z}_2^{(1)} \times \mathbb{Z}_2^{(2)}\) symmetry. We denote by \(\mathbb{Z}_2^{(1,2)}\) the \(\mathbb{Z}_2\) subgroup generated by the product of the generators of \(\mathbb{Z}_2^{(1)}\) and \(\mathbb{Z}_2^{(2)}\).
The phases are obtained by calculating the condensable algebras of the open SymTFT $\CZ\big({\rm Vec}_{\bbZ_2^{(1)}\times \bbZ_2^{(2)}\times \overline{\bbZ_2^{(1)}\times \bbZ_2^{(2)}}}\Big)$, whose $16\times 16$ simple anyons are labeled by
\be \big\{ e_1^{a_1} m_1^{a_2} e_2^{a_3} m_2^{a_4} \bar{e}_1^{a_5} \bar{m}_1^{a_6} \bar{e}_2^{a_7} \bar{m}_2^{a_8}  \big\} \ , \ a_i\in\bbZ_2   \ee
There are in total $16$ phases with 10 inherent mixed-state phases:
\begin{enumerate}
    \item The Lagrangian algebra
    \[\CL_S = \expval{ e_1,e_2,\bar{e}_1,\bar{e}_2 }    \]
    is assumed to be the topological boundary condition. Choosing it to be also the physical boundary condition would yield the pure-state phase of complete $\bbZ_2^{(1)}\times \bbZ_2^{(2)}$ SSB;
    \item The pure-state phase of $\bbZ_2^{(1)}$ SSB, $\bbZ_2^{(2)}$ SSB, and $\bbZ_2^{(1,2)}$ SSB are realized by condensable algebras on the physical boundary
    \begin{equation*}\ba
    \CA_1 &= \expval{e_1,m_2,\bar{e}_1,\bar{m}_2}  \\
    \CA_2 &= \expval{e_2,m_1,\bar{e}_2,\bar{m}_1} \\
    \CA_3 &= \expval{e_1e_2,m_1m_2,\bar{e}_1\bar{e}_2,\bar{m}_1\bar{m}_2} \ , \ea
    \end{equation*} 
    respectively.
    \item The pure-state phase of trivial symmetric state is given by all $m$-anyon condensation on the physical boundary,
    \[\CA_4 = \expval{m_1,m_2,\bar{m}_1,\bar{m}_2} \ . \]
    \item The pure-state phase of cluster state SPT is given by the physical boundary condition
    \[ \CA_5 = \expval{e_1m_2, e_2m_1, \bar{e}_1\bar{m}_2,\bar{e}_2\bar{m}_1} \ ; \]
    \item The SWSSB phase with symmetry broken down to $\bbZ_2^{(1),w}\times \bbZ_2^{(2),w}$ symmetry is given by condensing all the diagonal anyons,
    \[\CM_1 = \expval{e_1 \bar{e}_1, m_1 \bar{m}_1, e_2 \bar{e}_2, m_2\bar{m}_2} \ ; \]
    \item The SWSSB phase with symmetry broken down to $\bbZ_2^{(1),w}$ symmetry, to $\bbZ_2^{(2),w}$ symmetry, and to $\bbZ_2^{(1,2)),w}$ symmetry are given by anyon condensation
    \be\ba \CM_2 &= \expval{e_2,\bar{e}_2,e_1 \bar{e}_1, m_1\bar{m}_1,} \\
     \CM_3 &= \expval{e_1,\bar{e}_1,e_2 \bar{e}_2, m_2\bar{m}_2,} 
 \\
   \CM_4 &= \expval{ e_1e_2,\bar{e}_1\bar{e}_2,e_1\bar{e}_1,e_2\bar{e}_2,m_1m_2\bar{m}_1\bar{m}_2 } \ , \ea\ee
    respectively.
    \item The SWSSB phase with symmetry broken down to $\bbZ_2^{(1),s}\times \bbZ_2^{(2),w}$, the SWSSB phase with $\bbZ_2^{(2),s}\times \bbZ_2^{(1),w}$ symmetry, and the SWSSB phase with$\bbZ_2^{(1,2),w}\times \bbZ_2^{(1),s}$ symmetry respectively,
    \be\ba \CM_5 &= \expval{m_1,\bar{m}_1,e_2\bar{e}_2,m_2 \bar{m}_2} \\
     \CM_6 &= \expval{m_2,\bar{m}_2,e_1\bar{e}_1,m_1 \bar{m}_1} \ . \\
     \CM_7 &= \expval{m_1m_2,m_1\bar{m}_2,\bar{m}_1m_2,e_1e_2\bar{e}_1\bar{e}_2} \ ;\ea\ee
    \item The SWSSB phase with $\bbZ_2^{(1),s}\times \bbZ_2^{(2),w}$ symmetry and ASPT, and its $\bbZ_2^{(2),s}\times \bbZ_2^{(1),w}$ symmetric ASPT, $\bbZ_2^{(1,2),w}\times \bbZ_2^{(1),s}$ symmetric ASPT counterparts are given by
    \be\ba \CM_8 = &\expval{ e_2m_1,\bar{e}_2\bar{m}_1,e_2\bar{e}_2,m_1\bar{m}_1,e_1\bar{e}_1m_2\bar{m}_2}  \\
     \CM_9 = &\expval{ e_1m_2,\bar{e}_1\bar{m}_2,e_1\bar{e}_1,m_2\bar{m}_2,e_2\bar{e}_2m_1\bar{m}_1} \\
     \CM_{10} = &\langle e_1e_2m_1m_2,\bar{e}_1\bar{e}_2\bar{m}_1\bar{m}_2,e_1e_2\bar{e}_1\bar{e}_2,\\ &e_2\bar{e}_2 m_2\bar{m}_2,\bar{e}_1\bar{m}_1e_2m_2,e_1m_1\bar{e}_2\bar{m}_2 \rangle   
     \ea\ee
    respectively.
   
\end{enumerate}

\subsection{Mixed-state gapped phases with $S_3$ symmetry}
\label{sec:MsGappedS3}
Here we present the example of allowed $S_3$ mixed state phases. The group $S_3$ is generated by
\be
S_3 = \expval{a,b|a^3=b^2=e,ab=ba^2} \ .
\ee
According to the anyon classification of $\CZ({\rm Vec}_{S_3})$, the anyons are labeled by
    \be\ba \label{eq:anyons3}\Big\{ &([e],r_+),([e],r_-),([e],E),([a],\varphi_0),\\ &([a],\varphi_1), ([a],\varphi_2),([b],\phi_0),([b],\phi_1)  \Big\} \ , \ea\ee
where $r_+$ is the trivial representation of $S_3$, $r_-$ is the non-trivial 1-dimensional representation of $S_3$ and $E$ is the 2-dimensional irreducible representation of $S_3$; $\varphi_k$ labels the 1-dimensional representations of $\bbZ_3$, and $\phi_k$ labels the 1-dimensional representations of $\bbZ_2$. 
The modular data of $\CZ({\rm Vec}_{S_3})$ is presented in Appendix.\ref{app:moddataS3}. There are $8$ Lagrangian algebras satisfying the positivity constraints with 4 inherent mixed-state phases:
\begin{enumerate}
    \item The condensable algebra composed of all charge anyons
    \[\label{eq:S3LS}
    \mathcal{L}_S =\begin{pmatrix}1 & 1 & 2 & 0 & 0 & 0 & 0 & 0 \\ 1 & 1 & 2 & 0 & 0 & 0 & 0 & 0 \\ 2 & 2 & 4 & 0 & 0 & 0 & 0 & 0 \\ 0 & 0 & 0 & 0 & 0 & 0 & 0 & 0 \\ 0 & 0 & 0 & 0 & 0 & 0 & 0 & 0 \\ 0 & 0 & 0 & 0 & 0 & 0 & 0 & 0 \\ 0 & 0 & 0 & 0 & 0 & 0 & 0 & 0 \\ 0 & 0 & 0 & 0 & 0 & 0 & 0 & 0\end{pmatrix}
    \]
is assumed to be the topological boundary. Choosing it also to be the physical boundary would give a phase that completely breaks the entire $S_3$ symmetry, and is a pure-state phase.
\item The phase labeled by condensable algebra
    \[\label{eq:S3A1}
    \CA^{\rm{SSB }}_1= \begin{pmatrix}1 & 0 & 1 & 0 & 0 & 0 & 1 & 0 \\ 0 & 0 & 0 & 0 & 0 & 0 & 0 & 0 \\ 1 & 0 & 1 & 0 & 0 & 0 & 1 & 0 \\ 0 & 0 & 0 & 0 & 0 & 0 & 0 & 0 \\ 0 & 0 & 0 & 0 & 0 & 0 & 0 & 0 \\ 0 & 0 & 0 & 0 & 0 & 0 & 0 & 0 \\ 1 & 0 & 1 & 0 & 0 & 0 & 1 & 0 \\ 0 & 0 & 0 & 0 & 0 & 0 & 0 & 0\end{pmatrix}
    \]
retains a strong $\bbZ_2$ subgroup, while the $\bbZ_3$ subgroup is completely broken. This phase is also a pure-state phase, and can be obtained by $\CL_S$ phase gauging $\bbZ_2\times \overline{\bbZ_2}$ symmetry. 
\item The phase labeled by condensable algebra
    \[\label{eq:S3A2}
    \CA^{\rm{SSB }}_2 =\begin{pmatrix}1 & 1 & 0 & 2 & 0 & 0 & 0 & 0 \\ 1 & 1 & 0 & 2 & 0 & 0 & 0 & 0 \\ 0 & 0 & 0 & 0 & 0 & 0 & 0 & 0 \\ 2 & 2 & 0 & 4 & 0 & 0 & 0 & 0 \\ 0 & 0 & 0 & 0 & 0 & 0 & 0 & 0 \\ 0 & 0 & 0 & 0 & 0 & 0 & 0 & 0 \\ 0 & 0 & 0 & 0 & 0 & 0 & 0 & 0 \\ 0 & 0 & 0 & 0 & 0 & 0 & 0 & 0\end{pmatrix}
    \]
preserves the strong $\bbZ_3$ subgroups while breaks the $\bbZ_2$ subgroup completely. This is a pure-state phase, and can be obtained by $\CL_S$ phase gauging $\bbZ_3\times \overline{\bbZ_3}$ symmetry. 
\item    The phase labeled by condensable algebra  \[\label{eq:S3A3}
    \CA^{\rm{SPT }}_3 =\begin{pmatrix}1 & 0 & 0 & 1 & 0 & 0 & 1 & 0 \\ 0 & 0 & 0 & 0 & 0 & 0 & 0 & 0 \\ 0 & 0 & 0 & 0 & 0 & 0 & 0 & 0 \\ 1 & 0 & 0 & 1 & 0 & 0 & 1 & 0 \\ 0 & 0 & 0 & 0 & 0 & 0 & 0 & 0 \\ 0 & 0 & 0 & 0 & 0 & 0 & 0 & 0 \\ 1 & 0 & 0 & 1 & 0 & 0 & 1 & 0 \\ 0 & 0 & 0 & 0 & 0 & 0 & 0 & 0\end{pmatrix}
    \]
preserves the entire $S_3$ strong symmetry, and is a pure-state phase. This phase can be obtained by $\CL_S$ gauging $S_3\times \overline{S_3}$ symmetry.
\item    The phase labeled by condensable algebra   \[\label{eq:S3A4}
    \CM^{\rm{SWSSB }}_1= \begin{pmatrix}1 & 0 & 0 & 0 & 0 & 0 & 0 & 0 \\ 0 & 1 & 0 & 0 & 0 & 0 & 0 & 0 \\ 0 & 0 & 1 & 0 & 0 & 0 & 0 & 0 \\ 0 & 0 & 0 & 1 & 0 & 0 & 0 & 0 \\ 0 & 0 & 0 & 0 & 1 & 0 & 0 & 0 \\ 0 & 0 & 0 & 0 & 0 & 1 & 0 & 0 \\ 0 & 0 & 0 & 0 & 0 & 0 & 1 & 0 \\ 0 & 0 & 0 & 0 & 0 & 0 & 0 & 1\end{pmatrix}
    \]
    possesses $S_3$ weak symmetry, and can be obtained by $\CL_S$ gauging $S_3^d$ diagonal symmetry.
\item   The phase labeled by condensable algebra \[\label{eq:S3A5}
    \CM^{\rm{SWSSB }}_2= \begin{pmatrix}1 & 0 & 0 & 1 & 0 & 0 & 0 & 0 \\ 0 & 1 & 0 & 1 & 0 & 0 & 0 & 0 \\ 0 & 0 & 0 & 0 & 0 & 0 & 0 & 0 \\ 1 & 1 & 0 & 2 & 0 & 0 & 0 & 0 \\ 0 & 0 & 0 & 0 & 0 & 0 & 0 & 0 \\ 0 & 0 & 0 & 0 & 0 & 0 & 0 & 0 \\ 0 & 0 & 0 & 0 & 0 & 0 & 1 & 0 \\ 0 & 0 & 0 & 0 & 0 & 0 & 0 & 1\end{pmatrix}
    \]
    has SWSSB of the $\bbZ_2$ subgroup while $\bbZ_3$ subgroup remains a strong symmetry.
\item In the phase labeled by condensable algebra    \[\label{eq:S3A6}
    \CM^{\rm{SWSSB }}_3 =\begin{pmatrix}1 & 0 & 1 & 0 & 0 & 0 & 0 & 0 \\ 0 & 1 & 1 & 0 & 0 & 0 & 0 & 0 \\ 1 & 1 & 2 & 0 & 0 & 0 & 0 & 0 \\ 0 & 0 & 0 & 0 & 0 & 0 & 0 & 0 \\ 0 & 0 & 0 & 0 & 0 & 0 & 0 & 0 \\ 0 & 0 & 0 & 0 & 0 & 0 & 0 & 0 \\ 0 & 0 & 0 & 0 & 0 & 0 & 1 & 0 \\ 0 & 0 & 0 & 0 & 0 & 0 & 0 & 1\end{pmatrix}
    \]
    the $\bbZ_2$ symmetry is broken to weak symmetry, while the $\bbZ_3$ symmetry is broken completely. This phase can be obtained by $\CL_S$ gauging $\bbZ_2^d$ diagonal symmetry.

\item In the phase labeled by condensable algebra    \[\label{eq:S3A7}
    \CM^{\rm{SWSSB }}_{4}= \begin{pmatrix}1 & 1 & 0 & 0 & 0 & 0 & 0 & 0 \\ 1 & 1 & 0 & 0 & 0 & 0 & 0 & 0 \\ 0 & 0 & 2 & 0 & 0 & 0 & 0 & 0 \\ 0 & 0 & 0 & 2 & 0 & 0 & 0 & 0 \\ 0 & 0 & 0 & 0 & 2 & 0 & 0 & 0 \\ 0 & 0 & 0 & 0 & 0 & 2 & 0 & 0 \\ 0 & 0 & 0 & 0 & 0 & 0 & 0 & 0 \\ 0 & 0 & 0 & 0 & 0 & 0 & 0 & 0\end{pmatrix}
    \]
the $\bbZ_3$ symmetry is SWSSB, while the $\bbZ_2$ symmetry is completely broken. This phase can be obtained by $\CL_S$ gauging $\bbZ_3^d$ diagonal symmetry.

\end{enumerate}

Naively, one might expect a gapped phase with the \(\mathbb{Z}_2\) subgroup strongly symmetric and the \(\mathbb{Z}_3\) subgroup exhibiting SWSSB. However, such a symmetry breaking pattern is not allowed. Suppose a group element \(n\) is weak and another element \(g\) is strong; then it follows that the conjugate \(ngn^{-1}\) must also be strongly symmetric. In the case of the \(S_3\) group, keeping the \(\mathbb{Z}_3\) (generated by rotations $a$) weak while treating the reflection element \(b\) as strong leads to a contradiction: conjugation of $b$ by elements of the \(\mathbb{Z}_3\) subgroup generates the entire \(S_3\) group. This implies that the full group must be strongly symmetric, which contradicts the intended symmetry breaking pattern.

Another way to see this is that in the $\bbZ_3$ weak $\bbZ_2$ strong pattern, the remaining group is not closed. Suppose we strive for such an unbroken subgroup $H$ of $S_3\times \overline{S_3}$ with $J$-symmetry, the generators should be chosen as
\be H= \expval{a\bar{a},b,\bar{b}}   \ , \ee
then it's easy to verify that $\Big((b)(a\bar{a})\Big)^2 = \bar{a}^2$, thus $a,\bar{a}$ also belong to this subgroup, rendering the subgroup exactly $S_3\times \overline{S_3}$. Following this argument, all $D_{2n+1}$ symmetric systems cannot have $\bbZ_{2n+1} = \expval{a}$ SWSSB when $\bbZ_2 = \expval{b}$ is strongly symmetric.

More generically, to see if a symmetry pattern of $A\subset G$ strongly symmetric and $B\subset G$ weakly symmetric ($A\cap B = \{e\}$), we simply consider the minimal $J$-symmetric subgroup $S\subset G\times \overline{G}$ containing $A\times \bar{A}$ and diagonal $B^d$ as its subgroup. If
\be \big| S \big| >   \big| A \big|^2\cdot \big| B \big| \ , \ee
then this symmetry pattern is not viable.

\subsection{Mixed-state gapped phases with $D_4$ symmetry}
The $D_4$ symmetric phases are more complicated and also more fruitful than previous examples. In Appendix.\ref{app:D4sym}, we list all $38$ gapped phases in this symmetry class, their symmetry-breaking patterns and their corresponding Lagrangian algebras. Among the $38$ gapped phases, there are $11$ pure-state phases and $27$ inherently mixed-state phases involving SWSSB. Among the $27$ mixed-state phases, there are $11$ phases exhibiting ASPTs, and $3$ of them are intrinsic ASPT phases. In the following, we will present three examples of ASPT phases.

To clarify the notations, we note the MTC $\CZ({\rm Vec}_{D_4})$ contains $22$ simple anyons labeled and denoted \cite{Bhardwaj:2024qrf}
\begin{widetext}\be\ba   &1 \equiv ([e],1) \ , \ e_R\equiv([e],1_b)\ , \ e_G\equiv([e],1_{ab})\ , \ e_B\equiv([a^2],1_a)\ , \ e_{RG}\equiv([e],1_a)\ , \ e_{GB}\equiv([a^2],1_b)\ , \ e_{RB}\equiv([a^2],1_{ab})\ , \ \\
&e_{RGB}\equiv([a^2],1)\ , \ m_R\equiv([ab],+-)\ , \ f_R\equiv([ab],--)\ , \  m_G\equiv([b],+-), f_G\equiv([b],--)\ , \ m_B\equiv([e],E)\ , \ f_B\equiv([a^2],E)\ , \ \\ &m_{RG}\equiv ([a],1)\ , \
f_{RG}\equiv([a],-1)\ , \ m_{GB}\equiv ([b],++)\ , \ f_{GB}\equiv ([b],-+)\ , \ m_{RB}\equiv ([ab],++)\ , \ f_{RB}\equiv ([ab],-+)\ , \ \\ &s_{RGB}\equiv ([a],i)\ , \ s^*_{RGB}\equiv ([a],-i)  \ , 
\ea\ee\end{widetext}
where $1,1_a,1_b,1_{ab},E$ are the irreducible representations of $D_4$ leaving their subscript conjugacy class invariant, $E$ is the $2$-dimensional representation, $1,i,-1,-i$ are the $4$ irreducible representations of $\bbZ_4$ and $++,+-,--,-+$ are the $4$ irreducible representations of $\bbZ_2\times \bbZ_2$.
respectively. The RGB nomenclature comes from the construction of $\CZ({\rm Vec}_{D_4})\cong \CZ({\rm Vec}_{\bbZ_2\times \bbZ_2\times \bbZ_2}^\omega)$ for a certain non-trivial $\omega\in H^3\big(\bbZ_2^{\times 3},U(1)\big)$.

The first example we present is the phase given by
\be
\CM_{11}^{\rm SWSSB}  = \expval{e_G,e_B,\bar{e}_G,\bar{e}_B,e_R\bar{e}_R,m_R\bar{m}_R,f_R\bar{f}_R} \ .
 \ee
This phase has SWSSB such that the remaining symmetry is $\bbZ_{2,a^2}^s\times \bbZ_{2,ab}^w$ and exhibits ASPT. The ASPT feature is observed through the decorated domain wall approach. The strong symmetry flux is decorated with weak symmetry charge
\be \ba e_B &= \big( [a^2],1_a  \big) \\ e_{GB} &= \big( [a^2],1_b  \big)  \ea\ee
where $1_a(ab) = -1$ and $1_b(ab) = -1$; and the weak symmetry flux is decorated with strong symmetry charge
\be  m_R\bar{m}_R = \big( [ab] ,+- \big)\otimes \overline{\big( [ab] ,+- \big)}   \ee
where $+-(a^2) = -1$. This phase can be viewed as a descendant from the pure-state SPT phase given by $\CA_{4}$ in Appendix.\ref{app:D4sym}, and therefore not an intrinsic ASPT.

The second example is the phase given by
\be\ba
\CM_{25}^{\rm SWSSB} = \big\langle &e_{RG},e_{GB},e_{RB},\bar{e}_{RG},\bar{e}_{GB},\bar{e}_{RB},e_{R}\bar{e}_R,\\ &s_{RGB}\bar{s}_{RGB},s^*_{RGB}\bar{s}^*_{RGB} \big\rangle \ .
\ea\ee
This phase has $\bbZ_{4,a}^w$ weak symmetry and $\bbZ_{2,a^2}^s$ strong symmetry. This is an example where the strong and weak symmetries no longer form a (semi-)direct product. Observe that flux of $\bbZ_{2,a^2}$ carries the charge of $\bbZ_{4,a}$
\be  e_{GB} = \big([a^2],1_b  \big)    \ee
where $1_b(a) = -1$, and the $\bbZ_{4,a}$ flux carries $\bbZ_{2,a^2}$ charge
\be s_{RGB}\bar{s}_{RGB} = \big( [a],i  \big)\otimes \overline{\big( [a],i  \big)}  \ee
where $i(a^2) = -1$. This decorated domain wall demonstrate a non-trivial ASPT. This ASPT is actually intrinsic, it is not a descendant from any pure-state SPT phase.

The third example is an intrinsic ASPT given by condensable algebra
\be\ba
\CM^{\rm SWSSB}_{14} = \big\langle &e_{GB},\bar{e}_{GB},e_R\bar{e}_R,e_G\bar{e}_G,m_R\bar{m}_R \\  &m_{GB}\bar{m}_{GB}, s_{RGB}\bar{s}_{RGB} , s^*_{RGB}\bar{s}^*_{RGB} \big\rangle \ .
\ea\ee
This phase preserves a $D^w_4$ weak symmetry and $\bbZ_{2,a^2}^s$ strong symmetry. There is a different decorated domain wall pattern:
\begin{enumerate}
    \item the condensation of $s_{RGB}\bar{s}_{RGB}$ and $e_{GB}=([a^2],1_{b})$, we have $i(a^2) = -1$ and $1_{b}(a)= -1$.
    \item the condensation of 
    \be m_R\bar{m}_R = ([ab],+-)\otimes \overline{ ([ab],+-) }  \ee
    and $e_{GB}$ gives
    \be  +-(a^2) = -1 \ , \ 1_{b}(ab) = -1 \ . \ee
\end{enumerate} 
These data show that the flux associated with the strong $\bbZ_{2,a^2}$ symmetry is decorated with charges of both weak $\bbZ_{4,a}$ and $\bbZ_{2,b}$ symmetries. The decorated domain wall feature of this intrinsic ASPT phase resembles a combination of the previous two ASPTs.

\section{Phase Transitions}\label{sec:pt}

We now proceed to discuss phase transitions between the gapped mixed-state phases described above. Consider two such phases labeled by Lagrangian algebras \(\mathcal{A}_1\) and \(\mathcal{A}_2\). The critical point separating them is characterized by a non-Lagrangian condensable algebra given by their intersection, \(\mathcal{A}' = \mathcal{A}_1 \cap \mathcal{A}_2\). Importantly, since both \(\mathcal{A}_1\) and \(\mathcal{A}_2\) satisfy the positivity constraint, their intersection \(\mathcal{A}'\) automatically inherits this property. Condensing \(\mathcal{A}'\) leads to the confinement of a subset of anyons and leaves a residual subcategory of unconfined anyons, labeled by $\CC'$. This residual theory governs the gapless theory at the boundary.

To identify which CFT is compatible with a given transition point, one can follow the prescription in Ref.~\cite{Ji_2019,Ji:2019jhk}. Most importantly, the partition vector must be modular invariant under the combined action of the modular \(S\) and \(T\) matrices of both the bulk topological theory and the boundary CFT. Specifically, the partition vector \(\ket{Z(\tau,\bar{\tau})}\) must satisfy
\be
(S^{\mathrm{top}} \otimes S) \ket{Z(\tau,\bar{\tau})} = (T^{\mathrm{top}} \otimes T) \ket{Z(\tau,\bar{\tau})} = \ket{Z(\tau,\bar{\tau})},
\ee
where \(S^{\mathrm{top}}\) and \(T^{\mathrm{top}}\) are the modular matrices of the bulk MTC, and \(S\) and \(T\) are those of the boundary CFT. Each component of the partition vector corresponds to the partition function in a particular symmetry sector of the theory.

Given the modular data of the MTC and a candidate CFT, one can explicitly compute these symmetry-resolved partition functions in the doubled Hilbert space and match them to determine the correct critical theory. In the next section, we demonstrate this approach with a few examples involving a \(\mathbb{Z}_2\) symmetry, and construct a corresponding lattice model that realizes the associated phase transition.

One notable feature we observe is that if one of the phases involved in the transition is an inherent mixed-state phase, then the corresponding critical point is also inherent mixed-state in nature. Specifically, the remaining subcategory \(\mathcal{C}'\) describing the transition cannot be decomposed as a tensor product of MTCs associated with each individual copy of the Hilbert space. As a consequence, certain observables or correlation functions at the critical point cannot be fully accessed within a single copy -- they must either be measured in the doubled Hilbert space or measured through correlation functions that are nonlinear in the density matrix. This signifies the fundamentally mixed-state character of the transition.

\subsection{Mixed-state phase transitions with $\bbZ_2$ symmetry}
In the mixed-state \(\mathbb{Z}_2\) case, the doubled-space SymTFT is given by \(\mathcal{Z}(\mathrm{Vec}_{\mathbb{Z}_2 \times \overline{\mathbb{Z}}_2})\), whose condensation patterns and phase transitions have been analyzed in detail in~\cite{Chatterjee_2023}. In the mixed-state setting, the set of allowed gapped boundaries is restricted to three possibilities: \(\mathcal{L}_S\), \(\mathcal{A}\), and \(\mathcal{M}\), corresponding respectively to the fully symmetry-broken phase, the symmetric phase, and the SWSSB phase.

Let us first consider the transition between \(\mathcal{L}_S\) and \(\mathcal{M}\) -- that is, from a fully symmetry-breaking phase to an SWSSB phase. The corresponding condensable algebra at the transition point is given by their intersection: \(\mathcal{L}_S \cap \mathcal{M} = 1 \oplus e\bar{e}\). The SymTFT can be decomposed as
\[
\{1, e\bar{e}, m, f\bar{e} \} \boxtimes \{1, e, m\bar{m}, f\bar{m} \}.
\]
Condensing \(1 \oplus e\bar{e}\) confines the first factor, while the second factor remains unaffected, leaving a nontrivial critical sector. Since the remaining anyons form a toric code topological order, we know an Ising CFT is compatible on the boundary and the partition vection of the system can be written as 

\be \CM_{1\oplus e\bar{e}} =\mqty(1(1) \\ 1(e\bar{e}) \\ 0(m) \\ 0(f\bar{e}))\otimes \mqty(Z(\tau,\bar{\tau},1) (1)\\Z(\tau,\bar{\tau},e) (e)\\Z(\tau,\bar{\tau},m)(m\bar{m})\\Z(\tau,\bar{\tau},f)(f\bar{m})) , \ee
or equvalently 
\[
    \CM_{1\oplus e\bar{e}} =\left(
\begin{array}{cccc}
 Z(\tau,\bar{\tau},1) &  Z(\tau,\bar{\tau},e) & 0 & 0 \\
  Z(\tau,\bar{\tau},e) &  Z(\tau,\bar{\tau},1) & 0 & 0 \\
 0 & 0 &  Z(\tau,\bar{\tau},m) &  Z(\tau,\bar{\tau},f) \\
 0 & 0 &  Z(\tau,\bar{\tau},f) &  Z(\tau,\bar{\tau},m) \\
\end{array}
\right)
    \]
where
\be  \mqty(Z(\tau,\bar{\tau},1)\\Z(\tau,\bar{\tau},e)\\Z(\tau,\bar{\tau},m)\\Z(\tau,\bar{\tau},f)) = \mqty(\abs{\chi_1}^2 + \abs{\chi_\psi}^2 \\ \abs{\chi_\sigma}^2 \\ \abs{\chi_{\sigma}}^2 \\ \chi_1(\tau)\bar{\chi}_\psi(\bar{\tau})+\bar{\chi}_1(\bar{\tau})\chi_\psi(\tau) )  \ . \ee

Similarly, the phase transition point between $\CA$ and $\CM$, namely from trivial symmetric phase to SWSSB, can be described by the following partition vector,
\[\ba
    \CM_{1\oplus m\bar{m}}&=\mqty(1(1) \\ 0(e) \\ 1(m\bar{m}) \\ 0(f\bar{m}))\otimes \mqty(Z(\tau,\bar{\tau},1) (1)\\Z(\tau,\bar{\tau},e) (e\bar{e})\\Z(\tau,\bar{\tau},m)(m)\\Z(\tau,\bar{\tau},f)(f\bar{e})) \\
    &=\left(
\begin{array}{cccc}
 Z(\tau,\bar{\tau},1) & 0 & Z(\tau,\bar{\tau},m) & 0 \\
 0 & Z(\tau,\bar{\tau},e) & 0 & Z(\tau,\bar{\tau},f) \\
 Z(\tau,\bar{\tau},m) & 0 & Z(\tau,\bar{\tau},1) & 0 \\
 0 & Z(\tau,\bar{\tau},f) & 0 & Z(\tau,\bar{\tau},e) \\
\end{array}
\right).
\ea
\]

The prediction of the transition theory discussed above -- for example, the transition from the trivial phase to the SWSSB phase -- is that certain correlation functions in the canonical purified state exhibit power-law decay. Specifically, we expect
\be
\langle\!\langle \rho^c \big| W_{m}(x,y) \big| \rho^c \rangle\!\rangle = \langle\!\langle \rho^c \big| W_{\bar{m}}(x,y) \big| \rho^c \rangle\!\rangle \sim \frac{1}{|x - y|^{1/4}},
\label{eq:disorder}
\ee
and
\be
\langle\!\langle \rho^c \big| W_{e\bar{e}}(x) W_{e\bar{e}}(y) \big| \rho^c \rangle\!\rangle \sim \frac{1}{|x - y|^{1/4}}.
\label{eq:order}
\ee
Here, \( W_m(x, y) \) is the string operator corresponding to the \( m \) anyon, stretching horizontally and terminating at positions \( x \) and \( y \). \( W_{e\bar{e}}(x) \) is the string operator corresponding to the \( e\bar{e} \) anyon, which stretches vertically and terminates on both the topological and physical boundaries; \( x \) denotes its position along the horizontal direction.

These power-law behaviors are signatures of the underlying criticality at the phase transition point. Notably, the first correlation function involves operators acting only within a single Hilbert space. As such, it corresponds to a linear observable in the density matrix and can be accessed through conventional measurements. In contrast, the second correlation function probes cross-correlations between the two copies of the doubled space, and thus corresponds to a nonlinear function of the density matrix. These observables are not directly accessible from single-copy measurements. In the following section, we will discuss their realizations in a lattice model.

\subsection{Lattice realization of the mixed-state $\bbZ_2$ phase transitions}
\begin{figure}[!htbp]
    \centering
    \includegraphics[width=0.8\linewidth]{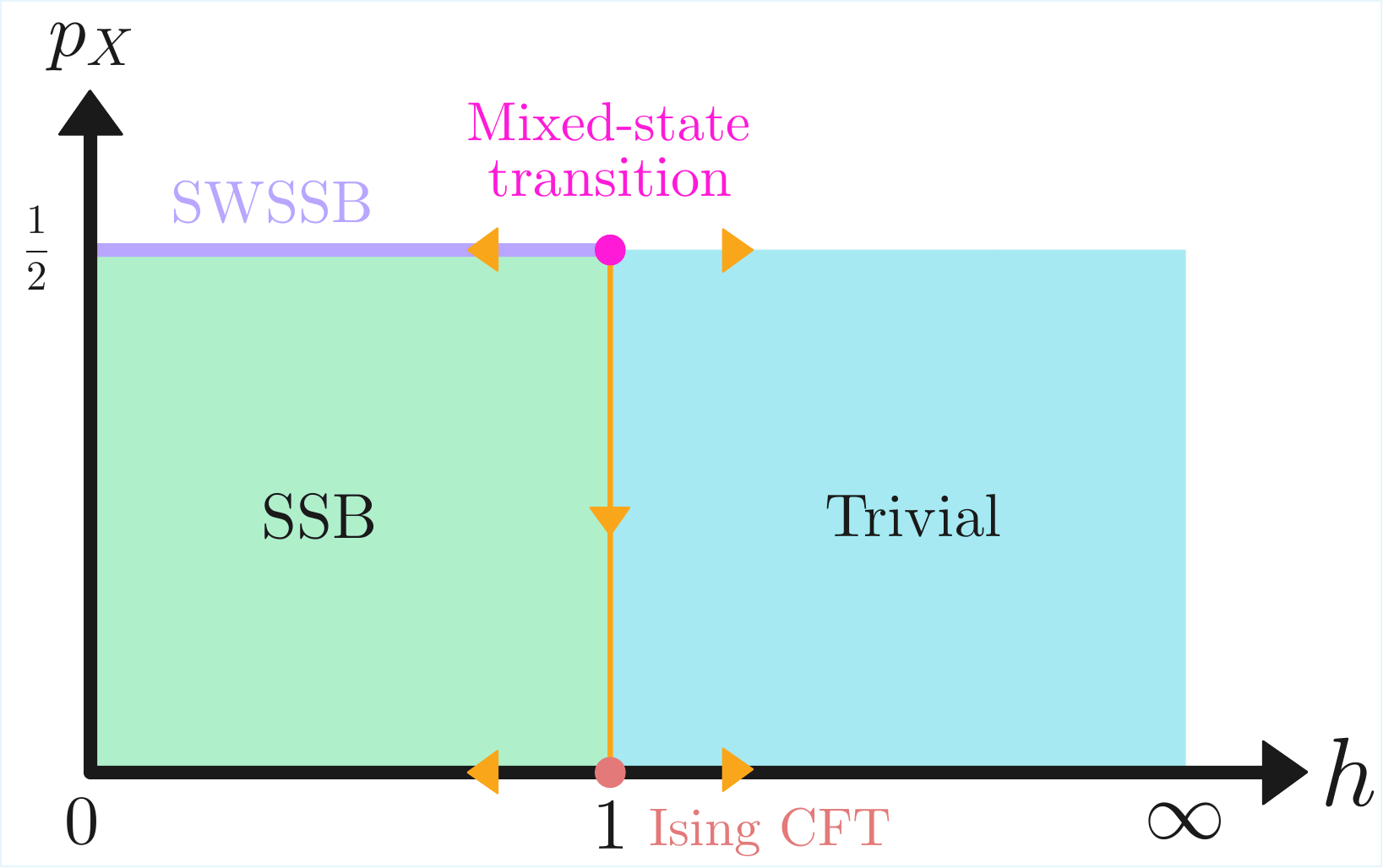}
    \caption{Phase diagram of the decohered transverse field Ising model.}\label{fig:dTFIM}
\end{figure}

In this section, we discuss a model that can realize the mixed-state phase transitions described above. The model is based on the \((1+1)\)D transverse field Ising model (TFIM),
\[
H_{\mathrm{TFIM}} = - \sum_j Z_j Z_{j+1} - h \sum_j X_j \, ,
\]
augmented with decoherence channels. We begin by reviewing the pure-state TFIM: as the transverse field \(h\) is tuned from \(0\) to \(\infty\), the ground state interpolates between a symmetry-breaking (SSB) phase and a symmetric product state. The critical point at \(h = 1\) is described by the Ising conformal field theory (CFT). We now consider extending the phase diagram by introducing X-decoherence channel on the ground state of the TFIM, implemented with strength controlled by a parameter \(p\). The channel acts as follows:
\[
\begin{aligned}
\mathcal{E}[\rho] &= \left( \prod_i \mathcal{E}_i \right)[\rho], \\
\mathcal{E}_i[\rho] &= (1 - p) \rho + p X_i \rho X_i \, .
\end{aligned}
\] 
We will focus on the strong decoherence limit, where \(p = 1/2\). The resulting phase diagram is shown in Fig.~\ref{fig:dTFIM}. In this limit, the symmetry-breaking (SSB) phase is transformed into a SWSSB phase, while the trivial symmetric phase remains unaffected by the \(X\)-decoherence channel. Consequently, the system at \(p = 1/2\) realizes a nontrivial transition between a trivial symmetric phase and a SWSSB phase. The corresponding critical point, arising from the Ising CFT deformed by decoherence, serves as a theory for this mixed-state phase transition.

It is clear that the trivial symmetric state remains unaffected by the \(X\)-decoherence channel. To justify our claim that the SSB phase is transformed into a SWSSB phase, we examine various correlation functions as a function of the decoherence strength \(p\). We focus on the \(h = 0\) limit, where the ground state of the pure TFIM is the GHZ state. In this setting, we can explicitly compute the two-point correlation function of \(Z\) operators in the decohered state $\rho=\CE(|\Psi_{GHZ}\rangle\langle\Psi_{GHZ}|)$:
\be
\mathrm{Tr}[\rho Z_i Z_j] = 1 - 4p(1 - p).
\ee
We observe that the correlation is nonzero for \(p < 1/2\), indicating spontaneous symmetry breaking. However, at \(p = 1/2\), the correlation vanishes, signaling the absence of conventional long-range order. Then we compute the Wightman-type correlation function at $p=1/2$ explicitly:
\be
\mathrm{Tr}[\sqrt{\rho} \, Z_i Z_j \, \sqrt{\rho} \, Z_i Z_j] = 1,
\ee
which indicates that the system exhibits SWSSB in the presence of maximal decoherence. 

In fact, for any \(p < \tfrac12\) the decoherence channel can be interpreted as a \emph{finite-time}, local, symmetry-preserving Lindbladian evolution. By the Lieb–Robinson bound\cite{LRBorginal, Poulin2010,Chen2023lrb}, such an evolution cannot destroy long-range order, so the system retains conventional SSB. At \(p = \tfrac12\), however, the channel corresponds to an \emph{infinite} effective evolution time: ordinary long-range order vanishes, yet SWSSB persists. The same argument implies that throughout the region \(p < \tfrac12\) and \(h < 1\) (see Fig.~\ref{fig:dTFIM}), the system remains in an SSB phase.

The most interesting physics appears along the \(p = \tfrac12\) line, corresponding to the strong-decoherence limit.  In this regime we can compute correlation functions of both order and disorder operators analytically.  Operationally, \(p = \tfrac12\) is equivalent to performing a \emph{projective measurement} of every \(X_i\) on the wavefunction.  Let
\(\,|\Psi(h)\rangle\) be the ground state of the transverse-field Ising model at field strength \(h\).  Expanding it in the \(X\)-basis,
\be
|\Psi(h)\rangle \;=\;
\sum_{\{X_i = \pm 1\}} c_{\{X\}}\,|\{X\}\rangle .
\ee
After a projective measurement of all \(X_i\) the state collapses to the mixed state
\be
\rho \;=\;
\sum_{\{X_i = \pm 1\}}
\bigl|c_{\{X\}}\bigr|^{2}\,
|\{X\}\rangle\!\langle\{X\}| .
\ee
Its canonical purification in the doubled Hilbert space is
\be
\ba
\bigl|\rho^{c}\bigr\rangle\!\bigr\rangle \;&=\;
\sum_{\{X_i = \pm 1\}}
\bigl|c_{\{X\}}\bigr|\,
|\{X\}\rangle\otimes
|\overline{\{X\}}\rangle \\
&=\sum_{\{X_i = \pm 1\}}
c_{\{X\}}\,
|\{X\}\rangle\otimes
|\overline{\{X\}}\rangle.
\ea
\ee
Interestingly, for the transverse-field Ising model the Perron–Frobenius theorem\footnote{To use the theorem, consider the following manipulation which does not change the eigenstates $H' = -H_{\rm TFIM} + c \mathds{1}$, where $c>0$ is a large enough so that entries of $H'$ are positive. The Perron–Frobenius theorem states that for a matrix with non-negative real components, the eigenvector with the largest eigenvalue also has non-negative real components.} guarantees that the ground-state coefficients \(c_{\{X\}}\) can be chosen real and non-negative; hence the absolute-value signs in the canonical purification can be dropped. With this expression, we can compute two key quantities: the correlation function of the disorder parameter in the original space, and the Wightman correlation function of the order parameter in the doubled space.

First let us consider the disorder parameter

\be\ba 
&\mathrm{Tr}[\rho \prod_{i= x}^y X_i]=\bbra{\rho^c} \prod_{i= x}^y X_i \kket{\rho^c} \\ =&\sum_{ \{X\},\{X'\} } c_{ \{X\} }c_{ \{X'\} }  \bra{ \{X\} } \prod_{i= x}^y X_i \ket{ \{X'\} } \langle \overline{\{X\}}  \ket{ \overline{\{X'\}} }  \\
= &\sum_{ \{X\} } c_{ \{X\} }^2 \cdot \big(\bra{ \{X\} } \prod_{i= x}^y X_i \ket{ \{X\} } \big)\\
= &\bra{\Psi(h)} \prod_{i= x}^y X_i\ket{\Psi(h)}.
\ea\ee  
This is identical to its value before decoherence, meaning that the quantity exhibits the same behavior as in the original TFIM: exponential decay in the SWSSB phase, long-range order in the trivial phase, and a characteristic power-law behavior $\sim \frac{1}{|x - y|^{1/4}}$ at the critical point.

For the Wightman correlation functions of $Z$ operators, we have
\be\ba
&\mathrm{Tr}[\sqrt{\rho} Z_i Z_j\sqrt{\rho}Z_iZ_j]= \bbra{\rho^c} Z_i \bar{Z}_iZ_j \bar{Z}_j \kket{\rho^c}   \\
= &\sum_{ \{X\},\{X'\} } c_{ \{X\} }c_{ \{X'\} }  \bra{ \{X\} } Z_iZ_j \ket{ \{X'\} }  \bra{ \overline{\{X\}} } \bar{Z}_i\bar{Z}_j \ket{ \overline{\{X'\}} } \\
= &\sum_{ \{X\},\{X'\} } c_{ \{X\} }c_{ \{X'\} } \delta_{\{X\} = Z_iZ_j \{X'\}}\\
=& \bra{\Psi(h)} Z_iZ_j\ket{\Psi(h)}.
\ea\ee

Interestingly, we find that the Wightman correlation function is equal to the original \(Z\)-correlation function in the pure state\cite{Weinstein2025wightman}. As a result, the Wightman correlator exhibits long-range order in the SWSSB phase and vanishes in the trivial phase, signaling a transition at \(h = 1\) characterized by the critical scaling behavior \(\sim |i - j|^{-1/4}\). In contrast, the ordinary correlation function \(\operatorname{Tr}[\rho Z_i Z_j]\) vanishes identically in the \(p = 1/2\) limit.

These discussions demonstrate that the mixed state \(\rho^c\) at \(h = 1\) exhibits critical behavior analogous to that of an Ising CFT. However, some sectors of this criticality are only detectable in the doubled (purified) state. These behaviors are consistent with the predictions from the SymTFT, as reflected in Eqs.~\ref{eq:disorder} and~\ref{eq:order}.

It was previously noted~\cite{ma2023measurement} that performing measurements in a symmetric basis over a finite duration in a \((1+1)\)D Ising CFT does not alter the scaling behavior of trajectory-averaged correlation functions -- whether linear or nonlinear in the density matrix -- as well as the von-Neumann entanglement entropy. Therefore, we expect that for \( p < 1/2 \), the system's behavior continues to be governed by the Ising CFT. The point \( p = 1/2 \) then marks a multicritical point within this model.

We also note that starting from the Ising chain, introducing a \(ZZ\) dephasing channel allows one to tune, in the \(p = \tfrac12\) limit, from a fully SSB phase to a SWSSB phase. This correspondence follows directly from applying Kramers–Wannier duality to the one-dimensional model. It is an intriguing open question to explore the effects of a noise channel that respects Kramers–Wannier duality, such as \( K \sim X + ZZ \), particularly in the strong decoherence limit \( p = 1/2 \). The resulting critical behavior remains to be fully understood.

\subsection{Generalizations}

The \(\mathbb{Z}_2\) case can be straightforwardly generalized to \(\mathbb{Z}_q\). For \(\mathbb{Z}_q\), the anyons in the open SymTFT can again be labeled by \(\{e^a m^b \bar{e}^c \bar{m}^d\}\) with $a,b,c,d=0,1,2,...,q-1$. Now we consider the following gapped phases: \(\CA_{\rm SSB} = \expval{e,\bar{e}}\), \(\CA_{\rm triv} = \expval{m,\bar{m}}\), and \(\CM_{\rm SWSSB} = \expval{e\bar{e}, m\bar{m}}\). The non-Lagrangian algebras corresponding to the phase transition points between these phases are
\be
\ba 
&\CM_{{\rm SSB-SWSSB}} = \CA_{\rm SSB} \cap \CM_{\rm SWSSB} = \expval{e\bar{e}} \\
&\CM_{{\rm triv-SWSSB}} = \CA_{\rm triv} \cap \CM_{\rm SWSSB} = \expval{m\bar{m}} 
\ea
\ee
In each case, the remaining fusion category is still isomorphic to \(\CZ({\rm Vec}_{\mathbb{Z}_q})\). The generating anyons, however, differ: they are \(\langle e\bar{e}, m\rangle\) for the trivial--SWSSB transition and \(\langle m\bar{m}, e\rangle\) for the SSB--SWSSB transition. Consequently, these mixed-state phase transitions are governed by the same CFT that appears in the pure-state SSB transition, but realized in a mixed-state setting. 

Lattice models that realize these transitions can likewise be constructed and analyzed in a straightforward manner. Consider the clock generalization of the $\bbZ_2$ spins. We consider the \(\mathbb{Z}_q\) clock \(Z\) and \(X\) operators satisfying
\be
Z^q = X^q = 1 \ , \quad ZX = \omega_q XZ \ , \quad \omega_q = \exp\left(\frac{2\pi i}{q}\right) \ .
\ee
Consider the analogue of the TFIM Hamiltonian:
\be
H = -\sum_i Z_i Z_{i+1}^\dagger - h \sum_i X_i + \text{h.c.}
\ee
This Hamiltonian has a \(\mathbb{Z}_q\) symmetry generated by $U = \prod_i X_i$.

For \( q \leq 4 \), the model undergoes a continuous phase transition at \( h = 1 \), while for \( q > 4 \), the transition becomes first order. One can then apply \( X \) or \( Z^\dagger Z \) noise to the ground state of the model to study the mixed-state phase diagram. This model shares several appealing features with the transverse field Ising model. In particular, the Perron–Frobenius theorem applies here as well, ensuring that the ground state wavefunction can be expressed in either the \( X \) or \( Z \) basis with all positive coefficients. In the projective measurement limit, this enables explicit calculations: one can show that the correlation functions of both the order and disorder parameters at the mixed-state phase transition -- captured by the purified state \(\kket{\rho^c}\) -- exhibit the same power-law scaling as predicted by the corresponding CFT.

For \( q > 4 \), although the model does not exhibit a continuous phase transition, this does not preclude the existence of a lattice realization of the corresponding critical theory suggested by the SymTFT analysis. In fact, there are known CFTs compatible with the \(\mathbb{Z}_q\) symmetry, but they possess multiple relevant primary fields. As a result, realizing these theories requires introducing additional terms in the Hamiltonian to allow for fine-tuning. The analysis of the decohered model in such cases becomes more complex and less tractable.


The pattern of condensable algebra at the transition between a trivial phase and a SWSSB phase, as well as between a fully symmetry-broken phase and a SWSSB phase, is general for any finite group \( G \). For a trivial-to-SWSSB transition, the condensable algebra at criticality consists of all bound states of fluxes with their conjugates in the doubled space. The remaining deconfined subcategory is generated by charge–conjugate pairs together with individual fluxes, and is isomorphic to the modular tensor category \(\mathcal{Z}(\mathrm{Vec}_G)\). Hence, the critical point can be described by any CFT that can appear as a consistent boundary of \(\mathcal{Z}(\mathrm{Vec}_G)\). The critical mixed state at the transition can be obtained by starting from the pure-state critical wavefunction and introducing decoherence that incoherently proliferates the anyon-conjugate pairs required for condensation. In our \(\mathbb{Z}_q\) example, we utilized special properties of the wavefunction -- specifically, that its coefficients are real and positive in a suitable basis -- to enable analytic calculations. However, we expect these particular features are not essential for capturing the universal long-distance physics, such as the scaling behavior of correlation functions. Although rigorously proving these behaviors in general settings is challenging, the underlying universality is expected to remain robust.

\subsection{Intrinsically gapless phases}

A particularly interesting class of non-Lagrangian algebras does not lie at a phase boundary between two gapped (Lagrangian) phases; instead, it realizes an intrinsically gapless symmetry-protected topological (igSPT) phase~\cite{Thorngren2021,Wen2023igsptbb,Wen2025igspt,huang2025topologicalholography}.  A well–studied pure-state example is the \(\mathbb{Z}_4\) igSPT, whose SymTFT description is a non-Lagrangian algebra $\mathcal{A}_{\mathrm{igSPT}} \;=\; 1 \,\oplus\, e^{2}m^{2}$.
Condensing \(\mathcal{A}_{\mathrm{igSPT}}\) leaves the bulk in the double-semion topological order, whose boundary is anomalous and gapless; hence the phase is intrinsically gapless. The notion extends naturally to mixed states.  We define a mixed-state igSPT as a non-Lagrangian algebra in the open SymTFT that cannot be realized as the intersection of two gapped condensable algebras. This definition automatically includes all pure-state igSPTs. A natural question is whether genuinely mixed-state examples of this type exist. For \(\mathbb{Z}_4\) symmetry, the answer is affirmative. One explicit example is
\[
\mathcal{M}_{\text{igSPT}}
  \;=\;
  (1 \oplus e^{2}m^{2})
  \,\otimes\,
  (1 \oplus \bar{e}^{2}\bar{m}^{2})
  \,\otimes\,
  (1 \oplus e m \bar{e} \bar{m}),
\]
and an alternative choice replaces the last factor by
\(1 \oplus e m^{3} \bar{e} \bar{m}^{3}\). When this set of anyons condenses in the doubled space, only a single copy of the double-semion topological order survives, leaving each individual copy in a chiral semion phase \cite{Ramanjit2025noisy,Ellison:2024svg}. The condensation preserves all symmetries, giving a truly mixed-state analogue of the pure-state igSPT. We also observe that this mixed-state igSPT can be obtained by subjecting the pure-state \(\mathbb{Z}_4\) igSPT to \(\mathbb{Z}_4\)-symmetric decoherence. Whether entirely new classes of mixed-state igSPTs -- arising from different mechanisms -- exist remains an open question.

\section{Gauging in open SymTFT}\label{sec:Gaugingopen}

The open SymTFT framework offers a natural venue for studying the gauging of strong and weak symmetries in mixed-state systems. In this setting, gauging is implemented by altering the topological boundary condition on the SymTFT slab. A new boundary condition changes which anyons can tunnel through the bulk, effectively redefining the ``electric" anyons and, in turn, modifying the global symmetry that admits local order parameters. Several equivalent definitions of gauging exist; for completeness, we review them in Appendix \ref{app:Gauging}.

\begin{figure}[htbp]
    \centering
    \includegraphics[width=\linewidth]{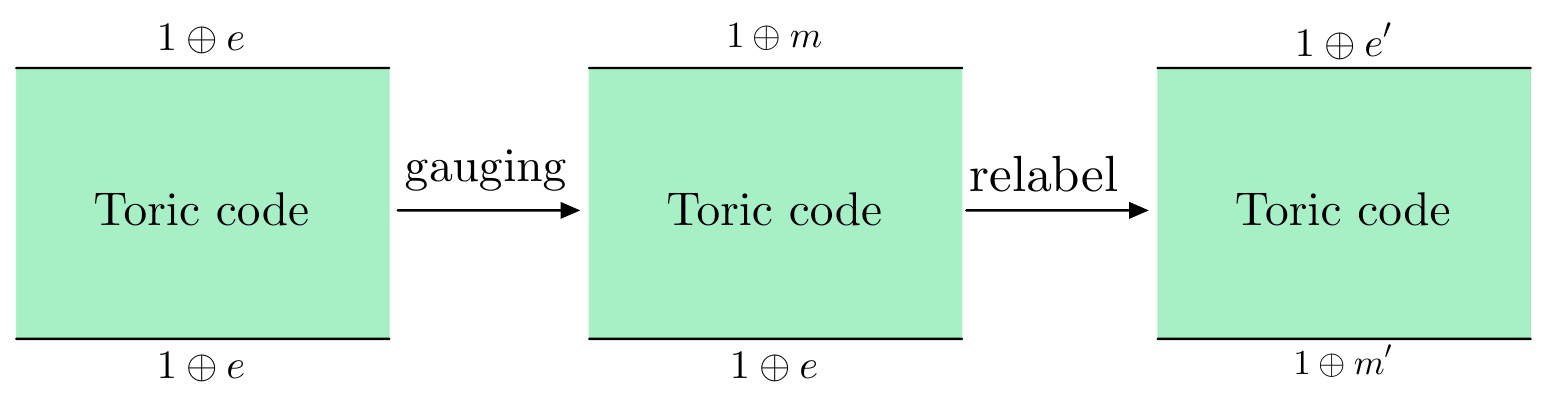}
    \caption{In the $\bbZ_2$ SymTFT, we show that gauging the SSB phase results in a trivial phase, and how to see it more explicitly through relabeling anyons.}
    \label{fig:gaugingrelabel}
\end{figure}

To keep track of how the “electric’’ sector changes under gauging, we will relabel the anyons after the new boundary condition is imposed. All anyons that condense on the topological boundary—those that can now tunnel freely and yield local order parameters—are referred to as electric (or charge) anyons. Conversely, anyons that braid non-trivially with this new charge sector and thus encode the symmetry action are labelled magnetic (or flux) anyons. For instance (see Fig.~\ref{fig:gaugingrelabel}), in a pure-state \(\mathbb{Z}_2\) SymTFT, gauging the symmetry amounts to changing the topological boundary condition from \(1 \oplus e\) to \(1 \oplus m\). After the boundary is switched, an \(m\) line can now tunnel through the slab and therefore serves as a local order parameter, while a horizontal \(e\) line implements the symmetry action. To reflect these new roles, we relabel the anyons as \(e' \equiv m\) (electric/charge) and \(m' \equiv e\) (magnetic/flux). In this relabelled form, the slab realizes exactly the same condensation pattern as the trivial symmetric phase in the ungauged construction. 

In the open SymTFT, the allowed patterns of gauging are further restricted due to the positivity constraint. A physical phase (meaning both boundary conditions satisfy the double space constraints) could become unphysical when changing the topological boundary condition into another allowed one. We will explicitly observe this in the $\bbZ_2$ case.

\subsection{Gauging with $\bbZ_2$}
\label{sec:gaugingZ2}
For the open SymTFT \(\CZ(\mathrm{Vec}_{\mathbb{Z}_2 \times \overline{\mathbb{Z}}_2})\) there are four distinct gauging options: gauging the left \(\mathbb{Z}_2\), gauging the right \(\overline{\mathbb{Z}}_2\), gauging the diagonal subgroup \(\mathbb{Z}_2^{d}\), and gauging the full \(\mathbb{Z}_2 \times \overline{\mathbb{Z}}_2\) symmetry.

Gauging the full \(\mathbb{Z}_2 \times \overline{\mathbb{Z}}_2\) is implemented by changing the topological boundary condition from
\(\mathcal{L}_{S}=1 \oplus e \oplus \bar{e} \oplus e\bar{e}\)
to
\(\mathcal{A}_{1}=1 \oplus m \oplus \bar{m} \oplus m\bar{m}
                 \equiv 1 \oplus e' \oplus \bar{e}' \oplus e'\bar{e}'\).
This gauging permutes the SSB and SPT phases while leaving the SWSSB phase unchanged:

\[
\begin{aligned}
&\text{Topological boundary:} \quad
\mathcal{A}_{1}=1\oplus e' \oplus \bar{e}' \oplus e'\bar{e}' \\[4pt]
&\text{SSB}: \quad
1\oplus e \oplus \bar{e} \oplus e\bar{e}
\;\; \longrightarrow\;\;
\text{SPT}:
1\oplus m' \oplus \bar{m}' \oplus m'\bar{m}' \\[4pt]
&\text{SPT}: \quad 
1\oplus m \oplus \bar{m} \oplus m\bar{m}
\;\; \longrightarrow\;\;
\text{SSB}:
1\oplus e' \oplus \bar{e}' \oplus e'\bar{e}' \\[4pt]
&\text{SWSSB}: \quad
1\oplus e\bar{e} \oplus m\bar{m} \oplus f\bar{f}
\;\; \longrightarrow\;\; \\
&\ \ \ \ \ \ \ \ \ \ \ \ \ \ \ \ \ \ \ \ \ \ \ \ \ \ \text{SWSSB}:
1\oplus m'\bar{m}' \oplus e'\bar{e}' \oplus f'\bar{f}' .
\end{aligned}
\]

\noindent
Remarkably, the SWSSB phase is invariant under this full gauging.

An interesting scenario is gauging the diagonal or weak symmetry \(\mathbb{Z}_2^{d}\). 
This is achieved by changing the topological boundary condition from the charge–condensed set 
\(\mathcal{L}_{S}=1\oplus e\oplus\bar{e}\oplus e\bar{e}\) 
to the Lagrangian algebra 
\(\mathcal{M}_{1}=1\oplus e\bar{e}\oplus m\bar{m}\oplus f\bar{f}\).
To make the new electric sector explicit, we relabel the anyons as 
\(e' \equiv m\bar{m}\) and \(\bar{e}' \equiv f\bar{f}\). An alternative relabeling, \(e' \equiv f\bar{f}\) and \(\bar{e}' \equiv m\bar{m}\), leads to an equivalent description. Since \( e\bar{e} \) is neutral under the weak symmetry, it remains invariant under the gauging procedure -- specifically, \( e'\bar{e}' = e\bar{e} \). The gauging procedure yields
\be \ba
&{\rm SSB}: \ 1\oplus e \oplus \bar{e} \oplus e\bar{e} \to {\rm SWSSB}: \ 1\oplus e'\bar{e}' \oplus  m'\bar{m}' \oplus f'\bar{f}' \\
&{\rm SWSSB}: \ 1\oplus e\bar{e} \oplus  m\bar{m} \oplus f\bar{f} \to {\rm SSB}:\  1\oplus e' \oplus \bar{e}' \oplus e'\bar{e}'
\ea \ee
we can see that through this gauging, the SSB phase becomes the SWSSB case and vice versa. However, gauging the $\bbZ_2^d$ symmetry in the trivial phase would differ with respect to the two choices: 1) $ 1\oplus m \oplus \bar{m} \oplus m\bar{m} \to  1\oplus e' \oplus \bar{m}' \oplus e'\bar{m}' $, 2) $ 1\oplus m \oplus \bar{m} \oplus m\bar{m} \to 1\oplus \bar{e}' \oplus m' \oplus m'\bar{e}' $. Both phases fail to satisfy the $J$ symmetry and are thus unphysical. This is physically reasonable: gauging the diagonal (weak) symmetry transforms the system into a state where that symmetry is spontaneously broken while preserving the off-diagonal (strong) symmetry. However, physically the strong and weak sectors are intertwined, and breaking the weak symmetry inevitably leads to the breaking of the strong symmetry as well. As a result, one should not expect a consistent or well-defined gauging procedure in this case. This highlights an interesting feature of mixed-state systems: certain gauging operations may be well-defined only for specific classes of states while failing for others.

\begin{figure}[h]
    \centering
    \includegraphics[width=0.9\linewidth]{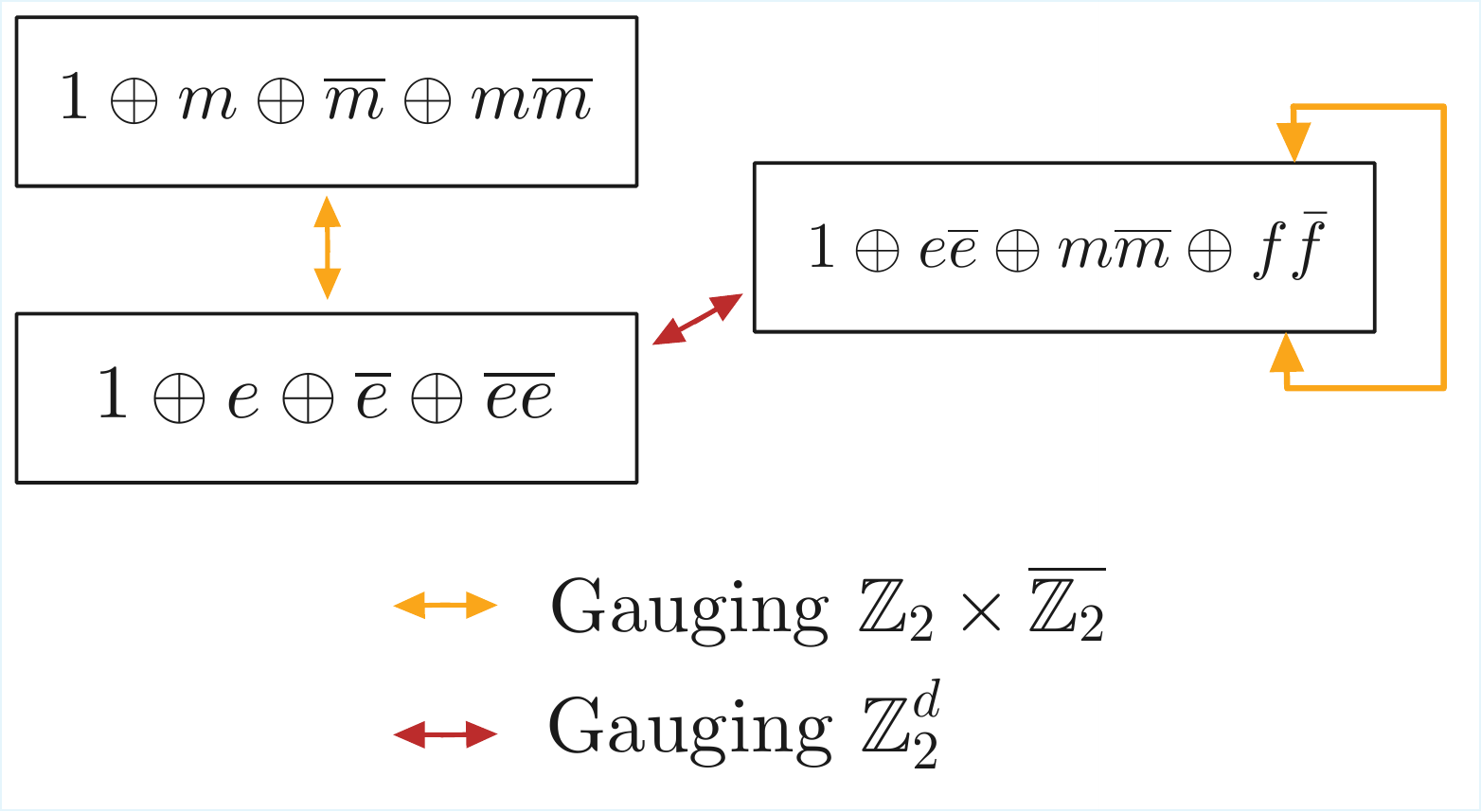}
    \caption{The web of gauging. The dark red arrows relate phases by gauging diagonal $\bbZ_2^d$; the orange arrows relate phases by gauging the entire $\bbZ_2\times \bbZ_2$ symmetry.}
    \label{fig:dualityweb}
\end{figure}

\subsection{Realization of gauging weak $\bbZ_2$ symmetry}
Other than the abstract discussion of gauging, we now examine the concrete transformation of the state under gauging within a one-dimensional lattice model in the doubled Hilbert space. The key intuition is that the Hilbert space sector neutral under the weak symmetry should remain invariant, while the transformation should act nontrivially on the sector charged under the weak symmetry. To this end, we implement a change of basis in the doubled state:
\be \ba \ket{\uparrow}^L_i\ket{\uparrow}^R_i \mapsto \ket{\uparrow}^L_i\ket{\uparrow}^M_i \ &, \  \ket{\uparrow}^L_i\ket{\downarrow}^R_i \mapsto \ket{\uparrow}^L_i\ket{\downarrow}^M_i \\ \ket{\downarrow}^L_i\ket{\uparrow}^R_i \mapsto \ket{\downarrow}^L_i\ket{\downarrow}^M_i \ &, \ 
\ket{\downarrow}^L_i\ket{\downarrow}^R_i \mapsto \ket{\downarrow}^L_i\ket{\uparrow}^M_i \ .
\ea \ee
Here, \( L \), \( R \), and \( M \) denote the left, right, and middle spaces, respectively. We perform a basis transformation from the \( L, R \) configuration to the \( L, M \) basis. The \( L \) (and similarly \( R \)) space carries nontrivial charges under both the strong and weak symmetries, whereas the \( M \) space is neutral under the weak symmetry but still carries a strong symmetry charge.

Gauging the diagonal $\bbZ_2^d$ symmetry amounts to do a Kramers-Wannier duality onto the $L$ part of Hilbert space while keeping $M$ invariant. Let us start from the SSB phase and show that this transformation leads to SWSSB state:

\be\ba \kket{\rho_{\rm SSB}} = &\frac{1}{2}\Big(\ket{\uparrow\cdots \uparrow}^L + \ket{\downarrow\cdots \downarrow}^L \Big) \otimes \Big(\ket{\uparrow\cdots \uparrow}^R+ \ket{\downarrow\cdots \downarrow}^R \Big)  \\
= & \frac{1}{2}\Big(\ket{\uparrow\cdots \uparrow}^L + \ket{\downarrow\cdots \downarrow}^L \Big) \otimes \Big(\ket{\uparrow\cdots \uparrow}^M+ \ket{\downarrow\cdots \downarrow}^M \Big) \\ 
\mapsto \kket{\tilde{\rho}} = &\Big(  \ket{+\cdots +}^L \Big)\otimes \sqrt{\frac{1}{2}} \Big(\ket{\uparrow\cdots \uparrow}^M+ \ket{\downarrow\cdots \downarrow}^M \Big) \\
= &\frac{1}{2^{\frac{N+1}{2}}} \Big( \sum_{\{\sigma_z\}} \ket{ \{\sigma_z\} }^L  \Big) \otimes \Big(\ket{\uparrow\cdots \uparrow}^M+ \ket{\downarrow\cdots \downarrow}^M \Big) \\ 
= & \frac{1}{2^{\frac{N+1}{2}}} \Big( \sum_{\{\sigma_z\}} \ket{ \{\sigma_z\} }^L\ket{ \{\sigma_z\} }^R +  \ket{ \{\sigma_z\} }^L\ket{ \{-\sigma_z\} }^R \Big) \\
= &\frac{1}{2^{\frac{N+3}{2}}}  \sum_{\{\sigma_z\}}  \big(\ket{\{\sigma_z\}} + \ket{\{-\sigma_z\}} \big)^L \otimes \big( \ket{\{\sigma_z\}} + \ket{\{-\sigma_z\}} \big)^R.\ea\ee
Reversing this procedure, the SWSSB phase can be transformed into an SSB phase by gauging the diagonal \(\mathbb{Z}_2^d\) symmetry.

However, applying the same transformation to a trivial symmetric state leads to a pathological outcome that does not respect the \(J\)-symmetry. This indicates that the gauging procedure is only well-defined for certain initial states and cannot be universally applied across all phases.

This transformation acts in the doubled Hilbert space. Ideally, however, one would prefer a transformation defined directly on the physical Hilbert space. Crucially, both the forward and backward processes must be quantum channels, as they map between pure and mixed states. From this viewpoint, the forward channel -- from an SSB phase to an SWSSB phase -- can be realized via projective measurements of the \( X \) operator on each site and tracing out the results, which classically proliferate domain walls. Reversing this process requires a quantum channel that eliminates domain walls, which can be achieved through a combination of measurement and feedback: when a domain wall is detected by measuring \( Z_i Z_{i+1} \), one applies a spin flip to one of the involved spins. This removes all domain walls and restores a pure GHZ state.

\subsection{Gauging weak non-Abelian symmetries}
In this section we explore gauging with non-Abelian symmetries in the mixed-state scenario. Specifically, we will focus on the case of $S_3$ symmetry. In Sec.\ref{sec:MsGappedS3}, we have classified the mixed-state gapped phases with $S_3$ symmetry, with the tool of open SymTFT, we can conveniently discuss gauging as switching topological boundary conditions. Through calculations presented in Appendix.\ref{app:gauginginS3}, we obtain the result that
\begin{enumerate}
    \item gauging the diagonal $\bbZ_2^d$ symmetry is equivalent to changing the topological boundary condition from $\CL_S$ into $\CM_6^{\rm SWSSB}$;
    \item gauging the diagonal $\bbZ_3^d$ symmetry is equivalent to changing the topological boundary condition from $\CL_S$ into $\CM_{7}^{\rm SWSSB}$;
    \item gauging the diagonal $S_3^d$ symmetry is equivalent to changing the topological boundary condition from $\CL_S$ into $\CM_{4}^{\rm SWSSB}$.
\end{enumerate}
The lattice transformation for gauging non-Abelian weak symmetries remains to be explored.

\section{Effective field theory from SymTFT}\label{sec:eft}

\subsection{Basic formalism}
We first introduce the schematics used in this section. For a discrete abelian group $G$, the 2+1d topological field theory action for $\CZ({\rm Vec}_{G\times \overline{G}})$ is
\be
\label{SymTFT_action}
S_{\rm SymTFT}=ik\int_{M_3} a\cup\delta\hat{a}-b\cup\delta\hat{b}\,,
\ee
where $a,\hat{a},b,\hat{b}\in C^1(M_3,G)$, where $M_3$ is the three-dimensional bulk with boundary $M_2=\partial M_3$. For example when $G=\mb{Z}_N$, the gauge fields $a,\hat{a},b,\hat{b}$ take value in $\{0,1,\dots,N-1\}$ and the coefficient $k=2\pi/N$. The anyon lines in $M_3$ would be generated by
\be
\ba
e&=\exp\left(\frac{2\pi i}{N}\int a\right)\ ,\ m=\exp\left(\frac{2\pi i}{N}\int \hat{a}\right)\cr
\bar{e}&=\exp\left(\frac{2\pi i}{N}\int b\right)\ ,\ \bar{m}=\exp\left(\frac{2\pi i}{N}\int \hat{b}\right)\,.
\ea
\ee
Due to that $M_3$ has a boundary, the action Eq. \ref{SymTFT_action} is not gauge invariant under gauge transformation $a\rightarrow a+\delta\Lambda$ $(\Lambda\in C^0(M_3,G))$, as
\be
\ba
\delta_\Lambda S_{\rm SymTFT}&=\frac{2\pi i}{N}\int_{M_3}\delta\Lambda\cup\delta\hat{a}\cr
&=\frac{2\pi i}{N}\int_{M_2}\Lambda\cup\delta\hat{a}\,.
\ea
\ee
Thus to guarantee gauge invariance of the theory with boundary, one should choose among a number of boundary conditions for the gauge fields. The simple choices of such boundary conditions for $(a,\hat{a})$ include:
\begin{enumerate}
\item Neumann boundary condition (NBC) for $a$, such that along the $[0,1]$ direction the differential $\delta \Lambda=0$, and $a$ is a dynamical gauge field on $M_2$. Dirichlet boundary condition (DBC) for $\hat{a}$, which means that $\hat{a}$ takes a fixed value $\hat{a}=A(x^0,x^1)$ and is no longer dynamical on $M_2$. In other words, the anyon line $m=\exp\left(\frac{2\pi i}{N}\int \hat{a}\right)$ is condensed. There is a dual 0-form $\widehat{\mb{Z}}_N$ global symmetry generated by the topological operator $e=\exp\left(\frac{2\pi i}{N}\int a\right)$. Nonetheless in our conventional this symmetry does not have local order parameters.

\item Neumann boundary condition for $\hat{a}$ and Dirichlet boundary condition for $a$. The 0-form $\mb{Z}_N$ global symmetry of $M_2$ is generated by the topological operator $m=\exp\left(\frac{2\pi i}{N}\int \hat{a}\right)$. The anyon line $e=\exp\left(\frac{2\pi i}{N}\int a\right)$ is condensed, and when it has a non-zero vacuum expectational value, the $\mb{Z}_N$ global symmetry is spontaneously broken, to be consistent with the previous notations. The gauging of $\mb{Z}_N$ 0-form symmetry in the first boundary condition is equivalent to swapping the boundary conditions for the gauge fields $a$ and $\hat{a}$.

\item Mixed boundary condition, i.e. Dirichlet boundary condition for $pa$ and $\frac{N}{p}\hat{a}$, where $p|N$. In this case, the anyon lines $e^p$ and $m^{N/p}$ are condensed in $M_3$. Nonetheless, the anyon line $m$ generates the $\mb{Z}_p$ 0-form symmetry acting on local operators, as $\hat{a}$ is free to take value (has Neumann boundary condition) among a subgroup $\mb{Z}_p=\{0,N/p,\dots,N-N/p\}$. The anyon line $e$ generates a $\widehat{\mb{Z}}_{N/p}$ dual 0-form symmetry, as $a$ is free to take value among a subgroup $\widehat{\mb{Z}}_{N/p}=\{0,p,\dots,N-p\}$.

\end{enumerate}

When we take into account of all the gauge fields $a,\hat{a},b,\hat{b}$, there are also mixed boundary conditions such as giving Dirichlet boundary condition to $a+b$. In such cases, it is convenient to rewrite the SymTFT action as a summation of BF terms with $a+b$. There also exists boundary conditions with a total derivative term $\int_{M_3}\delta I_2$ in the rewritten action, where the gauge fields in $I_2$ all have Neumann boundary conditions. In such case, after gauge transformation $\delta_\Lambda$, a non-vanishing term
\be
\int_{M_3}\delta(\delta_\Lambda I_2)=\int_{M_2}\delta_\Lambda I_2
\ee
appears on the boundary. To make the bulk/boundary system gauge invariant, we need to introduce a new topological term $-\int_{M_2}I_2$ into the physical action of the 2d theory. Such term is interpreted as the topological response term in certain phases, such as the ASPT phase.

Taking the simplest example of $\CZ({\rm Vec}_{\bbZ_2\times \overline{\bbZ_2}})$, the TQFT action and line operators are
\be\ba  &S = i\pi \int_{M_3} a \cup \delta \hat{a} + b\cup \delta \hat{b} \\  &a,\hat{a},b,\hat{b}\in C^1(M_3, \bbZ_2) \\ &e = \exp(i\pi \int a) \ , \ m =  \exp(i\pi \int \hat{a}) \\ &\bar{e}= \exp(i\pi \int b) \ , \ \bar{m} = \exp(i\pi \int\hat{b}) \\
&J: a\leftrightarrow b \ , \ \hat{a}\leftrightarrow \hat{b} \ .
\ea\ee
The $\CL_S$-condensed boundary condition is given by
\be
{\rm Dir: \ } a,b \ ; \ {\rm Neu: \ } \hat{a},\hat{b} \ .
\ee
To construct the $\CM_1$ condensed boundary condition, we rewrite the SymTFT action as
\be\ba  S &= i\pi \int_{M_3} a \cup \delta(\hat{a} + \hat{b}) + (a+b) \cup \delta \hat{b} \\ &= i\pi \int_{M_3}  b \cup \delta(\hat{a} + \hat{b}) + (a+b)\cup\delta\hat{a}  \ ,  \ea  \ee
where we impose Dirichlet boundary conditions on $a+b,\hat{a}+\hat{b}$. With these boundary conditions, no additional boundary terms are required, and the system exhibits no nontrivial topological response. 

We also comment on the case of condensing $1\oplus e\bar{m}\oplus m\bar{e}\oplus f\bar{f}$ (which fails to satisfy the positivity condition) and the resulting topological response. In this case, we know that $\hat{a}+b$ and $a+\hat{b}$ are set to be in Dir. boundary condition, therefore, we reorganize the bulk SymTFT action as
\be
S=i\pi\int_{M_3}(a\cup \delta(\hat{a}+b)+b\cup \delta(a+\hat{b}))+i\pi\int_{M_3}\delta(a\cup b)\,.
\ee
The last term indicates that an additional boundary term must be included to properly retain gauge invariance on the boundary: 
\be
i\pi\int_{M_3}\delta(a\cup b)=i\pi\int_{M_2}(a\cup b).
\ee
This is a topological term on the $2$D boundary $M_2$, where $a$ and $b$ are two dynamical $\mb{Z}_2$ gauge fields coming from doubled Hilbert space. Such topological term respects the $J$-symmetry explicitly but not the positivity condition \cite{Ma:2024kma,xue2024tensornetworkformulationsymmetry}.

\subsection{Effective field theory for $\bbZ_4$ cases}
\label{sec:gaugingZ4}

In this section we discuss the different topological boundary conditions for the SymTFT $\CZ(\text{Vec}_{\mb{Z}_4\times\overline{\mb{Z}}_4})$ from gauging. The TQFT action and line operators in this case are
\be\ba  \label{Z4-SymTFT}&S = \frac{i\pi}{2} \int_{M_3} a \cup \delta \hat{a} - b\cup \delta \hat{b} \\  &a,\hat{a},b,\hat{b}\in C^1(M_3, \bbZ_4) \\ &e = \exp(\frac{i\pi}{2} \int a) \ , \ m =  \exp(\frac{i\pi}{2} \int \hat{a}) \\ &\bar{e}= \exp(\frac{i\pi}{2} \int b) \ , \ \bar{m} = \exp(\frac{i\pi}{2} \int\hat{b}) \\
&J: a\leftrightarrow b \ , \ \hat{a}\leftrightarrow \hat{b} \ .
\ea\ee

\begin{enumerate}
\item Similar to before, the $\CL_S$-condensed boundary condition is given by
\be
{\rm Dir: \ } a,b \ ; \ {\rm Neu: \ } \hat{a},\hat{b} \ .
\ee

\item The strong symmetry breaking phase of $\mb{Z}_4\rightarrow\mb{Z}_2$ has the condensed anyons $\mc{A}_1=\expval{e^2,\bar{e}^2 , m^2 ,\bar{m}^2 }$. Hence the we impose Dirichlet boundary condition on $2a$, $2b$, $2\hat{a}$ and $2\hat{b}$, by gauging $\mb{Z}_2\times\overline{\mb{Z}_2}\subset\mb{Z}_4\times\overline{\mb{Z}_4}$ subgroup from the $\CL_S$ boundary condition.

\item The pure-state phase of trivial $\mb{Z}_4$ SPT, with condensed anyons $\mc{A}_2=\expval{m,\bar{m} }$. This is achieved by giving boundary conditions
\be
{\rm Dir: \ } \hat{a},\hat{b} \ ; \ {\rm Neu: \ } a,b \ ,
\ee
or equivalently gauging the whole $\mb{Z}_4\times\overline{\mb{Z}_4}$ group from the $\CL_S$ boundary condition.

\item The $\mb{Z}_4$ SWSSB phase, with  condensed anyons $\mc{M}_1=\expval{e\bar{e},m\bar{m} }$. This is achieved by giving Dirichlet boundary condition to $a+b$ and $\hat{a}+\hat{b}$. 

To see this boundary condition more clearly, we can write the TQFT action as
\be
S = \frac{i\pi}{2} \int_{M_3}(a+b)\cup\delta\hat{a}-b\cup\delta(\hat{a}+\hat{b})\,.
\ee
The boundary conditions are
\be
{\rm Dir: \ } a+b,\hat{a}+\hat{b} \ ; \ {\rm Neu: \ } \hat{a},b \ .
\ee

\item The SWSSB phase with $\mc{M}_2=\expval{e^2,e\bar{e},\bar{e}^2,m^2\bar{m}^2 }$.

To express this boundary condition in terms of gauge fields, we rewrite the TQFT action as
\be
S=\frac{i\pi}{2}\int_{M_3}a\cup\delta(\hat{a}+\hat{b})-(a+b)\cup\delta\hat{b}\,.
\ee
The boundary conditions are
\be
\ba
{\rm Dir: \ } a+b\ ,\ {\rm Neu: \ }\hat{b}\,,\cr
{\rm Dir: \ } 2a\,,2\hat{a}+2\hat{b}\,.
\ea
\ee

\item The SWSSB phase with $\mc{M}_3=\expval{m^2,m\bar{m},\bar{m}^2,e^2\bar{e}^2 }$.

To express this topological boundary condition in terms of gauge fields, we rewrite the TQFT action as
\be
S=\frac{i\pi}{2}\int_{M_3}(a+b)\cup\delta\hat{a}-b\cup\delta(\hat{a}+\hat{b})\,.
\ee
The boundary conditions are
\be
\ba
{\rm Dir: \ } \hat{a}+\hat{b}\ ,\ {\rm Neu: \ }b\,,\cr
{\rm Dir: \ }  2\hat{a}\,,2a+2b\,.
\ea
\ee

\item The SWSSB phase with intrinsic ASPT $\mc{M}_4=\expval{e^2\bar{e}^2,e^2\bar{m}^2 ,e^2 m^2,m^2\bar{e}^2,\bar{e}^2\bar{m}^2,em\bar{e}\bar{m}}$.

The Dirichlet boundary condition is imposed on $a+b+\hat{a}+\hat{b}$, as well as a $\mb{Z}_2\subset\mb{Z}_4$ subgroup of $a+b$ and $\hat{a}+b$. We rewrite the SymTFT action as
\be
\ba
S = \frac{i\pi}{2} \int_{M_3}& (a+b) \cup \delta (\hat{a} + b)-b\cup \delta (a+b+\hat{a}+\hat{b})\cr
&+\delta(a\cup b)\,. 
\ea
\ee

The boundary conditions are imposed as
\be
\ba
\label{Z4-bc7}
{\rm Dir: \ } a+b+\hat{a}+\hat{b}\ ,\ {\rm Neu: \ }b\,,\cr
{\rm Dir: \ } \ 2a+2b\,,2\hat{a}+2b\,.
\ea
\ee

Upon choosing this boundary condition, the TQFT is no longer gauge invariant due to the $\delta(a\cup b)$ term exhibiting on the boundary. To remedy the gauge invariance, we introduce a topological response term on the boundary
\be
\label{Sresponse}
S_{\text{response}}=-\frac{i\pi}{2}\int_{M_2=\ptl M_3}a\cup b\,.
\ee
Since we have imposed a Dirichlet boundary condition on the \(\mathbb{Z}_2 \subset \mathbb{Z}_4\) subgroup of \(a + b\), we can rewrite \(a = 2A - b\) on the boundary, where \(A\) is a \(\mathbb{Z}_2\) gauge field with Neumann boundary conditions, associated with the center of the \(\mathbb{Z}_4\) group. Replacing \( a \) by \( 2A - b \), we obtain the term \( i\pi \int_2 A \cup b \). In this expression, although \( b \) is a \( \mathbb{Z}_4 \) gauge field, only its mod 2 component contributes. Thus, by introducing a \( \mathbb{Z}_2 \) gauge field \( B = b \ (\mathrm{mod}\ 2) \), we can rewrite Eq.~\ref{Sresponse} as
\be
S_{\text{response}}=-i\pi\int_{M_2=\ptl M_3}A\cup B\,.
\ee
Crucially, \( A \) is the gauge field associated with the \( \mathbb{Z}_2 \) center of \( \mathbb{Z}_4 \), while \( B \) captures the mod 2 component of a \( \mathbb{Z}_4 \) gauge field. Both fields obey Neumann boundary conditions, and the topological term \( S_{\text{response}} \) encodes the response of the intrinsic ASPT phase.

One may wonder if we can replace Eq. \ref{Z4-bc7} by an equivalent set of Dirichlet boundary conditions for $a+b+\hat{a}+\hat{b}$, $2a+2b$ and $2\hat{a}+2a$. (This amounts to look at different set of generating anyons condensed on the boundary.) We will show that we can obtain exactly the same topological response term as physically expected. The new boundary conditions lead to the following equation
\be
a+b=2A\ ,\hat{a}+a=2C, \ a+b+\hat{a}+\hat{b}=D\,,
\ee
where $A, C\in\bbZ_2$ have Neumann boundary condition and $D\in\bbZ_4$ has Dirichlet boundary condition.

Plug back into Eq. \ref{Z4-SymTFT} in terms of $A$, $C$, $D$, $a$, we get the action (we retain only the terms that may generate nontrivial boundary contributions under the specified boundary conditions):
\be
\ba
S&=\frac{i\pi}{2}\int_{M_3}(a\cup \delta(2C-a)-(2A-a)\cup\delta(D-2C-2A+a))\cr
&=\frac{i\pi}{2}\int_{M_3}(-2A\cup\delta a-2a\cup\delta A)\cr
&=i\pi\int_{M_3}\delta(a\cup A)\cr
&=i\pi\int_{M_2=\ptl M_3}B\cup A\,,
\ea
\ee
where we only keep the terms with Neumann boundary condition. In the final step, we convert the \( a \in \mathbb{Z}_4 \) gauge field to its mod 2 component \( B = a \bmod 2 \), so that the resulting action matches exactly the expected form.

\end{enumerate}

\subsection{Mixed state SymTFT for $U(1)$ symmetry}\label{sec:U(1)}

The formalism of mixed state SymTFT developed in this paper can also be applied to the cases of continuous global symmetries, with the proposed SymTFT actions~\cite{Brennan:2024fgj,Antinucci:2024zjp,Apruzzi:2024htg,Bonetti:2024cjk,Antinucci:2024bcm,Gagliano:2024off,Tian:2025ooo}.

For the case of $U(1)$ 0-form symmetry, we can use the topological action
\be
S =i \int_{M_3} a\wedge \dd\hat{a} - b\wedge \dd\hat{b}\,,
\ee
where $a,b$ are non-compact $\mb{R}$ gauge fields while $\hat{a},\hat{b}$ are compact $U(1)$ gauge fields with $2\pi$ periodicity. The gauge invariant wilson loop operators (anyon lines) are
\be
\ba
m_\theta&=\exp\left(i\theta\int a\right)\ ,\ e=\exp\left(i\int\hat{a}\right)\cr
\bar{m}_\theta&=\exp\left(i\theta\int b\right)\ ,\ \bar{e}=\exp\left(i\int\hat{b}\right)\,
\ea
\ee
where $\int a, \int b\in \bbZ$, and $\theta\in[0,2\pi)$. The non-trivial correlation functions between linked loop operators are
\be
\ba
\label{U1-link}
\langle m_\theta(\mc{C})e(\mc{C}')\rangle &=\exp\left(i\theta\langle \mc{C},\mc{C}'\rangle_{M_3}\right)\,,\cr
\langle \bar{m}_\theta(\mc{C})\bar{e}(\mc{C}')\rangle &=\exp\left(-i\theta\langle \mc{C},\mc{C}'\rangle_{M_3}\right)\,,
\ea
\ee
where $\langle \mc{C},\mc{C}'\rangle_{M_3}$ denotes the linking number on the 3-manifold $M_3$. The $\theta$ parameter in $m_\theta(\mc{C})$ and $\bar{m}_\theta(\mc{C})$ has a periodicity of $2\pi$: $\theta\sim \theta+2\pi$, as the operators $m_\theta(\mc{C})$ ($\bar{m}_\theta(\mc{C})$) and $m_{\theta+2\pi}(\mc{C})$ ($\bar{m}_{\theta+2\pi}(\mc{C})$) are indistinguishable from the linking correlation functions Eq. \ref{U1-link}.

We give some examples of boundary conditions and discuss their physical implications
\begin{enumerate}

\item $\mc{L}_S$-condensed boundary condition with
\be
{\rm Dir: \ } \hat{a},\hat{b} \ ; \ {\rm Neu: \ } a,b \ .
\ee
Taking this as the topological boundary condition would give $U(1)$ strongly symmetric phases. We will assume this topological boundary condition unless specified otherwise.

\item Gauging the strong $U(1)$ symmetry is equivalent to switching the topological boundary condition into:
\be
{\rm Dir: \ } a,b \ ; \ {\rm Neu: \ } \hat{a},\hat{b} \ ,
\ee
which would result in phases with $\bbZ$ symmetry generated by anyon lines $e$, $\bar{e}$. 

Taking this as the physical boundary condition (while $\CL_S$ the topological boundary condition) gives a $U(1)$ strongly symmetric pure-state phase.

\item Gauging the $\bbZ_N$ strong symmetry is equivalent to switching the topological boundary condition from $\CL_S$-condensed into the mixed boundary condition where $m_{2\pi/N}$, $\bar{m}_{2\pi/N}$, $e^N$ and $\bar{e}^N$ are condensed. After such gauging the system shall have strong symmetry given by $U(1)/\mb{Z}_N\times \widehat{\mb{Z}}_N$, where only $U(1)/\mb{Z}_N\cong U(1)$ acts on local operators.

If we put this condensation on the physical boundary (while keep the original $\CL_S$ on the topological boundary), it produces a phase with SSB $U(1)\to \bbZ_{N}$.

\item SWSSB phase of $U(1)^s\rightarrow U(1)^w$: $e\bar{e}$ and $m_\theta\bar{m}_\theta$ are condensed on the physical boundary. 

To describe the boundary conditions we rewrite the TQFT action as
\be
S = i\int_{M_3} (a+b)\wedge \dd\hat{a} - b\wedge \dd(\hat{a}+\hat{b})\,,
\ee
and the boundary conditions are
\be
{\rm Dir: \ } a+b,\hat{a}+\hat{b} \ ; \ {\rm Neu: \ } \hat{a},b \ .
\ee

\item SWSSB with an intrinsic ASPT phase:
Consider a mixed-state scenario closely related to the \(\mathbb{Z}_4\) example discussed earlier. The anyon-condensation sequence proceeds in three steps. First, condense \(\langle e^{4}, \bar{e}^{4} \rangle\); this reduces the original \(U(1)\) strong symmetry to a strong \(\mathbb{Z}_4\).
Next, condense \(\langle e^{2} m_{\pi}, \bar{e}^{2} \bar{m}_{\pi} \rangle\); this transforms the SymTFT into the desired double-semion topological order, and the system is an igSPT state in pure state. Finally, condense \(\langle e^{2} \bar{e}^{2},\, e\,m_{\pi/2}\,\bar{e}\,\bar{m}_{\pi/2} \rangle\); this leaves the outer \(\mathbb{Z}_2\) factor weak while keeping the central \(\mathbb{Z}_2\) strong. The resulting phase is an SWSSB with intrinsic ASPT.
 
With these condensations in place, we can now derive the bulk topological response. 
After the first step, in which \(\langle e^{4}, \bar{e}^{4} \rangle\) is condensed, the remaining gauge fields satisfy 
\(\hat{a}, \hat{b} \in \tfrac{\pi}{2}\mathbb{Z}_4\).  
Write \(\hat{a} = \tfrac{\pi}{2}\hat{a}'\) and \(\hat{b} = \tfrac{\pi}{2}\hat{b}'\) with \(\hat{a}', \hat{b}' \in \mathbb{Z}_4\); 
substituting these into the BF action reproduces exactly the effective action given in Eq.~\ref{Z4-SymTFT}. 
Because the coefficient of the action is $\pi/2$, it is invariant under shifts by four. Therefore, the integrals \(\int \hat{a}\) and \(\int \hat{b}\) also take values in \(\mathbb{Z}_4\). The remainder of the analysis then follows the same steps as in the \(\mathbb{Z}_4\) intrinsic ASPT case.


\end{enumerate}

\section{Conclusion and Outlook}

In this work, we explored the categorical Landau paradigm for non-anomalous, group-like symmetries in \((1+1)\)D open quantum systems. By employing canonical purification for mixed states and their symmetries, we showed that the SymTFT construction naturally becomes a doubled copy of the corresponding pure-state phase with the same symmetry. To classify mixed-state phases using SymTFT, we derive additional constraints of Hermiticity and positivity on the condensable algebras that define boundary conditions. We presented a variety of mixed-state gapped phases, uncovering rich structures including SWSSB, ASPT, and intrinsic ASPT phases. Furthermore, we investigated mixed-state phase transition points for \(\mathbb{Z}_q\) symmetries (with prime \(q\)), supported by explicit lattice model realizations. Finally, we examined the gauging procedure for mixed-state phases from multiple perspectives -- including categorical analysis, lattice constructions, and TQFT approaches. The open SymTFT framework for mixed-state phases opens up a wide array of future directions.

Our framework based on positivity constraints provides a systematic way to determine which condensation patterns are possible in mixed states. However, it does not directly address the question of two-way channel connectivity -- i.e., whether different mixed states can be two-way connected through local finite-depth quantum channels. Recent studies\cite{Ellison:2024svg,Sang2024renormalization,Ramanjit2025noisy,Sang2025stability,sang2025mixedstatephases,yang2025topologicalmixedstates,lessa2025higherformanomaly} have approached the classification of mixed-state topological order from this channel-connectivity perspective and introduced the intriguing notion of intrinsic mixed-state topological order. It would be interesting to further explore how this perspective relates to our framework and whether a unified picture can emerge.

More fruitful results could be yielded by extending the calculations and discussions to higher-dimensional and higher categorical symmetries in mixed-state quantum systems\cite{Liu:2024znj,Luo:2023ive}. For example, in the $(2+1)$D case, we could consider a 2-group $\mathscr{G}$ and discuss possible phases through analyzing allowed boundaries of the doubled Drinfeld center of the symmetry category ~\cite{decoppet2024drinfeldcentersmoritaequivalence}
\be\CZ(2{\rm Vec}_{\mathscr{G}})\boxtimes \overline{\CZ(2{\rm Vec}_{\mathscr{G}})} \ee
with corresponding constraints from double space.

It would be natural to extend our discussions to anomalous symmetries. In \cite{Wang:2024vjl}, it is demonstrated that in $d$-dimensional theory, if the symmetry is $\Gamma = G\times K$, where $K$ is strong and $G$ is weak, then the anomaly is classified by
\be \bigoplus_{k = 0}^{d} H^{k}(G,H^{d+1-k}(K,U(1))) \ .  \ee
Applying the same SymTFT construction for categorical symmetries with anomalies will elucidate the anomalous phases and the effect of anomaly on the possible SymTFT boundary conditions. The Morita equivalence is also to be discussed in the mixed-state setting. We can already consider such scenario using the results of $D_4$ symmetry by considering
\be \CZ({\rm Vec}_{D_4}) = \CZ\Big({\rm Vec}_{(\bbZ_2)^3}^\omega\Big)  \ee
for a certain $\omega\in H^3\big((\bbZ_2)^3,U(1)\big)$.

It is straightforward to extend our formalism to discuss fermionic phases and their phase transition points by considering fermionic SymTFT \cite{Wen:2024udn,Chen:2024ulc}. It would be intriguing for future works to find lattice models to realize these mixed-state fermionic phases and dualities.

It is curious to see if it is possible to extend the discussion to continuous symmetries and achieve a systematic classification of mixed-state phases with continuous symmetries. We have demonstrated the mixed-state SymTFT construction for $U(1)$ symmetry and given boundary conditions with SWSSB and ASPT in Sec.\ref{sec:U(1)}. Several formulations of SymTFT for non-Abelian continuous symmetries has also been proposed~\cite{Bonetti:2024cjk,Tian:2025ooo}. It is tempting to apply the SymTFT method to develop Goldstone counting rules for SWSSB in physical systems\cite{Huang2025}.

To incorporate the information of $G$-ality and their corresponding defects, the SymTFT $\CZ(\CC)$ can be extended by a group $G$ into a SymSET \cite{Lu:2025gpt}. Constructing a SymSET for mixed-state systems and realizing them on lattice systems will provide a more in-depth perspective on into the dualities of mixed-state phases.

\section*{Acknowledgement}
We thank Meng Cheng, Sheng-Jie Huang, Wenjie Ji, Qiang Jia, Zhian Jia,  Ruochen Ma, Carolyn Zhang, Gen Yue, Xiao-Gang Wen for various helpful discussions, and Boyan Liu for the early stage collaboration. RL especially thanks Gen Yue for illuminating discussions into the details of modular tensor categories. RL and YNW are supported by National Natural Science Foundation of China under Grant No. 12175004, No. 12422503 and by Young Elite Scientists Sponsorship Program by CAST (2024QNRC001). This research was supported in part by Perimeter Institute for Theoretical Physics. Research at Perimeter Institute is supported by the Government of Canada through the Department of Innovation, Science and Economic Development and by the Province of Ontario through the Ministry of Colleges and Universities. ZB acknowledge support from a startup fund from the Pennsylvania State University and the NSF CAREER Grant DMR-2339319.

\textbf{Note added:} While finalizing this manuscript, we became aware of several related studies~\cite{SymTaco, ZhangSymTFT} with potentially overlapping results, which are expected to appear in the same arXiv posting.

\bibliography{Ref}

\appendix
\onecolumngrid
\section{Categorical Landau Paradigm}
\label{app:CatLanPar}
In this appendix we briefly review the categorical Landau paradigm for $(1+1)$D phases~\cite{Bhardwaj:2023fca,Xu:2022rtj,Bhardwaj:2023idu,Hai:2023osv,Bhardwaj:2024qrf,Chen:2024ulc}. The original Landau paradigm classifies phases of matter based on the presence of local order parameters that break symmetries. While this framework proved highly successful, it fell short with the discovery of new topological phases—such as symmetry-protected topological (SPT) phases, symmetry-enriched topological (SET) phases, and topological order (TO). The categorical Landau paradigm extends this framework by employing SymTFT to capture more refined symmetry behaviors across different phases. This leads to a more comprehensive classification of phases of matter that encompasses both conventional and topological phases.

\begin{figure}[h]
    \centering
    \includegraphics[width=0.5\linewidth]{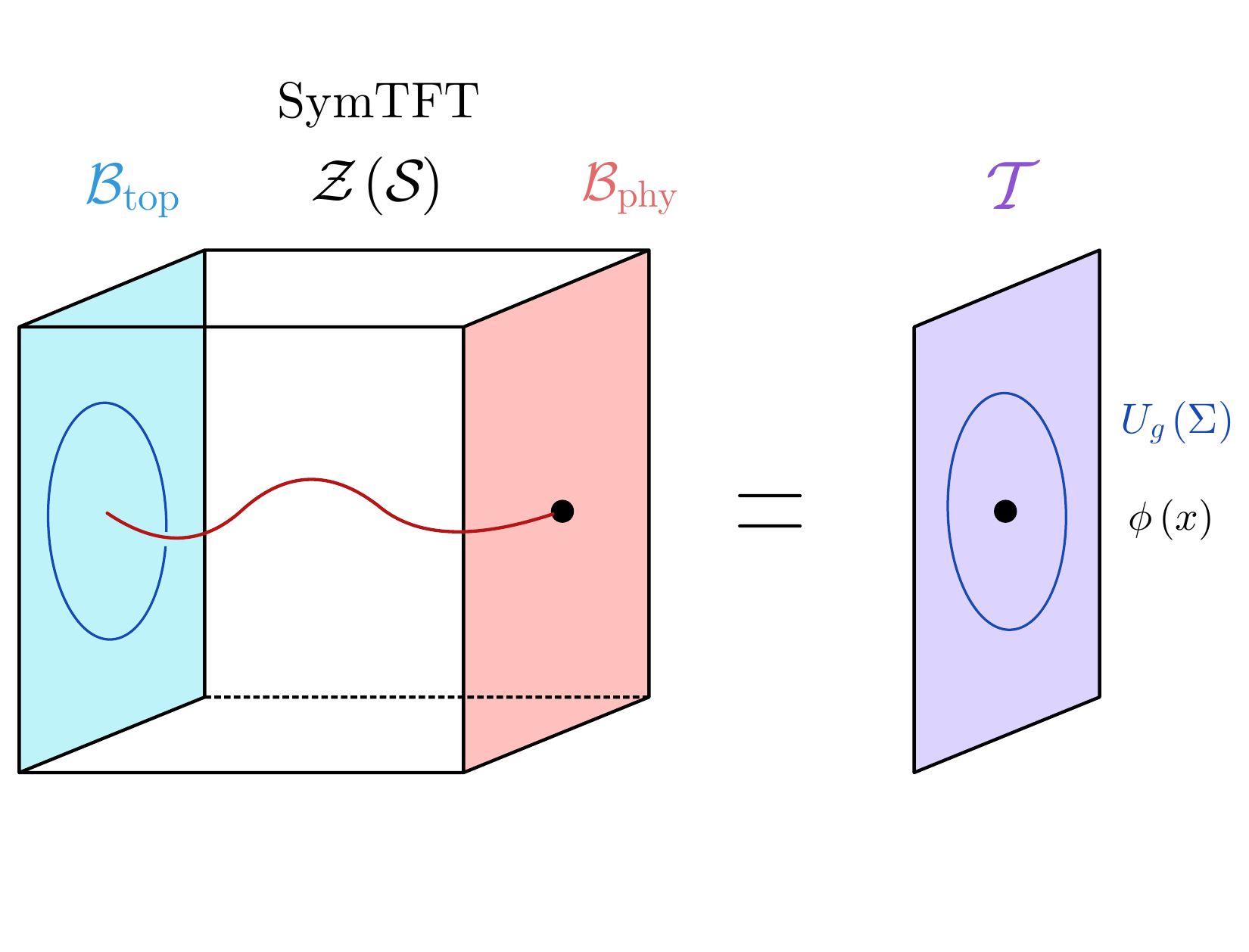}
    \caption{Picture of SymTFT. On the right is a theory $\CT$ on $M^d$ with symmetry operator $U_g(\Sigma)$ acting on operator $\phi(x)$ by linking. On the left we have a TQFT on $M^d\times [0,1]$, with two boundaries $\CB_{\rm top}$ in cyan and $\CB_{\rm phy}$ in red. The $\phi(x)$ operator charged under the symmetry remains on $\CB_{\rm phy}$, while the dark blue symmetry operator is pushed to $\CB_{\rm top}$, linking with the Wilson line attached to $\phi(x)$.}
    \label{fig:SymTFT}
\end{figure}

With a modernized understanding of symmetry as topological defects, predecessors developed the method to extract the information of topological operators in theory $\CT$ on $M^d$ into a TQFT on $M^d\times [0,1]$. This TQFT in named SymTFT, graphically shown as Fig.\ref{fig:SymTFT}, and it has two boundaries 
\begin{enumerate}
    \item Topological boundary $\CT_{\rm top}$ on $M^d \times \{ 0 \} $, where all information of topological operators are pushed to;
    \item Physical boundary $\CT_{\rm phy}$ on $M^d \times \{ 1 \}$, where all the dynamical information stays.
\end{enumerate}
The topological boundary determines the symmetry category and the charges (representations) of the symmetry, while the physical boundary determines which charge operators can tunnel through the SymTFT, leading to the spontaneous breaking of the symmetry in the IR sector.

Specifically, the $(2+1)$D SymTFT of a $(1+1)$D theory can be described by the Drinfeld center of the symmetry category $\CS$ of $\CT$. For the case considered in this article, the symmetry categories of non-anomalous 0-form symmetries are taken to be of the form ${\rm Vec}_G$, the category of $G$-graded vector spaces. The Drinfeld center $\CZ({\rm Vec}_G)$ is a modular tensor category with the following data\cite{Coste:2000tq}:
\begin{enumerate}
    \item Simple anyons are labeled by $([g],\rho)$, a conjugacy class in $G$ and an irreducible representation of its centralizer (denoted $N(C_{g})$);
    \item An $S$-matrix giving the information of anyon braiding calculated by
    \be  S_{([g],\rho),([h],\sigma)} = \frac{1}{\abs{G}} \sum_{a\in [g]} \sum_{b\in [h]\cap N(C_a)} \Tr_\rho(xbx^{-1})^* \cdot \Tr_{\sigma}(yay^{-1})^* \ , \ee
    where $x$ and $y$ are solutions to $a = x^{-1} g x$ and $b = y^{-1} h y$.
    \item A $T$-matrix giving the information of anyon spin, calculated by
    \be T_{([g],\rho),([h],\sigma)}  = \delta_{[g],[h]} \delta_{\rho,\sigma} \frac{\Tr_\rho(g)}{\Tr_\rho(e)} \ . \ee
\end{enumerate}
The fusion rule of anyon lines is given by ($\alpha,\beta,\gamma$ are labels of simple anyons, $0= ([e],\mathds{1})$ labels the trivial anyon)
\be\ba  \label{eq:fusionrule}
&W_{\alpha}\otimes W_{\beta} = \sum_\gamma \CN_{\alpha \beta}^\gamma W_\gamma\\
&\CN_{\alpha \beta}^\gamma=\sum_{\delta=0}^{N-1} \frac{S_{\delta \alpha}^* S_{\delta \beta}^* S_{\delta \gamma}}{S_{0 \delta}} \ . \ea\ee
The quantum dimension of each simple anyon is
\be d_{([g],\rho)} = \frac{S_{([g],\rho),([e],\mathds{1})}}{S_{([e],\mathds{1}),([e],\mathds{1})}} = \Big|[g]\Big|\cdot {\rm dim}(\rho) \ee

The two boundary conditions are given by anyon condensations. Anyon condensations are classified by commutative separable algebras (also named condensable algebras). Condensable algebra is equivalently defined as an algebra
\be \CA =  \bigoplus_{\alpha} N_\alpha \alpha  \ee
invariant under $S$ and $T$ matrix
\be\ba
S\CA &= \bigoplus_{\alpha,\beta} S_{\alpha,\beta} N_\beta \beta = \CA \\
T\CA &= \bigoplus_{\alpha,\beta} T_{\alpha,\beta} N_\beta \beta = \CA \ .
\ea\ee
Lagrangian algebra is defined to be a ``maximal" condensable algebra $\CA =  \bigoplus_{\alpha} N_\alpha \alpha$, fulfilling the quantum dimension constraint
\be
N_\alpha d_\alpha  = \sqrt{\sum_\alpha d_\alpha^2} \equiv{\rm Qdim}(\CZ({\rm Vec}_G)) \ .
\ee

The categorical Landau paradigm utilize the anyon condensation on both the topological and physical boundaries to generate the information of phases. For the purpose of phase classification, the topological boundary condition is taken to be condensation of all charge anyons $\{ ([e],\rho): \rho \in {\rm Irrep}(G)\}$
\be  \CL_\CS = \bigoplus_{\rho \in {\rm Irrep}(G)} {\rm dim}(\rho)  ([e],\rho) \ . \ee
Each gapped phase corresponds to condensing a Lagrangian algebra $\CL$ on the physical boundary,
\begin{enumerate}
    \item if $\CL$ contains any charge anyon, this is a spontaneously symmetry breaking (SSB) phase;
    \item if $\CL$ contains no charge anyon, this is an SPT phase.
\end{enumerate}
Each gapless phase corresponds to a non-Lagrangian condensable algebra $\CA$ on the physical boundary
\begin{enumerate}
    \item if $\CA$ contains any charge anyon, this is a gapless SSB (gSSB) phase;
    \item if $\CA$ contains charge anyon and cannot be deformed into gapped SSB phase, then this is an intrinsically gapless SSB (igSSB) phase;
    \item if $\CA$ contains no charge anyon, this is a gapless SPT (gSPT) phase;
    \item if the phase $\CA$ cannot be deformed into a gapped SPT phase, then this is an intrinsically gapless SPT (igSPT) phase.
\end{enumerate}

\section{Modular Data}
In this section we present the modular data in the modular tensor categories we encountered in this paper.

\subsection{$\CZ({\rm Vec}_{\bbZ_2})$}
There are four simple anyons labelled
\be \Big\{ ([0],r_+),([0],r_-),([1],r_+),([1],r_-)   \Big\}  \ee
where $r_+$ is the trivial $1$-dimensional representation and $r_-$ is the non-trivial one. They are more frequently referred to as $\{ 1,e,m,f \}$ respectively.

The $S$ matrix is
\be S = \frac{1}{2}\mqty( 1 & 1 & 1 & 1 \\ 1 & 1 & -1 & -1 \\ 1 & -1 & 1 & -1 \\ 1 & -1 & -1 & 1) \ , \ee
and the $T$ matrix is
\be T = {\rm Diag}(1,1,1,-1) \ . \ee

\subsection{$\CZ({\rm Vec}_{\bbZ_4})$}
There are 16 simple anyons, labelled by
\be \{ e^am^b \} \ , \ a,b\in \bbZ_4 \ ,  \ee
where $e = ([0],\phi_1), \  \phi_1:1\mapsto e^{\pi i /2})$ and $m = ([1],\phi_0)$ with $\phi_0$ being the trivial representation.
The modular $S$ and $T$ matrices are
\be  S = \frac{1}{4}\mqty(1 & 1 & 1 & 1 & 1 & 1 & 1 & 1 & 1 & 1 & 1 & 1 & 1 & 1 & 1 & 1 \\ 1 & 1 & 1 & 1 & -i & -i & -i & -i & -1 & -1 & -1 & -1 & i & i & i & i \\ 1 & 1 & 1 & 1 & -1 & -1 & -1 & -1 & 1 & 1 & 1 & 1 & -1 & -1 & -1 & -1 \\ 1 & 1 & 1 & 1 & i & i & i & i & -1 & -1 & -1 & -1 & -i & -i & -i & -i \\ 1 & -i & -1 & i & 1 & -i & -1 & i & 1 & -i & -1 & i & 1 & -i & -1 & i \\ 1 & -i& -1 & i & -i & -1 & i & 1 & -1 & i & 1 & -i & i & 1 & -i & -1 \\ 1 & -i & -1 & i & -1 & i & 1 & -i & 1 & -i & -1 & i & -1 & i & 1 & -i \\ 1 & -i & -1 & i & i & 1 & -i & -1 & -1 & i & 1 & -i & -i & -1 & i & 1 \\ 1 & -1 & 1 & -1 & 1 & -1 & 1 & -1 & 1 & -1 & 1 & -1 & 1 & -1 & 1 & -1 \\ 1 & -1 & 1 & -1 & -i & i & -i & i & -1 & 1 & -1 & 1 & i & -i & i & -i\\ 1 & -1 & 1 & -1 & -1 & 1 & -1 & 1 & 1 & -1 & 1 & -1 & -1 & 1 & -1 & 1 \\ 1 & -1 & 1 & -1 & i & -i & i & -i & -1 & 1 & -1 & 1 & -i & i & -i & i \\ 1 & i & -1 & -i & 1 & i & -1 & -i & 1 & i & -1 & -i & 1 & i & -1 & -i \\ 1 & i & -1 & -i & -i & 1 & i & -1 & -1 & -i & 1 & i & i & -1 & -i & 1 \\ 1 & i & -1 & -i & -1 & -i & 1 & i & 1 & i & -1 & -i & -1 & -i & 1 & i \\ 1 & i & -1 & -i & i & -1 & -i & 1 & -1 & -i & 1 & i & -i & 1 & i & -1)  \ , \ee \be T ={\rm Diag}(1,1,1,1,1,i,-1,-i,1,-1,1,-1,1,-i,-1,i) \ . \ee
\subsection{$\CZ({\rm Vec}_{S_3})$}
\label{app:moddataS3}
There are $8$ simple anyons are labelled
\be \Big\{ ([e],r_+),([e],r_-),([e],E),([a],\varphi_0),([a],\varphi_1), ([a],\varphi_2),([b],\phi_0),([b],\phi_1)  \Big\} \ , \ee
and referred to as
\be
\Big\{ W_1,W_2,W_3,W_4,W_5,W_6,W_7,W_8 \Big\} \ .
\ee

The $S$ matrix is
\be  S =  \frac{1}{6} \mqty( 1 & 1 & 2 & 2 & 2 & 2 & 3 & 3 \\ 1 & 1 & 2 & 2 & 2 & 2 & -3 & -3 \\ 2 & 2 & 4 & -2 & -2 & -2 & 0 & 0 \\ 2 & 2 & -2 & 4 & -2 & -2 & 0 & 0 \\ 2 & 2 & -2 & -2 & -2 & 4 & 0 & 0 \\ 2 & 2 & -2 & -2 & 4 & -2 & 0 & 0 \\ 3 & -3 & 0 & 0 & 0 & 0 & 3 & -3 \\ 3 & -3 & 0 & 0 & 0 & 0 & -3 & 3 ) \ , \ee
and the $T$ matrix is
\be T = {\rm Diag}(1,1,1,1,\omega,\omega^2,1,-1) \ ,  \ee
where $\omega = \exp(2\pi i/3)$.

\subsection{$\CZ({\rm Vec}_{D_4})$}\label{app:d4modular}
The group $D_4$ is presented as
\be D_4 = \expval{a,b|a^4 = b^2 = e, ab = ba^3}  \ .\ee
There are $22$ simple anyons labeled \cite{Bhardwaj:2024qrf}

\be\ba \Big\{  &([e],1),([e],1_b),([e],1_{ab}),([a^2],1_a),([e],1_a),([a^2],1_b),([a^2],1_{ab}),([a^2],1),([ab],+-),([ab],--), ([b],+-), \\ &([b],--),([e],E),([a^2],E), ([a],1),([a],-1),([b],++),([b],-+),([ab],++),([ab],-+),([a],i),([a],-i)\Big\}  \ , \ea\ee
where $1,1_a,1_b,1_{ab},E$ are the irreducible representations of $D_4$ leaving their subscript conjugacy class invariant, $E$ is the $2$-dimensional representation, $1,i,-1,-i$ are the $4$ irreducible representations of $\bbZ_4$ and $++,+-,--,-+$ are the $4$ irreducible representations of $\bbZ_2\times \bbZ_2$. These anyons are referred to as \cite{Bhardwaj:2024qrf}
\be  \big\{ 
1 , e_R ,e_G, e_B ,e_{RG} ,e_{GB}, e_{RB}, e_{RGB}, m_R ,f_R, m_G, f_G, m_B ,f_B, m_{RG} ,f_{RG}, m_{GB}, f_{GB}, m_{RB}, f_{RB}, s_{RGB} ,s^*_{RGB}
\big\} \ . \ee
The nomenclature comes from the construction of $\CZ({\rm Vec}_{D_4})\cong \CZ({\rm Vec}_{\bbZ_2\times \bbZ_2\times \bbZ_2}^\omega)$.

The $S$-matrix is
\be
S = \frac{1}{8}\mqty(1 & 1 & 1 & 1 & 1 & 1 & 1 & 1 & 2 & 2 & 2 & 2 & 2 & 2 & 2 & 2 & 2 & 2 & 2 & 2 & 2 & 2 \\ 
1 & 1 & 1 & 1 & 1 & 1 & 1 & 1 & -2 & -2 & 2 & 2 & 2 & 2 & -2 & -2 & 2 & 2 & -2 & -2 & -2 & -2 \\ 
1 & 1 & 1 & 1 & 1 & 1 & 1 & 1 & 2 & 2 & -2 & -2 & 2 & 2 & -2 & -2 & -2 & -2 & 2 & 2 & -2 & -2 \\ 
1 & 1 & 1 & 1 & 1 & 1 & 1 & 1 & 2 & 2 & 2 & 2 & -2 & -2 & 2 & 2 & -2 & -2 & -2 & -2 & -2 & -2 \\ 
1 & 1 & 1 & 1 & 1 & 1 & 1 & 1 & -2 & -2 & -2 & -2 & 2 & 2 & 2 & 2 & -2 & -2 & -2 & -2 & 2 & 2 \\ 
1 & 1 & 1 & 1 & 1 & 1 & 1 & 1 & 2 & 2 & -2 & -2 & -2 & -2 & -2 & -2 & 2 & 2 & -2 & -2 & 2 & 2 \\ 
1 & 1 & 1 & 1 & 1 & 1 & 1 & 1 & -2 & -2 & 2 & 2 & -2 & -2 & -2 & -2 & -2 & -2 & 2 & 2 & 2 & 2 \\ 
1 & 1 & 1 & 1 & 1 & 1 & 1 & 1 & -2 & -2 & -2 & -2 & -2 & -2 & 2 & 2 & 2 & 2 & 2 & 2 & -2 & -2 \\ 
2 & -2 & 2 & 2 & -2 & 2 & -2 & -2 & 4 & -4 & 0 & 0 & 0 & 0 & 0 & 0 & 0 & 0 & 0 & 0 & 0 & 0 \\ 
2 & -2 & 2 & 2 & -2 & 2 & -2 & -2 & -4 & 4 & 0 & 0 & 0 & 0 & 0 & 0 & 0 & 0 & 0 & 0 & 0 & 0 \\ 
2 & 2 & -2 & 2 & -2 & -2 & 2 & -2 & 0 & 0 & 4 & -4 & 0 & 0 & 0 & 0 & 0 & 0 & 0 & 0 & 0 & 0 \\ 
2 & 2 & -2 & 2 & -2 & -2 & 2 & -2 & 0 & 0 & -4 & 4 & 0 & 0 & 0 & 0 & 0 & 0 & 0 & 0 & 0 & 0 \\ 
2 & 2 & 2 & -2 & 2 & -2 & -2 & -2 & 0 & 0 & 0 & 0 & 4 & -4 & 0 & 0 & 0 & 0 & 0 & 0 & 0 & 0 \\ 
2 & 2 & 2 & -2 & 2 & -2 & -2 & -2 & 0 & 0 & 0 & 0 & -4 & 4 & 0 & 0 & 0 & 0 & 0 & 0 & 0 & 0 \\ 
2 & -2 & -2 & 2 & 2 & -2 & -2 & 2 & 0 & 0 & 0 & 0 & 0 & 0 & 4 & -4 & 0 & 0 & 0 & 0 & 0 & 0 \\ 
2 & -2 & -2 & 2 & 2 & -2 & -2 & 2 & 0 & 0 & 0 & 0 & 0 & 0 & -4 & 4 & 0 & 0 & 0 & 0 & 0 & 0 \\ 
2 & 2 & -2 & -2 & -2 & 2 & -2 & 2 & 0 & 0 & 0 & 0 & 0 & 0 & 0 & 0 & 4 & -4 & 0 & 0 & 0 & 0 \\ 
2 & 2 & -2 & -2 & -2 & 2 & -2 & 2 & 0 & 0 & 0 & 0 & 0 & 0 & 0 & 0 & -4 & 4 & 0 & 0 & 0 & 0 \\ 
2 & -2 & 2 & -2 & -2 & -2 & 2 & 2 & 0 & 0 & 0 & 0 & 0 & 0 & 0 & 0 & 0 & 0 & 4 & -4 & 0 & 0 \\ 
2 & -2 & 2 & -2 & -2 & -2 & 2 & 2 & 0 & 0 & 0 & 0 & 0 & 0 & 0 & 0 & 0 & 0 & -4 & 4 & 0 & 0 \\ 
2 & -2 & -2 & -2 & 2 & 2 & 2 & -2 & 0 & 0 & 0 & 0 & 0 & 0 & 0 & 0 & 0 & 0 & 0 & 0 & 4 & -4 \\ 
2 & -2 & -2 & -2 & 2 & 2 & 2 & -2 & 0 & 0 & 0 & 0 & 0 & 0 & 0 & 0 & 0 & 0 & 0 & 0 & -4 & 4) \ ,
\ee
and the $T$-matrix is
\be T = {\rm Diag}(1, 1, 1, 1, 1, 1, 1, 1, 1, -1, 1, -1, 1, -1, 1, -1, 1, -1, 1, -1, i, -i)\ee

\section{Gauging}
\label{app:Gauging}
Within the SymTFT, we can define gauging in the following equivalent ways\cite{Chen:2024ulc}:
\begin{enumerate}
    \item changing topological boundary condition;
    \item summing over defect sectors.
\end{enumerate}
To explicitly relate the two definitions, we derive the symmetry sectors of the partition vector for the physical theory, and observe how it transforms under gauging. In some cases where gauging maintains the symmetry category invariant, gauging can be realized by an automorphism of anyons in the modular tensor category.

\begin{figure}
    \centering
    \includegraphics[width=0.4\linewidth]{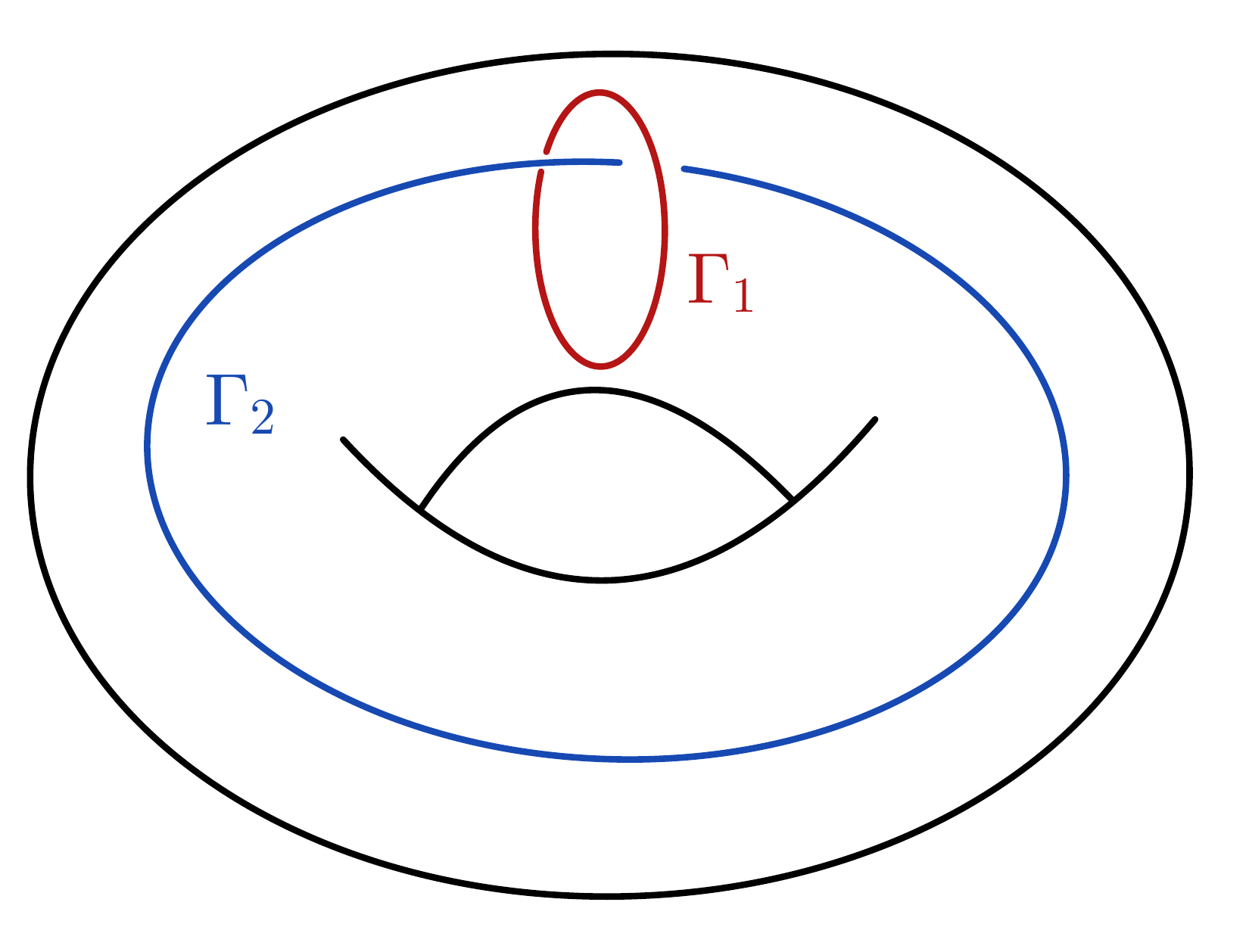}
    \caption{The image of solid torus. We label the two directions of loops as the above. The dark red loop gives $\Gamma_1$, and the dark blue loop gives $\Gamma_2$. The topological boundary is parallel to $\Gamma_2$.}
    \label{fig:solidtorus}
\end{figure}

Consider placing the theory and SymTFT on solid torus $D^2\times S^1$: physical boundary torus on the outside and topological boundary on the inside, shrunk as a line. Then the boundary conditions are vectors in $\CH(T^2)$, spanned by simple anyons. A condensation is represented as
\be
\CA = \bigoplus_\alpha N_\alpha \alpha \mapsto \ket{\CA} = \sum_\alpha N_\alpha \ket{\alpha} \ .
\ee
Anyon loops can wind around two directions $\Gamma_{1,2}$, shown as Fig.\ref{fig:solidtorus}. 
Anyon loops act on $\CH(T^2)$ as
\be W_\alpha(\Gamma_1) \ket{\beta}= \frac{S_{\beta\alpha}}{S_{\beta 0}} \ket{\beta} \ , \ W_\alpha(\Gamma_2) \ket{\beta}= \sum_{\gamma} \CN_{\alpha\beta}^\gamma \ket{\gamma} \ .\ee
From the topological boundary condition $\ket{\CL_\CS}$ acted by anyon lines, we can construct another basis of $\CH(T^2)$, which we label by $\{\ket{i}_{\CL_\CS}\}$. This basis is often not orthonormal, and metric of this basis is given by
\be g_{ij} = _{\CL_\CS}\expval{i|j}_{\CL_\CS} \ . \ee
The the physical boundary condition can be written in both the anyon basis and the $\CL_S$ basis
\be \ket{\chi}_\CT = \sum_{\alpha} \chi_\alpha \ket{\alpha} =\sum_{ij}g^{ij}Z_{\CT}[i]\ket{j}_{\CL_\CS} \ . \ee
The $Z_{\CT}[i]$ is interpreted as the partition vector component w.r.t. the symmetry sector given by $[i]$.
Gauging changes the topological boundary condition into $\ket{\CL'}$, and thus generates a different basis $\{\ket{i'}_{\CL'}\}$. We can again expand the physical boundary condition as
\be \ket{\chi}_\CT =\sum_{ij}g^{ij}Z_{\CT}[i]\ket{j}_{\CL_\CS} =\sum_{i'j'}g^{i'j'}Z_{\CT}[i']\ket{j'}_{\CL'}  \ . \ee
The gauging pattern as summing over defect sectors can be explicitly observed from expressing $Z_\CT[i']$ in terms of $Z_\CT[i]$. The partition vector components in the gauged symmetry sectors can be obtained by linear sums of those in the ungauged symmetry sectors,
\be  Z_\CT[i'] = \sum_{j} c_{\CL_S,\CL}^{i',j} Z_\CT[j]  \ .  \ee
The non-zero coefficients $c$ entails the information of how summing over sectors with different defect configurations contribute to the partition functions of symmetry sector $[i']$ after gauging.

In the following sections, we explicitly demonstrate the details of gauging in $\CS = {\rm Vec}_{\bbZ_2\times \overline{\bbZ_2}}$ case, and perform the calculation for the $S_3\times \overline{S_3}$ case.

\subsection{Gauging in $\bbZ_2\times\overline{\bbZ_2}$}
\label{sec:gaugeZ2Z2}

\subsubsection{Gauging $\bbZ_2\times\overline{\bbZ_2}$}
As previously stated, there are three ways to describe this gauging, the first two are
\begin{enumerate}
    \item change topological boundary from $\CL_\CS = 1\oplus e \oplus \bar{e} \oplus e\bar{e}$ to $\CA_1 = 1\oplus m\oplus \bar{m} \oplus m\bar{m}$;
    \item automorphism of MTC, by relabeling simple anyons
    \be 1\leftrightarrow 1 \ , \ e  \leftrightarrow m \ , \ \bar{e} \leftrightarrow \bar{m} \ , \ee
    and thus $\CL_\CS \mapsto \CA_1$.
\end{enumerate}

For the third perspective, we first construct the sectors and physical partition vector. Since
\be\CZ({\rm Vec}_{G\times G})\cong \CZ({\rm Vec}_{G})\boxtimes \overline{\CZ({\rm Vec}_{G})} \ , \ee 
we can do the decomposition
\be\ba
\ket{\chi_\CT} = &\Big( Z_\CT[00]\ket{00}+ Z_\CT[01]\ket{01} +Z_\CT[10] \ket{10} + Z_\CT[11]\ket{11}  \Big) \\ &\otimes \Big(Z_\CT[\overline{00}]\ket{\bar{0}\bar{0}}+ Z_\CT[\overline{01}]\ket{\bar{0}\bar{1}} +Z_\CT[\overline{10}] \ket{\bar{1}\bar{0}} + Z_\CT[\overline{11}]\ket{\bar{1}\bar{1}}  \Big)
\ea\ee
where we denote
\be\ba &\ket{00} = \ket{1} + \ket{e} \ , \ &\ket{01} =  \ket{1}-\ket{e} =W_m[\Gamma_1]\ket{00} \\ &\ket{10} = \ket{m} + \ket{f} =W_m[\Gamma_2]\ket{00}\ , \ &\ket{11} = \ket{m}-\ket{f}=TW_m[\Gamma_2]\ket{00} \  .\ea\ee
Through gauging $\bbZ_2\times \overline{\bbZ_2}$, the new basis becomes (the new anyons after gauging are given by $e'= m$ and $m'=e$)
\be\ba
&\ket{00}' = \ket{1} + \ket{m} = \ket{1}' + \ket{e}' \ , \  &\ket{01}' =  \ket{1}-\ket{m} =\ket{1}' -\ket{e}'= W_e[\Gamma_1]\ket{00}' \\ &\ket{10}' = \ket{e} + \ket{f} = \ket{m}' + \ket{f}'= W_e[\Gamma_2]\ket{00}'\ , \ &\ket{11}' = \ket{e}-\ket{f} = \ket{m}' + \ket{f}'=TW_e[\Gamma_2]\ket{00}'\ .
\ea\ee
Thus the new partition vector components become
\be Z_{\CT/\bbZ_2\times \overline{\bbZ_2}}[a'b'\overline{c'd'}] \equiv Z_{\CT/\bbZ_2\times \overline{\bbZ_2}}[a'b']Z_{\CT/\bbZ_2\times \overline{\bbZ_2}}[\overline{c'd'}] = \Bigg( \frac{1}{2}\sum_{ab}(-1)^{ab' + a'b}Z_\CT[ab]  \Bigg)\cdot \Bigg( \frac{1}{2}\sum_{cd}(-1)^{\overline{cd'} + \overline{c'}\overline{d}}Z_\CT[\overline{cd}]  \Bigg)  \ . \ee

Now we calculate how the three gapped phases of our concern transform under this gauging
\be \ket{1,e,\bar{e},e\bar{e}} = \ket{00}\otimes \ket{\overline{00}} \Rightarrow Z_\CT[ab\overline{cd}] = \delta_{a,0}\delta_{b,0}\delta_{\bar{c},0}\delta_{\bar{d},0} \Rightarrow Z_{\CT/\bbZ_2\times \overline{\bbZ_2}} [a'b'\overline{c'd'}] = \frac{1}{4}
\ee
which sums up to be $\ket{\CA}' = \ket{1,m',\bar{m}',m'\bar{m}'}$.
The SWSSB phase $\ket{\CM} = 1\oplus e\bar{e}\oplus m\bar{m} \oplus f\bar{f}$ is invariant under such transformation
\be\ba
&\ket{\CM} = \frac{1}{2}\big(\ket{00\overline{01}}+\ket{01\overline{00}}+\ket{10\overline{11}}+\ket{11\overline{10}}\big) \\
&\Rightarrow Z_\CT[00\overline{01}]= Z_\CT[01\overline{00}] =Z_\CT[10\overline{11}]=Z_\CT[11\overline{10}] = \frac{1}{2} \\
&\Rightarrow Z_{\CT/\bbZ_2\times \overline{\bbZ_2}}[00\overline{01}]= Z_{\CT/\bbZ_2\times \overline{\bbZ_2}}[01\overline{00}] =Z_{\CT/\bbZ_2\times \overline{\bbZ_2}}[10\overline{11}]\\ &=Z_{\CT/\bbZ_2\times \overline{\bbZ_2}}[11\overline{10}] = \frac{1}{2} \ ,
\ea\ee
where the ignored coefficients are zero. We see that the partition vector do not change, signifying the invariance of SWSSB phase under gauging $\bbZ_2\times \overline{\bbZ_2}$.

\subsubsection{Gauging $\bbZ_2^d$}
\label{app:gaugeZ2d}
For the convenience of gauging diagonal $\bbZ_2^d\subset \bbZ_2\times \overline{\bbZ_2}$, we regroup the anyons as
\be  \{ 1,e,m\bar{m},f\bar{m}  \} \otimes \{ 1,e\bar{e},\bar{m},e\bar{f} \} \ .  \ee
Thus the gauging process only happens on the first set of anyons. The physical boundary before gauging read
\be\ba \ket{\chi}  &= \Big(\sum_{a,b\in\{0,1\}} Z[ab]\ket{(a,b)} \Big)\otimes \Big( \sum_{x,y\in\{0,1\}} Z[xy]\ket{(x,y)}  \Big) \ .  \ea\ee
where the sectors are defined by
\be\ba
\ket{(0,1)} &\equiv W_{(0,1)}[\Gamma_1]\ket{\Omega}_{\rm top} \\
\ket{(1,0)} &\equiv W_{(0,1)}[\Gamma_2]\ket{\Omega}_{\rm top}
\\
\ket{(1,1)} &\equiv T W_{(0,1)}[\Gamma_2]\ket{\Omega}_{\rm top} 
\ea  \ee
and thus applying to $a,b$
\be\ba
\ket{(0,0)} = \ket{1}+ \ket{e}\ &, \ \ket{(0,1)} = \ket{1} - \ket{e} \\
\ket{(1,0)} = \ket{m\bar{m}}+ \ket{f\bar{m}}\ &, \ \ket{(1,1)} = \ket{m\bar{m}}- \ket{f\bar{m}}
\ea\ee
applying to $x,y$
\be\ba
\ket{(0,0)} = \ket{1}+ \ket{e\bar{e}}\ &, \ \ket{(0,1)} = \ket{1} - \ket{e\bar{e}} \\
\ket{(1,0)} = \ket{\bar{m}}+ \ket{e\bar{f}}\ &, \ \ket{(1,1)} = \ket{\bar{m}}- \ket{e\bar{f}}
\ea\ee
Gauging changes the topological boundary, and affects only the first portion after regrouping,
\be  \ket{a'b'xy} = \frac{1}{2}\sum_{a,b} (-1)^{a'b+ab'}\ket{abxy} \Rightarrow Z_{\CT/\bbZ_2^d} [a'b'xy] =  \frac{1}{2}\sum_{a,b} (-1)^{a'b+ab'}Z_{\CT}[abxy]  \ . \ee
We long to see how this plays out for the case of our concern
\be\ba
&\ket{\chi_{\rm phy}} = \ket{1,e\bar{e},m\bar{m},f\bar{f}}= \frac{1}{2}(\ket{0000} + \ket{0100} + \ket{1000} + \ket{1100})\\
&\Rightarrow Z[0000] = Z[0100] = Z[1000] = Z[1100] = \frac{1}{2}
\ea\ee
which under transformation becomes
\be \ba
Z_{\CT/\bbZ_2^d}[0'0'00] &= \frac{1}{2} (Z[0000] + Z[0100] + Z[1000] + Z[1100]) = 1 \\ Z_{\CT/\bbZ_2^d}[0'1'00] &= \frac{1}{2} (Z[0000] + Z[0100] - Z[1000] - Z[1100]) = 0 \\
Z_{\CT/\bbZ_2^d}[1'0'00] &= \frac{1}{2} (Z[0000] - Z[0100] + Z[1000] - Z[1100]) = 0 \\ Z_{\CT/\bbZ_2^d}[1'1'00] &= \frac{1}{2} (Z[0000] + Z[0100] - Z[1000] + Z[1100]) = 0 \ .
\ea   \ee
This aligns with our expectation, since
\be  \ket{0'0'00}  = (\ket{1} + \ket{e'})\otimes (\ket{1} + \ket{e\bar{e}}) = \ket{1,e,\bar{e},e\bar{e}} \ . \ee

The original SSB phase
\be  \ket{1,e,\bar{e},e\bar{e}}  = \ket{0000} \Rightarrow Z[0000] = 1 \ . \ee
It transforms into
\be\ba  Z_{\CT/\bbZ_2^d}[0'0'00] &= \frac{1}{2} (Z[0000] + Z[0100] + Z[1000] + Z[1100]) = \frac{1}{2} \\ Z_{\CT/\bbZ_2^d}[0'1'00] &= \frac{1}{2} (Z[0000] + Z[0100] - Z[1000] - Z[1100]) = \frac{1}{2} \\
Z_{\CT/\bbZ_2^d}[1'0'00] &= \frac{1}{2} (Z[0000] - Z[0100] + Z[1000] - Z[1100]) = \frac{1}{2} \\ Z_{\CT/\bbZ_2^d}[1'1'00] &= \frac{1}{2} (Z[0000] + Z[0100] - Z[1000] + Z[1100]) = \frac{1}{2} \ . \ea \ee
which is exactly $\ket{1,e'\bar{e'},m'\bar{m'},f'\bar{f'}}$.

The original trivial SPT phase
\be  \ket{1,m,\bar{m},m\bar{m}} = \frac{1}{4} \sum_{a,b,x,y}\ket{abxy} \Rightarrow Z[abxy] = \frac{1}{4} \ .\ee
It transforms into
\be  Z_{\CT/\bbZ_2^d}[0'0'00] = Z_{\CT/\bbZ_2^d}[0'0'01] =Z_{\CT/\bbZ_2^d}[0'0'10] = Z_{\CT/\bbZ_2^d}[0'0'11] = \frac{1}{2} \ee
meaning
\be \ket{1,e',\bar{m'},e'\bar{m'}}  \ee
which is not $J$-symmetric. This result is changed when we choose 
\be\{ 1,\bar{e},m\bar{m},m\bar{f}  \} \otimes \{ 1,e\bar{e},m,f\bar{e} \} \ee
which is equally valid for our purpose of gauging $\bbZ_2^d$. This is due to the choice of boundary condition, which we analyze in Sec.\ref{sec:gaugingZ2}.

\subsection{Gauging in $S_3\times \overline{S_3}$}
\label{app:gauginginS3}
The gauging of non-Abelian symmetries is more complicated than Abelian ones, for it cannot be seen as automorphisms of MTC anymore. As an example we demonstrate the procedure of gauging $\bbZ_3^d$. In this specific case, the previous technique of regrouping anyons does not apply either, since $S_3^d$ is no longer a normal subgroup of $S_3\times \overline{S_3}$. Here we present three examples of mixed-state gauging: gauging $\bbZ_2^d$, gauging $\bbZ_3^d$ and gauging $S_3^d$.

The simple anyons of $\CZ({\rm Vec}_{S_3})$ labeled by conjugacy classes and irreducible representations of their centralizers
    \be \big\{ ([e],r_+),([e],r_-),([e],E),([a],\varphi_0),([a],\varphi_1), ([a],\varphi_2),([b],\phi_0),([b],\phi_1)  \big\}  \ee
are labelled as
\be
\{ W_1,W_2,W_3,W_4,W_5,W_6,W_7 ,W_8\} \ .
\ee

For the pure-state theories with $S_3$ symmetry, the physical boundary partition vector can be decomposed into $8$ symmetry sectors
\be\ba  \ket{\chi_\CT} = & Z_\CT[\mathds{1}] \ \ket{\CL_S} + Z_\CT[W_4[\Gamma_1]]\ W_4[\Gamma_1]\ket{\CL_S} + Z_\CT[W_4[\Gamma_2]]\ W_4[\Gamma_2]\ket{\CL_S} \\ &+ Z_\CT[T(W_4[\Gamma_2])] \ T W_4[\Gamma_2]\ket{\CL_S}+
Z_\CT[T^2(W_4[\Gamma_2])]\ T^2 W_4[\Gamma_2]\ket{\CL_S} \\ &+ Z_\CT[W_7[\Gamma_1]] \ W_7[\Gamma_1]\ket{\CL_S} +Z_\CT[W_7[\Gamma_2]] \ W_7[\Gamma_2]\ket{\CL_S} +Z_\CT[T(W_7[\Gamma_2])] \ T W_7[\Gamma_2]\ket{\CL_S} \ . \ea\ee
Then the mixed-state partition function is merely the tensor product of two copies $\ket{\chi_\CT}\otimes \overline{\ket{\chi_\CT}}$, with $64$ symmetry sectors.

\paragraph{Gauging $\bbZ_3^d$.} Gauging $\bbZ_3^d$ is equivalent to switching the topological boundary condition from 
\[
    \mathcal{L}_S = \begin{pmatrix}1 & 1 & 2 & 0 & 0 & 0 & 0 & 0 \\ 1 & 1 & 2 & 0 & 0 & 0 & 0 & 0 \\ 2 & 2 & 4 & 0 & 0 & 0 & 0 & 0 \\ 0 & 0 & 0 & 0 & 0 & 0 & 0 & 0 \\ 0 & 0 & 0 & 0 & 0 & 0 & 0 & 0 \\ 0 & 0 & 0 & 0 & 0 & 0 & 0 & 0 \\ 0 & 0 & 0 & 0 & 0 & 0 & 0 & 0 \\ 0 & 0 & 0 & 0 & 0 & 0 & 0 & 0\end{pmatrix}
    \]
into
\[
    \CM^{\rm{SWSSB }}_{11}= \begin{pmatrix}1 & 1 & 0 & 0 & 0 & 0 & 0 & 0 \\ 1 & 1 & 0 & 0 & 0 & 0 & 0 & 0 \\ 0 & 0 & 2 & 0 & 0 & 0 & 0 & 0 \\ 0 & 0 & 0 & 2 & 0 & 0 & 0 & 0 \\ 0 & 0 & 0 & 0 & 2 & 0 & 0 & 0 \\ 0 & 0 & 0 & 0 & 0 & 2 & 0 & 0 \\ 0 & 0 & 0 & 0 & 0 & 0 & 0 & 0 \\ 0 & 0 & 0 & 0 & 0 & 0 & 0 & 0\end{pmatrix}
    \]
This can be seen from the change of basis
\be\ba  Z_{\CT/\bbZ_3^d}[\mathds{1}] = \frac{1}{3}\Bigg( &Z_\CT [\mathds{1}] + 2Z_\CT[W_4\otimes \overline{W_4}[\Gamma_1]] +  2Z_\CT[W_4\otimes \overline{W_4}[\Gamma_2]] \\ &+ 2Z_\CT[T(W_4\otimes \overline{W_4}[\Gamma_2])] + 2Z_\CT[T^2(W_4\otimes \overline{W_4}[\Gamma_2])] \Bigg) \ ,  \ea\ee
the trivial symmetry sector after gauging is the result of summing over all sectors composed of $W_4\otimes \overline{W_4}$, which on the boundary gives the group elements generating the $\bbZ_3^d$ symmetry.

\paragraph{Gauging $\bbZ_2^d$.} Gauging $\bbZ_2^d$ is equivalent to switching the topological boundary condition from
\[
    \mathcal{L}_S = \begin{pmatrix}1 & 1 & 2 & 0 & 0 & 0 & 0 & 0 \\ 1 & 1 & 2 & 0 & 0 & 0 & 0 & 0 \\ 2 & 2 & 4 & 0 & 0 & 0 & 0 & 0 \\ 0 & 0 & 0 & 0 & 0 & 0 & 0 & 0 \\ 0 & 0 & 0 & 0 & 0 & 0 & 0 & 0 \\ 0 & 0 & 0 & 0 & 0 & 0 & 0 & 0 \\ 0 & 0 & 0 & 0 & 0 & 0 & 0 & 0 \\ 0 & 0 & 0 & 0 & 0 & 0 & 0 & 0\end{pmatrix}
    \]
into
\[
\CM^{\rm SWSSB}_7=    \begin{pmatrix}1 & 0 & 1 & 0 & 0 & 0 & 0 & 0 \\ 0 & 1 & 1 & 0 & 0 & 0 & 0 & 0 \\ 1 & 1 & 2 & 0 & 0 & 0 & 0 & 0 \\ 0 & 0 & 0 & 0 & 0 & 0 & 0 & 0 \\ 0 & 0 & 0 & 0 & 0 & 0 & 0 & 0 \\ 0 & 0 & 0 & 0 & 0 & 0 & 0 & 0 \\ 0 & 0 & 0 & 0 & 0 & 0 & 1 & 0 \\ 0 & 0 & 0 & 0 & 0 & 0 & 0 & 1\end{pmatrix}.
    \]
The basis change reads
\be\ba
Z_{\CT/\bbZ_3^d}[\mathds{1}] = \frac{1}{2}\Bigg( Z_\CT[\mathds{1}] + Z_\CT[W_7\otimes \overline{W_7}[\Gamma_1]] + Z_\CT[W_7\otimes \overline{W_7}[\Gamma_2]] + Z_\CT[T(W_7\otimes \overline{W_7}[\Gamma_2])]  \Bigg) \ ,
\ea\ee
the trivial sector of the gauged theory is a summation of sectors with lines generating the diagonal $\bbZ_2$ symmetry.

\paragraph{Gauging $S_3^d$.} Gauging $S_3^d$ is equivalent to switching the topological boundary condition from
\[
    \mathcal{L}_S = \begin{pmatrix}1 & 1 & 2 & 0 & 0 & 0 & 0 & 0 \\ 1 & 1 & 2 & 0 & 0 & 0 & 0 & 0 \\ 2 & 2 & 4 & 0 & 0 & 0 & 0 & 0 \\ 0 & 0 & 0 & 0 & 0 & 0 & 0 & 0 \\ 0 & 0 & 0 & 0 & 0 & 0 & 0 & 0 \\ 0 & 0 & 0 & 0 & 0 & 0 & 0 & 0 \\ 0 & 0 & 0 & 0 & 0 & 0 & 0 & 0 \\ 0 & 0 & 0 & 0 & 0 & 0 & 0 & 0\end{pmatrix}
    \]
into
\[
    \CM_6^{\rm SWSSB}=\begin{pmatrix}1 & 0 & 0 & 0 & 0 & 0 & 0 & 0 \\ 0 & 1 & 0 & 0 & 0 & 0 & 0 & 0 \\ 0 & 0 & 1 & 0 & 0 & 0 & 0 & 0 \\ 0 & 0 & 0 & 1 & 0 & 0 & 0 & 0 \\ 0 & 0 & 0 & 0 & 1 & 0 & 0 & 0 \\ 0 & 0 & 0 & 0 & 0 & 1 & 0 & 0 \\ 0 & 0 & 0 & 0 & 0 & 0 & 1 & 0 \\ 0 & 0 & 0 & 0 & 0 & 0 & 0 & 1\end{pmatrix}.
    \]
The basis change reads
\be\ba
Z_{\CT/\bbZ_3^d}[\mathds{1}] = \frac{1}{6}\Bigg( &Z_\CT [\mathds{1}] + 2Z_\CT[W_4\otimes \overline{W_4}[\Gamma_1]] +  2Z_\CT[W_4\otimes \overline{W_4}[\Gamma_2]] \\ &+ 2Z_\CT[T(W_4\otimes \overline{W_4}[\Gamma_2])] + 2Z_\CT[T^2(W_4\otimes \overline{W_4}[\Gamma_2])] \\ &  + 3Z_\CT[W_7\otimes \overline{W_7}[\Gamma_1]]  +3 Z_\CT[W_7\otimes \overline{W_7}[\Gamma_2]] + 3Z_\CT[T(W_7\otimes \overline{W_7}[\Gamma_2])]  \Bigg) \ ,
\ea\ee
the trivial sector of the gauged theory is a summation of sectors with lines generating the diagonal $S_3$ symmetry.

\section{Mixed-State Gapped Phases of $D_4$ Symmetry}
\label{app:D4sym}
Here we present the mixed-state gapped phases with $D_4$ symmetry. These phases are described by the following condensable algebras of $\CZ({\rm Vec}_{D_4\times \overline{D_4}})$ satisfying the mixed-state conditions we imposed. There are $38$ phases in total, with $11$ phases being essentially pure-state phases and $27$ phases involves strong-to-weak symmetry breaking. Among the SWSSB phases, $11$ of them formulates ASPT from the charged domain walls, among which $3$ are the intrinsic ASPT phases. Specifically, phases given by Lagrangian algebras
\be  \{ \CM_{ 11 } , \CM_{12  },\CM_{14  },\CM_{ 16 },\CM_{ 18 },\CM_{ 25 },\CM_{ 28 },\CM_{ 29 },\CM_{ 32 },\CM_{ 33 },\CM_{ 36 }\}  \ee
exhibit non-trivial ASPT features, and phases given by
\be  \{ \CM_{14},\CM_{18},\CM_{25}  \}  \ee
are intrinsic ASPT phases.

In the following cases with spontaneous symmetry breaking, we only state the preserved symmetries and their generators, the rest part of the symmetries are automatically broken.

\begin{enumerate}
\item The pure-state phase of $D_4$ symmetry SSB, given by the Lagrangian algebra
\be
\mathcal{L}_S = \left(
\begin{array}{cccccccccccccccccccccc}
        1 & \bar{e}_R & \bar{e}_G & 0 & \bar{e}_{\text{RG}} & 0 & 0 & 0 & 0 & 0 & 0 & 0 & 2 \bar{m}_B & 0 & 0 & 0 & 0 & 0 & 0 & 0 & 0 & 0 \\
        e_R & e_R \bar{e}_R & e_R \bar{e}_G & 0 & e_R \bar{e}_{\text{RG}} & 0 & 0 & 0 & 0 & 0 & 0 & 0 & 2 e_R \bar{m}_B & 0 & 0 & 0 & 0 & 0 & 0 & 0 & 0 & 0 \\
        e_G & e_G \bar{e}_R & e_G \bar{e}_G & 0 & e_G \bar{e}_{\text{RG}} & 0 & 0 & 0 & 0 & 0 & 0 & 0 & 2 e_G \bar{m}_B & 0 & 0 & 0 & 0 & 0 & 0 & 0 & 0 & 0 \\
        0 & 0 & 0 & 0 & 0 & 0 & 0 & 0 & 0 & 0 & 0 & 0 & 0 & 0 & 0 & 0 & 0 & 0 & 0 & 0 & 0 & 0 \\
        e_{\text{RG}} & e_{\text{RG}} \bar{e}_R & e_{\text{RG}} \bar{e}_G & 0 & e_{\text{RG}} \bar{e}_{\text{RG}} & 0 & 0 & 0 & 0 & 0 & 0 & 0 & 2 e_{\text{RG}} \bar{m}_B & 0 & 0 & 0 & 0 & 0 & 0 & 0 & 0 & 0 \\
        0 & 0 & 0 & 0 & 0 & 0 & 0 & 0 & 0 & 0 & 0 & 0 & 0 & 0 & 0 & 0 & 0 & 0 & 0 & 0 & 0 & 0 \\
        0 & 0 & 0 & 0 & 0 & 0 & 0 & 0 & 0 & 0 & 0 & 0 & 0 & 0 & 0 & 0 & 0 & 0 & 0 & 0 & 0 & 0 \\
        0 & 0 & 0 & 0 & 0 & 0 & 0 & 0 & 0 & 0 & 0 & 0 & 0 & 0 & 0 & 0 & 0 & 0 & 0 & 0 & 0 & 0 \\
\end{array}
    \right) \ .
\ee

\item The pure-state phase with $\bbZ_{2,a^2} \equiv \{e,a^2\}\subset D_4$ symmetry preserved, given by the Lagrangian algebra
\begin{equation}
\adjustbox{width = 0.92\textwidth}{$
    \CA_1^{\rm{SSB} } =\left(
\begin{array}{cccccccccccccccccccccc}
 1 & \bar{e}_R & \bar{e}_G & \bar{e}_B & \bar{e}_{\text{RG}} & \bar{e}_{\text{GB}} & \bar{e}_{\text{RB}} & \bar{e}_{\text{RGB}} & 0 & 0 & 0 & 0 & 0 & 0 & 0 & 0 & 0 & 0 & 0 & 0 & 0 & 0 \\
 e_R & e_R \bar{e}_R & e_R \bar{e}_G & e_R \bar{e}_B & e_R \bar{e}_{\text{RG}} & e_R \bar{e}_{\text{GB}} & e_R \bar{e}_{\text{RB}} & e_R \bar{e}_{\text{RGB}} & 0 & 0 & 0 & 0 & 0 & 0 & 0 & 0 & 0 & 0 & 0 & 0 & 0 & 0 \\
 e_G & e_G \bar{e}_R & e_G \bar{e}_G & e_G \bar{e}_B & e_G \bar{e}_{\text{RG}} & e_G \bar{e}_{\text{GB}} & e_G \bar{e}_{\text{RB}} & e_G \bar{e}_{\text{RGB}} & 0 & 0 & 0 & 0 & 0 & 0 & 0 & 0 & 0 & 0 & 0 & 0 & 0 & 0 \\
 e_B & e_B \bar{e}_R & e_B \bar{e}_G & e_B \bar{e}_B & e_B \bar{e}_{\text{RG}} & e_B \bar{e}_{\text{GB}} & e_B \bar{e}_{\text{RB}} & e_B \bar{e}_{\text{RGB}} & 0 & 0 & 0 & 0 & 0 & 0 & 0 & 0 & 0 & 0 & 0 & 0 & 0 & 0 \\
 e_{\text{RG}} & e_{\text{RG}} \bar{e}_R & e_{\text{RG}} \bar{e}_G & e_{\text{RG}} \bar{e}_B & e_{\text{RG}} \bar{e}_{\text{RG}} & e_{\text{RG}} \bar{e}_{\text{GB}} & e_{\text{RG}} \bar{e}_{\text{RB}} & e_{\text{RG}} \bar{e}_{\text{RGB}} & 0 & 0 & 0 & 0 & 0 & 0 & 0 & 0 & 0 & 0 & 0 & 0 & 0 & 0 \\
 e_{\text{GB}} & e_{\text{GB}} \bar{e}_R & e_{\text{GB}} \bar{e}_G & e_{\text{GB}} \bar{e}_B & e_{\text{GB}} \bar{e}_{\text{RG}} & e_{\text{GB}} \bar{e}_{\text{GB}} & e_{\text{GB}} \bar{e}_{\text{RB}} & e_{\text{GB}} \bar{e}_{\text{RGB}} & 0 & 0 & 0 & 0 & 0 & 0 & 0 & 0 & 0 & 0 & 0 & 0 & 0 & 0 \\
 e_{\text{RB}} & e_{\text{RB}} \bar{e}_R & e_{\text{RB}} \bar{e}_G & e_{\text{RB}} \bar{e}_B & e_{\text{RB}} \bar{e}_{\text{RG}} & e_{\text{RB}} \bar{e}_{\text{GB}} & e_{\text{RB}} \bar{e}_{\text{RB}} & e_{\text{RB}} \bar{e}_{\text{RGB}} & 0 & 0 & 0 & 0 & 0 & 0 & 0 & 0 & 0 & 0 & 0 & 0 & 0 & 0 \\
\end{array}
\right)
$} \ .\ee

\item The pure-state phase with $\bbZ_{2,a^2}\times \bbZ_{2,b}$ symmetry preserved, given by Lagrangian algebra
\be
    \CA_2^{\rm{SSB }}=\left(
\begin{array}{cccccccccccccccccccccc}
 1 & \bar{e}_R & 0 & 0 & 0 & \bar{e}_{\text{GB}} & 0 & \bar{e}_{\text{RGB}} & 0 & 0 & 0 & 0 & 0 & 0 & 0 & 0 & 2 \bar{m}_{\text{GB}} & 0 & 0 & 0 & 0 & 0 \\
 e_R & e_R \bar{e}_R & 0 & 0 & 0 & e_R \bar{e}_{\text{GB}} & 0 & e_R \bar{e}_{\text{RGB}} & 0 & 0 & 0 & 0 & 0 & 0 & 0 & 0 & 2 e_R \bar{m}_{\text{GB}} & 0 & 0 & 0 & 0 & 0 \\
 0 & 0 & 0 & 0 & 0 & 0 & 0 & 0 & 0 & 0 & 0 & 0 & 0 & 0 & 0 & 0 & 0 & 0 & 0 & 0 & 0 & 0 \\
 0 & 0 & 0 & 0 & 0 & 0 & 0 & 0 & 0 & 0 & 0 & 0 & 0 & 0 & 0 & 0 & 0 & 0 & 0 & 0 & 0 & 0 \\
 0 & 0 & 0 & 0 & 0 & 0 & 0 & 0 & 0 & 0 & 0 & 0 & 0 & 0 & 0 & 0 & 0 & 0 & 0 & 0 & 0 & 0 \\
 e_{\text{GB}} & e_{\text{GB}} \bar{e}_R & 0 & 0 & 0 & e_{\text{GB}} \bar{e}_{\text{GB}} & 0 & e_{\text{GB}} \bar{e}_{\text{RGB}} & 0 & 0 & 0 & 0 & 0 & 0 & 0 & 0 & 2 e_{\text{GB}} \bar{m}_{\text{GB}} & 0 & 0 & 0 & 0 & 0 \\
 0 & 0 & 0 & 0 & 0 & 0 & 0 & 0 & 0 & 0 & 0 & 0 & 0 & 0 & 0 & 0 & 0 & 0 & 0 & 0 & 0 & 0 \\
 e_{\text{RGB}} & e_{\text{RGB}} \bar{e}_R & 0 & 0 & 0 & e_{\text{RGB}} \bar{e}_{\text{GB}} & 0 & e_{\text{RGB}} \bar{e}_{\text{RGB}} & 0 & 0 & 0 & 0 & 0 & 0 & 0 & 0 & 2 e_{\text{RGB}} \bar{m}_{\text{GB}} & 0 & 0 & 0 & 0 & 0 \\
\end{array}
\right)
\ee

\item The pure-state phase with $\bbZ_{2,b}\equiv \{e,b\}$ symmetry preserved, given by the Lagrangian algebra
\be
    \CA_3^{\rm{SSB }}=\left(
\begin{array}{cccccccccccccccccccccc}
 1 & \bar{e}_R & 0 & 0 & 0 & 0 & 0 & 0 & 0 & 0 & \bar{m}_G & 0 & \bar{m}_B & 0 & 0 & 0 & \bar{m}_{\text{GB}} & 0 & 0 & 0 & 0 & 0 \\
 e_R & e_R \bar{e}_R & 0 & 0 & 0 & 0 & 0 & 0 & 0 & 0 & e_R \bar{m}_G & 0 & e_R \bar{m}_B & 0 & 0 & 0 & e_R \bar{m}_{\text{GB}} & 0 & 0 & 0 & 0 & 0 \\
 0 & 0 & 0 & 0 & 0 & 0 & 0 & 0 & 0 & 0 & 0 & 0 & 0 & 0 & 0 & 0 & 0 & 0 & 0 & 0 & 0 & 0 \\
 0 & 0 & 0 & 0 & 0 & 0 & 0 & 0 & 0 & 0 & 0 & 0 & 0 & 0 & 0 & 0 & 0 & 0 & 0 & 0 & 0 & 0 \\
 0 & 0 & 0 & 0 & 0 & 0 & 0 & 0 & 0 & 0 & 0 & 0 & 0 & 0 & 0 & 0 & 0 & 0 & 0 & 0 & 0 & 0 \\
 0 & 0 & 0 & 0 & 0 & 0 & 0 & 0 & 0 & 0 & 0 & 0 & 0 & 0 & 0 & 0 & 0 & 0 & 0 & 0 & 0 & 0 \\
 0 & 0 & 0 & 0 & 0 & 0 & 0 & 0 & 0 & 0 & 0 & 0 & 0 & 0 & 0 & 0 & 0 & 0 & 0 & 0 & 0 & 0 \\
 0 & 0 & 0 & 0 & 0 & 0 & 0 & 0 & 0 & 0 & 0 & 0 & 0 & 0 & 0 & 0 & 0 & 0 & 0 & 0 & 0 & 0 \\
\end{array}
\right)
\ee

\item The pure-state phase of non-trivial $\bbZ_{2,a^2}\times \bbZ_{2,ab}$ SPT, given by the Lagrangian algebra
\be
    \CA_4^{\rm{SSB }}=\left(
\begin{array}{cccccccccccccccccccccc}
 1 & 0 & \bar{e}_G & \bar{e}_B & 0 & \bar{e}_{\text{GB}} & 0 & 0 & 2 \bar{m}_R & 0 & 0 & 0 & 0 & 0 & 0 & 0 & 0 & 0 & 0 & 0 & 0 & 0 \\
 0 & 0 & 0 & 0 & 0 & 0 & 0 & 0 & 0 & 0 & 0 & 0 & 0 & 0 & 0 & 0 & 0 & 0 & 0 & 0 & 0 & 0 \\
 e_G & 0 & e_G \bar{e}_G & e_G \bar{e}_B & 0 & e_G \bar{e}_{\text{GB}} & 0 & 0 & 2 e_G \bar{m}_R & 0 & 0 & 0 & 0 & 0 & 0 & 0 & 0 & 0 & 0 & 0 & 0 & 0 \\
 e_B & 0 & e_B \bar{e}_G & e_B \bar{e}_B & 0 & e_B \bar{e}_{\text{GB}} & 0 & 0 & 2 e_B \bar{m}_R & 0 & 0 & 0 & 0 & 0 & 0 & 0 & 0 & 0 & 0 & 0 & 0 & 0 \\
 0 & 0 & 0 & 0 & 0 & 0 & 0 & 0 & 0 & 0 & 0 & 0 & 0 & 0 & 0 & 0 & 0 & 0 & 0 & 0 & 0 & 0 \\
 e_{\text{GB}} & 0 & e_{\text{GB}} \bar{e}_G & e_{\text{GB}} \bar{e}_B & 0 & e_{\text{GB}} \bar{e}_{\text{GB}} & 0 & 0 & 2 e_{\text{GB}} \bar{m}_R & 0 & 0 & 0 & 0 & 0 & 0 & 0 & 0 & 0 & 0 & 0 & 0 & 0 \\
 0 & 0 & 0 & 0 & 0 & 0 & 0 & 0 & 0 & 0 & 0 & 0 & 0 & 0 & 0 & 0 & 0 & 0 & 0 & 0 & 0 & 0 \\
 0 & 0 & 0 & 0 & 0 & 0 & 0 & 0 & 0 & 0 & 0 & 0 & 0 & 0 & 0 & 0 & 0 & 0 & 0 & 0 & 0 & 0 \\
\end{array}
\right)
\ee

\item The pure-state phase of $\bbZ_{2,a^2}\times \bbZ_{2,ab} $ trivial SPT, given by the Lagrangian algebra
\be
    \CA_5^{\rm{SSB }}=\left(
\begin{array}{cccccccccccccccccccccc}
 1 & 0 & \bar{e}_G & 0 & 0 & 0 & \bar{e}_{\text{RB}} & \bar{e}_{\text{RGB}} & 0 & 0 & 0 & 0 & 0 & 0 & 0 & 0 & 0 & 0 & 2 \bar{m}_{\text{RB}} & 0 & 0 & 0 \\
 0 & 0 & 0 & 0 & 0 & 0 & 0 & 0 & 0 & 0 & 0 & 0 & 0 & 0 & 0 & 0 & 0 & 0 & 0 & 0 & 0 & 0 \\
 e_G & 0 & e_G \bar{e}_G & 0 & 0 & 0 & e_G \bar{e}_{\text{RB}} & e_G \bar{e}_{\text{RGB}} & 0 & 0 & 0 & 0 & 0 & 0 & 0 & 0 & 0 & 0 & 2 e_G \bar{m}_{\text{RB}} & 0 & 0 & 0 \\
 0 & 0 & 0 & 0 & 0 & 0 & 0 & 0 & 0 & 0 & 0 & 0 & 0 & 0 & 0 & 0 & 0 & 0 & 0 & 0 & 0 & 0 \\
 0 & 0 & 0 & 0 & 0 & 0 & 0 & 0 & 0 & 0 & 0 & 0 & 0 & 0 & 0 & 0 & 0 & 0 & 0 & 0 & 0 & 0 \\
 0 & 0 & 0 & 0 & 0 & 0 & 0 & 0 & 0 & 0 & 0 & 0 & 0 & 0 & 0 & 0 & 0 & 0 & 0 & 0 & 0 & 0 \\
 e_{\text{RB}} & 0 & e_{\text{RB}} \bar{e}_G & 0 & 0 & 0 & e_{\text{RB}} \bar{e}_{\text{RB}} & e_{\text{RB}} \bar{e}_{\text{RGB}} & 0 & 0 & 0 & 0 & 0 & 0 & 0 & 0 & 0 & 0 & 2 e_{\text{RB}} \bar{m}_{\text{RB}} & 0 & 0 & 0 \\
 e_{\text{RGB}} & 0 & e_{\text{RGB}} \bar{e}_G & 0 & 0 & 0 & e_{\text{RGB}} \bar{e}_{\text{RB}} & e_{\text{RGB}} \bar{e}_{\text{RGB}} & 0 & 0 & 0 & 0 & 0 & 0 & 0 & 0 & 0 & 0 & 2 e_{\text{RGB}} \bar{m}_{\text{RB}} & 0 & 0 & 0 \\
\end{array}
\right)
\ee

\item The pure-state phase with $\bbZ_{2,ab} $ symmetry, given by the Lagrangian algebra
\be
    \CA_6^{\rm{SSB }}=\left(
\begin{array}{cccccccccccccccccccccc}
 1 & 0 & \bar{e}_G & 0 & 0 & 0 & 0 & 0 & \bar{m}_R & 0 & 0 & 0 & \bar{m}_B & 0 & 0 & 0 & 0 & 0 & \bar{m}_{\text{RB}} & 0 & 0 & 0 \\
 0 & 0 & 0 & 0 & 0 & 0 & 0 & 0 & 0 & 0 & 0 & 0 & 0 & 0 & 0 & 0 & 0 & 0 & 0 & 0 & 0 & 0 \\
 e_G & 0 & e_G \bar{e}_G & 0 & 0 & 0 & 0 & 0 & e_G \bar{m}_R & 0 & 0 & 0 & e_G \bar{m}_B & 0 & 0 & 0 & 0 & 0 & e_G \bar{m}_{\text{RB}} & 0 & 0 & 0 \\
 0 & 0 & 0 & 0 & 0 & 0 & 0 & 0 & 0 & 0 & 0 & 0 & 0 & 0 & 0 & 0 & 0 & 0 & 0 & 0 & 0 & 0 \\
 0 & 0 & 0 & 0 & 0 & 0 & 0 & 0 & 0 & 0 & 0 & 0 & 0 & 0 & 0 & 0 & 0 & 0 & 0 & 0 & 0 & 0 \\
 0 & 0 & 0 & 0 & 0 & 0 & 0 & 0 & 0 & 0 & 0 & 0 & 0 & 0 & 0 & 0 & 0 & 0 & 0 & 0 & 0 & 0 \\
 0 & 0 & 0 & 0 & 0 & 0 & 0 & 0 & 0 & 0 & 0 & 0 & 0 & 0 & 0 & 0 & 0 & 0 & 0 & 0 & 0 & 0 \\
 0 & 0 & 0 & 0 & 0 & 0 & 0 & 0 & 0 & 0 & 0 & 0 & 0 & 0 & 0 & 0 & 0 & 0 & 0 & 0 & 0 & 0 \\
\end{array}
\right)
\ee

\item The pure-state phase with $\bbZ_{4,a}$ symmetric state

\be
    \CA_7^{\rm{SSB }}=\left(
\begin{array}{cccccccccccccccccccccc}
 1 & 0 & 0 & \bar{e}_B & \bar{e}_{\text{RG}} & 0 & 0 & \bar{e}_{\text{RGB}} & 0 & 0 & 0 & 0 & 0 & 0 & 2 \bar{m}_{\text{RG}} & 0 & 0 & 0 & 0 & 0 & 0 & 0 \\
 0 & 0 & 0 & 0 & 0 & 0 & 0 & 0 & 0 & 0 & 0 & 0 & 0 & 0 & 0 & 0 & 0 & 0 & 0 & 0 & 0 & 0 \\
 0 & 0 & 0 & 0 & 0 & 0 & 0 & 0 & 0 & 0 & 0 & 0 & 0 & 0 & 0 & 0 & 0 & 0 & 0 & 0 & 0 & 0 \\
 e_B & 0 & 0 & e_B \bar{e}_B & e_B \bar{e}_{\text{RG}} & 0 & 0 & e_B \bar{e}_{\text{RGB}} & 0 & 0 & 0 & 0 & 0 & 0 & 2 e_B \bar{m}_{\text{RG}} & 0 & 0 & 0 & 0 & 0 & 0 & 0 \\
 e_{\text{RG}} & 0 & 0 & e_{\text{RG}} \bar{e}_B & e_{\text{RG}} \bar{e}_{\text{RG}} & 0 & 0 & e_{\text{RG}} \bar{e}_{\text{RGB}} & 0 & 0 & 0 & 0 & 0 & 0 & 2 e_{\text{RG}} \bar{m}_{\text{RG}} & 0 & 0 & 0 & 0 & 0 & 0 & 0 \\
 0 & 0 & 0 & 0 & 0 & 0 & 0 & 0 & 0 & 0 & 0 & 0 & 0 & 0 & 0 & 0 & 0 & 0 & 0 & 0 & 0 & 0 \\
 0 & 0 & 0 & 0 & 0 & 0 & 0 & 0 & 0 & 0 & 0 & 0 & 0 & 0 & 0 & 0 & 0 & 0 & 0 & 0 & 0 & 0 \\
 e_{\text{RGB}} & 0 & 0 & e_{\text{RGB}} \bar{e}_B & e_{\text{RGB}} \bar{e}_{\text{RG}} & 0 & 0 & e_{\text{RGB}} \bar{e}_{\text{RGB}} & 0 & 0 & 0 & 0 & 0 & 0 & 2 e_{\text{RGB}} \bar{m}_{\text{RG}} & 0 & 0 & 0 & 0 & 0 & 0 & 0 \\
\end{array}
\right)
\ee

\item The non-trivial $D_4$ SPT phase, given by the Lagrangian algebra
\be
    \CA_8^{\rm{SPT }}=\left(
\begin{array}{cccccccccccccccccccccc}
 1 & 0 & 0 & \bar{e}_B & 0 & 0 & 0 & 0 & \bar{m}_R & 0 & \bar{m}_G & 0 & 0 & 0 & \bar{m}_{\text{RG}} & 0 & 0 & 0 & 0 & 0 & 0 & 0 \\
 0 & 0 & 0 & 0 & 0 & 0 & 0 & 0 & 0 & 0 & 0 & 0 & 0 & 0 & 0 & 0 & 0 & 0 & 0 & 0 & 0 & 0 \\
 0 & 0 & 0 & 0 & 0 & 0 & 0 & 0 & 0 & 0 & 0 & 0 & 0 & 0 & 0 & 0 & 0 & 0 & 0 & 0 & 0 & 0 \\
 e_B & 0 & 0 & e_B \bar{e}_B & 0 & 0 & 0 & 0 & e_B \bar{m}_R & 0 & e_B \bar{m}_G & 0 & 0 & 0 & e_B \bar{m}_{\text{RG}} & 0 & 0 & 0 & 0 & 0 & 0 & 0 \\
 0 & 0 & 0 & 0 & 0 & 0 & 0 & 0 & 0 & 0 & 0 & 0 & 0 & 0 & 0 & 0 & 0 & 0 & 0 & 0 & 0 & 0 \\
 0 & 0 & 0 & 0 & 0 & 0 & 0 & 0 & 0 & 0 & 0 & 0 & 0 & 0 & 0 & 0 & 0 & 0 & 0 & 0 & 0 & 0 \\
 0 & 0 & 0 & 0 & 0 & 0 & 0 & 0 & 0 & 0 & 0 & 0 & 0 & 0 & 0 & 0 & 0 & 0 & 0 & 0 & 0 & 0 \\
 0 & 0 & 0 & 0 & 0 & 0 & 0 & 0 & 0 & 0 & 0 & 0 & 0 & 0 & 0 & 0 & 0 & 0 & 0 & 0 & 0 & 0 \\
\end{array}
\right)
\ee

\item The pure-state phase of $D_4$ trivial SPT, given by the Lagrangian algebra
\be
    \CA_9^{\rm{SPT }}=\left(
\begin{array}{cccccccccccccccccccccc}
 1 & 0 & 0 & 0 & 0 & 0 & 0 & \bar{e}_{\text{RGB}} & 0 & 0 & 0 & 0 & 0 & 0 & \bar{m}_{\text{RG}} & 0 & \bar{m}_{\text{GB}} & 0 & \bar{m}_{\text{RB}} & 0 & 0 & 0 \\
 0 & 0 & 0 & 0 & 0 & 0 & 0 & 0 & 0 & 0 & 0 & 0 & 0 & 0 & 0 & 0 & 0 & 0 & 0 & 0 & 0 & 0 \\
 0 & 0 & 0 & 0 & 0 & 0 & 0 & 0 & 0 & 0 & 0 & 0 & 0 & 0 & 0 & 0 & 0 & 0 & 0 & 0 & 0 & 0 \\
 0 & 0 & 0 & 0 & 0 & 0 & 0 & 0 & 0 & 0 & 0 & 0 & 0 & 0 & 0 & 0 & 0 & 0 & 0 & 0 & 0 & 0 \\
 0 & 0 & 0 & 0 & 0 & 0 & 0 & 0 & 0 & 0 & 0 & 0 & 0 & 0 & 0 & 0 & 0 & 0 & 0 & 0 & 0 & 0 \\
 0 & 0 & 0 & 0 & 0 & 0 & 0 & 0 & 0 & 0 & 0 & 0 & 0 & 0 & 0 & 0 & 0 & 0 & 0 & 0 & 0 & 0 \\
 0 & 0 & 0 & 0 & 0 & 0 & 0 & 0 & 0 & 0 & 0 & 0 & 0 & 0 & 0 & 0 & 0 & 0 & 0 & 0 & 0 & 0 \\
 e_{\text{RGB}} & 0 & 0 & 0 & 0 & 0 & 0 & e_{\text{RGB}} \bar{e}_{\text{RGB}} & 0 & 0 & 0 & 0 & 0 & 0 & e_{\text{RGB}} \bar{m}_{\text{RG}} & 0 & e_{\text{RGB}} \bar{m}_{\text{GB}} & 0 & e_{\text{RGB}} \bar{m}_{\text{RB}} & 0 & 0 & 0 \\
\end{array}
\right)
\ee

\item The pure-state phase of non-trivial $\bbZ_{2,a^2}\times \bbZ_{2,b}$  SPT, given by the Lagrangian algebra
\be
    \CA_{10}^{\rm{SSB }}=\left(
\begin{array}{cccccccccccccccccccccc}
 1 & \bar{e}_R & 0 & \bar{e}_B & 0 & 0 & \bar{e}_{\text{RB}} & 0 & 0 & 0 & 2 \bar{m}_G & 0 & 0 & 0 & 0 & 0 & 0 & 0 & 0 & 0 & 0 & 0 \\
 e_R & e_R \bar{e}_R & 0 & e_R \bar{e}_B & 0 & 0 & e_R \bar{e}_{\text{RB}} & 0 & 0 & 0 & 2 e_R \bar{m}_G & 0 & 0 & 0 & 0 & 0 & 0 & 0 & 0 & 0 & 0 & 0 \\
 0 & 0 & 0 & 0 & 0 & 0 & 0 & 0 & 0 & 0 & 0 & 0 & 0 & 0 & 0 & 0 & 0 & 0 & 0 & 0 & 0 & 0 \\
 e_B & e_B \bar{e}_R & 0 & e_B \bar{e}_B & 0 & 0 & e_B \bar{e}_{\text{RB}} & 0 & 0 & 0 & 2 e_B \bar{m}_G & 0 & 0 & 0 & 0 & 0 & 0 & 0 & 0 & 0 & 0 & 0 \\
 0 & 0 & 0 & 0 & 0 & 0 & 0 & 0 & 0 & 0 & 0 & 0 & 0 & 0 & 0 & 0 & 0 & 0 & 0 & 0 & 0 & 0 \\
 0 & 0 & 0 & 0 & 0 & 0 & 0 & 0 & 0 & 0 & 0 & 0 & 0 & 0 & 0 & 0 & 0 & 0 & 0 & 0 & 0 & 0 \\
 e_{\text{RB}} & e_{\text{RB}} \bar{e}_R & 0 & e_{\text{RB}} \bar{e}_B & 0 & 0 & e_{\text{RB}} \bar{e}_{\text{RB}} & 0 & 0 & 0 & 2 e_{\text{RB}} \bar{m}_G & 0 & 0 & 0 & 0 & 0 & 0 & 0 & 0 & 0 & 0 & 0 \\
 0 & 0 & 0 & 0 & 0 & 0 & 0 & 0 & 0 & 0 & 0 & 0 & 0 & 0 & 0 & 0 & 0 & 0 & 0 & 0 & 0 & 0 \\
\end{array}
\right)
\ee

\item The SWSSB phase with $\bbZ_{2,a^2}^s\times \bbZ_{2,ab}^w$ ASPT, given by the Lagrangian algebra
\begin{equation}
\adjustbox{width = 0.92\textwidth}{$
    \CM_{11}^{\rm{SWSSB }}=\left(
\begin{array}{cccccccccccccccccccccc}
 1 & 0 & \bar{e}_G & \bar{e}_B & 0 & \bar{e}_{\text{GB}} & 0 & 0 & 0 & 0 & 0 & 0 & 0 & 0 & 0 & 0 & 0 & 0 & 0 & 0 & 0 & 0 \\
 0 & e_R \bar{e}_R & 0 & 0 & e_R \bar{e}_{\text{RG}} & 0 & e_R \bar{e}_{\text{RB}} & e_R \bar{e}_{\text{RGB}} & 0 & 0 & 0 & 0 & 0 & 0 & 0 & 0 & 0 & 0 & 0 & 0 & 0 & 0 \\
 e_G & 0 & e_G \bar{e}_G & e_G \bar{e}_B & 0 & e_G \bar{e}_{\text{GB}} & 0 & 0 & 0 & 0 & 0 & 0 & 0 & 0 & 0 & 0 & 0 & 0 & 0 & 0 & 0 & 0 \\
 e_B & 0 & e_B \bar{e}_G & e_B \bar{e}_B & 0 & e_B \bar{e}_{\text{GB}} & 0 & 0 & 0 & 0 & 0 & 0 & 0 & 0 & 0 & 0 & 0 & 0 & 0 & 0 & 0 & 0 \\
 0 & e_{\text{RG}} \bar{e}_R & 0 & 0 & e_{\text{RG}} \bar{e}_{\text{RG}} & 0 & e_{\text{RG}} \bar{e}_{\text{RB}} & e_{\text{RG}} \bar{e}_{\text{RGB}} & 0 & 0 & 0 & 0 & 0 & 0 & 0 & 0 & 0 & 0 & 0 & 0 & 0 & 0 \\
 e_{\text{GB}} & 0 & e_{\text{GB}} \bar{e}_G & e_{\text{GB}} \bar{e}_B & 0 & e_{\text{GB}} \bar{e}_{\text{GB}} & 0 & 0 & 0 & 0 & 0 & 0 & 0 & 0 & 0 & 0 & 0 & 0 & 0 & 0 & 0 & 0 \\
 0 & e_{\text{RB}} \bar{e}_R & 0 & 0 & e_{\text{RB}} \bar{e}_{\text{RG}} & 0 & e_{\text{RB}} \bar{e}_{\text{RB}} & e_{\text{RB}} \bar{e}_{\text{RGB}} & 0 & 0 & 0 & 0 & 0 & 0 & 0 & 0 & 0 & 0 & 0 & 0 & 0 & 0 \\
 0 & e_{\text{RGB}} \bar{e}_R & 0 & 0 & e_{\text{RGB}} \bar{e}_{\text{RG}} & 0 & e_{\text{RGB}} \bar{e}_{\text{RB}} & e_{\text{RGB}} \bar{e}_{\text{RGB}} & 0 & 0 & 0 & 0 & 0 & 0 & 0 & 0 & 0 & 0 & 0 & 0 & 0 & 0 \\
\end{array}
\right)
 $}\end{equation}

\item The SWSSB phase with $D_4^w$ within which $\bbZ_{2,a^2}^s$ is strong with ASPT, given by Lagrangian algebra
\begin{equation}
\adjustbox{width = 0.92\textwidth}{$
    \CM_{12}^{\rm{SWSSB }}=\left(
\begin{array}{cccccccccccccccccccccc}
 1 & 0 & 0 & \bar{e}_B & 0 & 0 & 0 & 0 & 0 & 0 & 0 & 0 & 0 & 0 & 0 & 0 & 0 & 0 & 0 & 0 & 0 & 0 \\
 0 & e_R \bar{e}_R & 0 & 0 & 0 & 0 & e_R \bar{e}_{\text{RB}} & 0 & 0 & 0 & 0 & 0 & 0 & 0 & 0 & 0 & 0 & 0 & 0 & 0 & 0 & 0 \\
 0 & 0 & e_G \bar{e}_G & 0 & 0 & e_G \bar{e}_{\text{GB}} & 0 & 0 & 0 & 0 & 0 & 0 & 0 & 0 & 0 & 0 & 0 & 0 & 0 & 0 & 0 & 0 \\
 e_B & 0 & 0 & e_B \bar{e}_B & 0 & 0 & 0 & 0 & 0 & 0 & 0 & 0 & 0 & 0 & 0 & 0 & 0 & 0 & 0 & 0 & 0 & 0 \\
 0 & 0 & 0 & 0 & e_{\text{RG}} \bar{e}_{\text{RG}} & 0 & 0 & e_{\text{RG}} \bar{e}_{\text{RGB}} & 0 & 0 & 0 & 0 & 0 & 0 & 0 & 0 & 0 & 0 & 0 & 0 & 0 & 0 \\
 0 & 0 & e_{\text{GB}} \bar{e}_G & 0 & 0 & e_{\text{GB}} \bar{e}_{\text{GB}} & 0 & 0 & 0 & 0 & 0 & 0 & 0 & 0 & 0 & 0 & 0 & 0 & 0 & 0 & 0 & 0 \\
 0 & e_{\text{RB}} \bar{e}_R & 0 & 0 & 0 & 0 & e_{\text{RB}} \bar{e}_{\text{RB}} & 0 & 0 & 0 & 0 & 0 & 0 & 0 & 0 & 0 & 0 & 0 & 0 & 0 & 0 & 0 \\
 0 & 0 & 0 & 0 & e_{\text{RGB}} \bar{e}_{\text{RG}} & 0 & 0 & e_{\text{RGB}} \bar{e}_{\text{RGB}} & 0 & 0 & 0 & 0 & 0 & 0 & 0 & 0 & 0 & 0 & 0 & 0 & 0 & 0 \\
\end{array}
\right)
$}\ee

\item The SWSSB phase with $\bbZ_{2,a^2}^w\times \bbZ_{2,ab}^w$ symmetry, given by Lagrangian algebra
\begin{equation}
\adjustbox{width = 0.92\textwidth}{$
    \CM_{13}^{\rm{SWSSB }}=\left(
\begin{array}{cccccccccccccccccccccc}
 1 & 0 & \bar{e}_G & 0 & 0 & 0 & 0 & 0 & 0 & 0 & 0 & 0 & 0 & 0 & 0 & 0 & 0 & 0 & 0 & 0 & 0 & 0 \\
 0 & e_R \bar{e}_R & 0 & 0 & e_R \bar{e}_{\text{RG}} & 0 & 0 & 0 & 0 & 0 & 0 & 0 & 0 & 0 & 0 & 0 & 0 & 0 & 0 & 0 & 0 & 0 \\
 e_G & 0 & e_G \bar{e}_G & 0 & 0 & 0 & 0 & 0 & 0 & 0 & 0 & 0 & 0 & 0 & 0 & 0 & 0 & 0 & 0 & 0 & 0 & 0 \\
 0 & 0 & 0 & e_B \bar{e}_B & 0 & e_B \bar{e}_{\text{GB}} & 0 & 0 & 0 & 0 & 0 & 0 & 0 & 0 & 0 & 0 & 0 & 0 & 0 & 0 & 0 & 0 \\
 0 & e_{\text{RG}} \bar{e}_R & 0 & 0 & e_{\text{RG}} \bar{e}_{\text{RG}} & 0 & 0 & 0 & 0 & 0 & 0 & 0 & 0 & 0 & 0 & 0 & 0 & 0 & 0 & 0 & 0 & 0 \\
 0 & 0 & 0 & e_{\text{GB}} \bar{e}_B & 0 & e_{\text{GB}} \bar{e}_{\text{GB}} & 0 & 0 & 0 & 0 & 0 & 0 & 0 & 0 & 0 & 0 & 0 & 0 & 0 & 0 & 0 & 0 \\
 0 & 0 & 0 & 0 & 0 & 0 & e_{\text{RB}} \bar{e}_{\text{RB}} & e_{\text{RB}} \bar{e}_{\text{RGB}} & 0 & 0 & 0 & 0 & 0 & 0 & 0 & 0 & 0 & 0 & 0 & 0 & 0 & 0 \\
 0 & 0 & 0 & 0 & 0 & 0 & e_{\text{RGB}} \bar{e}_{\text{RB}} & e_{\text{RGB}} \bar{e}_{\text{RGB}} & 0 & 0 & 0 & 0 & 0 & 0 & 0 & 0 & 0 & 0 & 0 & 0 & 0 & 0 \\
\end{array}
\right)
$}\ee

\item The SWSSB phase with $D_4^w$ weak symmetry, within which $\bbZ_{2,a^2}$ is strong, and formulates an intrinsic ASPT. This phase is given by Lagrangian algebra
\begin{equation}
\adjustbox{width = 0.92\textwidth}{$
    \CM_{14}^{\rm{SWSSB }}=\left(
\begin{array}{cccccccccccccccccccccc}
 1 & 0 & 0 & 0 & 0 & \bar{e}_{\text{GB}} & 0 & 0 & 0 & 0 & 0 & 0 & 0 & 0 & 0 & 0 & 0 & 0 & 0 & 0 & 0 & 0 \\
 0 & e_R \bar{e}_R & 0 & 0 & 0 & 0 & 0 & e_R \bar{e}_{\text{RGB}} & 0 & 0 & 0 & 0 & 0 & 0 & 0 & 0 & 0 & 0 & 0 & 0 & 0 & 0 \\
 0 & 0 & e_G \bar{e}_G & e_G \bar{e}_B & 0 & 0 & 0 & 0 & 0 & 0 & 0 & 0 & 0 & 0 & 0 & 0 & 0 & 0 & 0 & 0 & 0 & 0 \\
 0 & 0 & e_B \bar{e}_G & e_B \bar{e}_B & 0 & 0 & 0 & 0 & 0 & 0 & 0 & 0 & 0 & 0 & 0 & 0 & 0 & 0 & 0 & 0 & 0 & 0 \\
 0 & 0 & 0 & 0 & e_{\text{RG}} \bar{e}_{\text{RG}} & 0 & e_{\text{RG}} \bar{e}_{\text{RB}} & 0 & 0 & 0 & 0 & 0 & 0 & 0 & 0 & 0 & 0 & 0 & 0 & 0 & 0 & 0 \\
 e_{\text{GB}} & 0 & 0 & 0 & 0 & e_{\text{GB}} \bar{e}_{\text{GB}} & 0 & 0 & 0 & 0 & 0 & 0 & 0 & 0 & 0 & 0 & 0 & 0 & 0 & 0 & 0 & 0 \\
 0 & 0 & 0 & 0 & e_{\text{RB}} \bar{e}_{\text{RG}} & 0 & e_{\text{RB}} \bar{e}_{\text{RB}} & 0 & 0 & 0 & 0 & 0 & 0 & 0 & 0 & 0 & 0 & 0 & 0 & 0 & 0 & 0 \\
 0 & e_{\text{RGB}} \bar{e}_R & 0 & 0 & 0 & 0 & 0 & e_{\text{RGB}} \bar{e}_{\text{RGB}} & 0 & 0 & 0 & 0 & 0 & 0 & 0 & 0 & 0 & 0 & 0 & 0 & 0 & 0 \\
\end{array}
\right)
$}\ee

\item The mixed-state phase with $D_4^w$ symmetry
\begin{equation}
\adjustbox{width = 0.92\textwidth}{$
    \CM_{15}^{\rm{SWSSB }}=\left(
\begin{array}{cccccccccccccccccccccc}
 1 & 0 & 0 & 0 & 0 & 0 & 0 & 0 & 0 & 0 & 0 & 0 & 0 & 0 & 0 & 0 & 0 & 0 & 0 & 0 & 0 & 0 \\
 0 & e_R \bar{e}_R & 0 & 0 & 0 & 0 & 0 & 0 & 0 & 0 & 0 & 0 & 0 & 0 & 0 & 0 & 0 & 0 & 0 & 0 & 0 & 0 \\
 0 & 0 & e_G \bar{e}_G & 0 & 0 & 0 & 0 & 0 & 0 & 0 & 0 & 0 & 0 & 0 & 0 & 0 & 0 & 0 & 0 & 0 & 0 & 0 \\
 0 & 0 & 0 & e_B \bar{e}_B & 0 & 0 & 0 & 0 & 0 & 0 & 0 & 0 & 0 & 0 & 0 & 0 & 0 & 0 & 0 & 0 & 0 & 0 \\
 0 & 0 & 0 & 0 & e_{\text{RG}} \bar{e}_{\text{RG}} & 0 & 0 & 0 & 0 & 0 & 0 & 0 & 0 & 0 & 0 & 0 & 0 & 0 & 0 & 0 & 0 & 0 \\
 0 & 0 & 0 & 0 & 0 & e_{\text{GB}} \bar{e}_{\text{GB}} & 0 & 0 & 0 & 0 & 0 & 0 & 0 & 0 & 0 & 0 & 0 & 0 & 0 & 0 & 0 & 0 \\
 0 & 0 & 0 & 0 & 0 & 0 & e_{\text{RB}} \bar{e}_{\text{RB}} & 0 & 0 & 0 & 0 & 0 & 0 & 0 & 0 & 0 & 0 & 0 & 0 & 0 & 0 & 0 \\
 0 & 0 & 0 & 0 & 0 & 0 & 0 & e_{\text{RGB}} \bar{e}_{\text{RGB}} & 0 & 0 & 0 & 0 & 0 & 0 & 0 & 0 & 0 & 0 & 0 & 0 & 0 & 0 \\
\end{array}
\right)
$}\ee

\item The SWSSB phase with $\bbZ_{2,a^2}^s \times \bbZ_{2,b}^w$ formulating an ASPT, given by Lagrangian algebra
\begin{equation}
\adjustbox{width = 0.92\textwidth}{$
    \CM_{16}^{\rm{SWSSB }}=\left(
\begin{array}{cccccccccccccccccccccc}
 1 & \bar{e}_R & 0 & \bar{e}_B & 0 & 0 & \bar{e}_{\text{RB}} & 0 & 0 & 0 & 0 & 0 & 0 & 0 & 0 & 0 & 0 & 0 & 0 & 0 & 0 & 0 \\
 e_R & e_R \bar{e}_R & 0 & e_R \bar{e}_B & 0 & 0 & e_R \bar{e}_{\text{RB}} & 0 & 0 & 0 & 0 & 0 & 0 & 0 & 0 & 0 & 0 & 0 & 0 & 0 & 0 & 0 \\
 0 & 0 & e_G \bar{e}_G & 0 & e_G \bar{e}_{\text{RG}} & e_G \bar{e}_{\text{GB}} & 0 & e_G \bar{e}_{\text{RGB}} & 0 & 0 & 0 & 0 & 0 & 0 & 0 & 0 & 0 & 0 & 0 & 0 & 0 & 0 \\
 e_B & e_B \bar{e}_R & 0 & e_B \bar{e}_B & 0 & 0 & e_B \bar{e}_{\text{RB}} & 0 & 0 & 0 & 0 & 0 & 0 & 0 & 0 & 0 & 0 & 0 & 0 & 0 & 0 & 0 \\
 0 & 0 & e_{\text{RG}} \bar{e}_G & 0 & e_{\text{RG}} \bar{e}_{\text{RG}} & e_{\text{RG}} \bar{e}_{\text{GB}} & 0 & e_{\text{RG}} \bar{e}_{\text{RGB}} & 0 & 0 & 0 & 0 & 0 & 0 & 0 & 0 & 0 & 0 & 0 & 0 & 0 & 0 \\
 0 & 0 & e_{\text{GB}} \bar{e}_G & 0 & e_{\text{GB}} \bar{e}_{\text{RG}} & e_{\text{GB}} \bar{e}_{\text{GB}} & 0 & e_{\text{GB}} \bar{e}_{\text{RGB}} & 0 & 0 & 0 & 0 & 0 & 0 & 0 & 0 & 0 & 0 & 0 & 0 & 0 & 0 \\
 e_{\text{RB}} & e_{\text{RB}} \bar{e}_R & 0 & e_{\text{RB}} \bar{e}_B & 0 & 0 & e_{\text{RB}} \bar{e}_{\text{RB}} & 0 & 0 & 0 & 0 & 0 & 0 & 0 & 0 & 0 & 0 & 0 & 0 & 0 & 0 & 0 \\
 0 & 0 & e_{\text{RGB}} \bar{e}_G & 0 & e_{\text{RGB}} \bar{e}_{\text{RG}} & e_{\text{RGB}} \bar{e}_{\text{GB}} & 0 & e_{\text{RGB}} \bar{e}_{\text{RGB}} & 0 & 0 & 0 & 0 & 0 & 0 & 0 & 0 & 0 & 0 & 0 & 0 & 0 & 0 \\
\end{array}
\right)
$}\ee

\item The SWSSB phase with $\bbZ_{2,a^2}^s\times \bbZ_{2,b}^w$ symmetry, given by Lagrangian algebra 
\begin{equation}
\adjustbox{width = 0.92\textwidth}{$
    \CM_{17}^{\rm{SWSSB }}=\left(
\begin{array}{cccccccccccccccccccccc}
 1 & \bar{e}_R & 0 & 0 & 0 & 0 & 0 & 0 & 0 & 0 & 0 & 0 & 0 & 0 & 0 & 0 & 0 & 0 & 0 & 0 & 0 & 0 \\
 e_R & e_R \bar{e}_R & 0 & 0 & 0 & 0 & 0 & 0 & 0 & 0 & 0 & 0 & 0 & 0 & 0 & 0 & 0 & 0 & 0 & 0 & 0 & 0 \\
 0 & 0 & e_G \bar{e}_G & 0 & e_G \bar{e}_{\text{RG}} & 0 & 0 & 0 & 0 & 0 & 0 & 0 & 0 & 0 & 0 & 0 & 0 & 0 & 0 & 0 & 0 & 0 \\
 0 & 0 & 0 & e_B \bar{e}_B & 0 & 0 & e_B \bar{e}_{\text{RB}} & 0 & 0 & 0 & 0 & 0 & 0 & 0 & 0 & 0 & 0 & 0 & 0 & 0 & 0 & 0 \\
 0 & 0 & e_{\text{RG}} \bar{e}_G & 0 & e_{\text{RG}} \bar{e}_{\text{RG}} & 0 & 0 & 0 & 0 & 0 & 0 & 0 & 0 & 0 & 0 & 0 & 0 & 0 & 0 & 0 & 0 & 0 \\
 0 & 0 & 0 & 0 & 0 & e_{\text{GB}} \bar{e}_{\text{GB}} & 0 & e_{\text{GB}} \bar{e}_{\text{RGB}} & 0 & 0 & 0 & 0 & 0 & 0 & 0 & 0 & 0 & 0 & 0 & 0 & 0 & 0 \\
 0 & 0 & 0 & e_{\text{RB}} \bar{e}_B & 0 & 0 & e_{\text{RB}} \bar{e}_{\text{RB}} & 0 & 0 & 0 & 0 & 0 & 0 & 0 & 0 & 0 & 0 & 0 & 0 & 0 & 0 & 0 \\
 0 & 0 & 0 & 0 & 0 & e_{\text{RGB}} \bar{e}_{\text{GB}} & 0 & e_{\text{RGB}} \bar{e}_{\text{RGB}} & 0 & 0 & 0 & 0 & 0 & 0 & 0 & 0 & 0 & 0 & 0 & 0 & 0 & 0 \\
\end{array}
\right)
$}\ee

\item The SWSSB phase with $D_4^w$ weak symmetry, within which the $\bbZ_{2,a^2}$ is strong, and formulates an intrinsic ASPT. This phase is given by the Lagrangian algebra
\begin{equation}
\adjustbox{width = 0.92\textwidth}{$
    \CM_{18}^{\rm{SWSSB }}=\left(
\begin{array}{cccccccccccccccccccccc}
 1 & 0 & 0 & 0 & 0 & 0 & \bar{e}_{\text{RB}} & 0 & 0 & 0 & 0 & 0 & 0 & 0 & 0 & 0 & 0 & 0 & 0 & 0 & 0 & 0 \\
 0 & e_R \bar{e}_R & 0 & e_R \bar{e}_B & 0 & 0 & 0 & 0 & 0 & 0 & 0 & 0 & 0 & 0 & 0 & 0 & 0 & 0 & 0 & 0 & 0 & 0 \\
 0 & 0 & e_G \bar{e}_G & 0 & 0 & 0 & 0 & e_G \bar{e}_{\text{RGB}} & 0 & 0 & 0 & 0 & 0 & 0 & 0 & 0 & 0 & 0 & 0 & 0 & 0 & 0 \\
 0 & e_B \bar{e}_R & 0 & e_B \bar{e}_B & 0 & 0 & 0 & 0 & 0 & 0 & 0 & 0 & 0 & 0 & 0 & 0 & 0 & 0 & 0 & 0 & 0 & 0 \\
 0 & 0 & 0 & 0 & e_{\text{RG}} \bar{e}_{\text{RG}} & e_{\text{RG}} \bar{e}_{\text{GB}} & 0 & 0 & 0 & 0 & 0 & 0 & 0 & 0 & 0 & 0 & 0 & 0 & 0 & 0 & 0 & 0 \\
 0 & 0 & 0 & 0 & e_{\text{GB}} \bar{e}_{\text{RG}} & e_{\text{GB}} \bar{e}_{\text{GB}} & 0 & 0 & 0 & 0 & 0 & 0 & 0 & 0 & 0 & 0 & 0 & 0 & 0 & 0 & 0 & 0 \\
 e_{\text{RB}} & 0 & 0 & 0 & 0 & 0 & e_{\text{RB}} \bar{e}_{\text{RB}} & 0 & 0 & 0 & 0 & 0 & 0 & 0 & 0 & 0 & 0 & 0 & 0 & 0 & 0 & 0 \\
 0 & 0 & e_{\text{RGB}} \bar{e}_G & 0 & 0 & 0 & 0 & e_{\text{RGB}} \bar{e}_{\text{RGB}} & 0 & 0 & 0 & 0 & 0 & 0 & 0 & 0 & 0 & 0 & 0 & 0 & 0 & 0 \\
\end{array}
\right)
$}\ee

\item The SWSSB phase with $\bbZ_{2,a^2}^w$ symmetry, given by Lagrangian algebra
\begin{equation}
\adjustbox{width = 0.92\textwidth}{$
    \CM_{19}^{\rm{SWSSB }}=\left(
\begin{array}{cccccccccccccccccccccc}
 1 & \bar{e}_R & \bar{e}_G & 0 & \bar{e}_{\text{RG}} & 0 & 0 & 0 & 0 & 0 & 0 & 0 & 0 & 0 & 0 & 0 & 0 & 0 & 0 & 0 & 0 & 0 \\
 e_R & e_R \bar{e}_R & e_R \bar{e}_G & 0 & e_R \bar{e}_{\text{RG}} & 0 & 0 & 0 & 0 & 0 & 0 & 0 & 0 & 0 & 0 & 0 & 0 & 0 & 0 & 0 & 0 & 0 \\
 e_G & e_G \bar{e}_R & e_G \bar{e}_G & 0 & e_G \bar{e}_{\text{RG}} & 0 & 0 & 0 & 0 & 0 & 0 & 0 & 0 & 0 & 0 & 0 & 0 & 0 & 0 & 0 & 0 & 0 \\
 0 & 0 & 0 & e_B \bar{e}_B & 0 & e_B \bar{e}_{\text{GB}} & e_B \bar{e}_{\text{RB}} & e_B \bar{e}_{\text{RGB}} & 0 & 0 & 0 & 0 & 0 & 0 & 0 & 0 & 0 & 0 & 0 & 0 & 0 & 0 \\
 e_{\text{RG}} & e_{\text{RG}} \bar{e}_R & e_{\text{RG}} \bar{e}_G & 0 & e_{\text{RG}} \bar{e}_{\text{RG}} & 0 & 0 & 0 & 0 & 0 & 0 & 0 & 0 & 0 & 0 & 0 & 0 & 0 & 0 & 0 & 0 & 0 \\
 0 & 0 & 0 & e_{\text{GB}} \bar{e}_B & 0 & e_{\text{GB}} \bar{e}_{\text{GB}} & e_{\text{GB}} \bar{e}_{\text{RB}} & e_{\text{GB}} \bar{e}_{\text{RGB}} & 0 & 0 & 0 & 0 & 0 & 0 & 0 & 0 & 0 & 0 & 0 & 0 & 0 & 0 \\
 0 & 0 & 0 & e_{\text{RB}} \bar{e}_B & 0 & e_{\text{RB}} \bar{e}_{\text{GB}} & e_{\text{RB}} \bar{e}_{\text{RB}} & e_{\text{RB}} \bar{e}_{\text{RGB}} & 0 & 0 & 0 & 0 & 0 & 0 & 0 & 0 & 0 & 0 & 0 & 0 & 0 & 0 \\
 0 & 0 & 0 & e_{\text{RGB}} \bar{e}_B & 0 & e_{\text{RGB}} \bar{e}_{\text{GB}} & e_{\text{RGB}} \bar{e}_{\text{RB}} & e_{\text{RGB}} \bar{e}_{\text{RGB}} & 0 & 0 & 0 & 0 & 0 & 0 & 0 & 0 & 0 & 0 & 0 & 0 & 0 & 0 \\
\end{array}
\right)
$}\ee

\item The SWSSB phase with $\bbZ_{4,a}^w$ symmetry, given by Lagrangian algebra
\begin{equation}
\adjustbox{width = 0.92\textwidth}{$
    \CM_{20}^{\rm{SWSSB }}=\left(
\begin{array}{cccccccccccccccccccccc}
 1 & 0 & 0 & 0 & \bar{e}_{\text{RG}} & 0 & 0 & 0 & 0 & 0 & 0 & 0 & 0 & 0 & 0 & 0 & 0 & 0 & 0 & 0 & 0 & 0 \\
 0 & e_R \bar{e}_R & e_R \bar{e}_G & 0 & 0 & 0 & 0 & 0 & 0 & 0 & 0 & 0 & 0 & 0 & 0 & 0 & 0 & 0 & 0 & 0 & 0 & 0 \\
 0 & e_G \bar{e}_R & e_G \bar{e}_G & 0 & 0 & 0 & 0 & 0 & 0 & 0 & 0 & 0 & 0 & 0 & 0 & 0 & 0 & 0 & 0 & 0 & 0 & 0 \\
 0 & 0 & 0 & e_B \bar{e}_B & 0 & 0 & 0 & e_B \bar{e}_{\text{RGB}} & 0 & 0 & 0 & 0 & 0 & 0 & 0 & 0 & 0 & 0 & 0 & 0 & 0 & 0 \\
 e_{\text{RG}} & 0 & 0 & 0 & e_{\text{RG}} \bar{e}_{\text{RG}} & 0 & 0 & 0 & 0 & 0 & 0 & 0 & 0 & 0 & 0 & 0 & 0 & 0 & 0 & 0 & 0 & 0 \\
 0 & 0 & 0 & 0 & 0 & e_{\text{GB}} \bar{e}_{\text{GB}} & e_{\text{GB}} \bar{e}_{\text{RB}} & 0 & 0 & 0 & 0 & 0 & 0 & 0 & 0 & 0 & 0 & 0 & 0 & 0 & 0 & 0 \\
 0 & 0 & 0 & 0 & 0 & e_{\text{RB}} \bar{e}_{\text{GB}} & e_{\text{RB}} \bar{e}_{\text{RB}} & 0 & 0 & 0 & 0 & 0 & 0 & 0 & 0 & 0 & 0 & 0 & 0 & 0 & 0 & 0 \\
 0 & 0 & 0 & e_{\text{RGB}} \bar{e}_B & 0 & 0 & 0 & e_{\text{RGB}} \bar{e}_{\text{RGB}} & 0 & 0 & 0 & 0 & 0 & 0 & 0 & 0 & 0 & 0 & 0 & 0 & 0 & 0 \\
\end{array}
\right)
$}\ee

\item The SWSSB phase with $\bbZ_{4,a}^w$ within which the $\bbZ_{2,a^2}^s$ is strong, given by Lagrangian algebra
\begin{equation}
\adjustbox{width = 0.92\textwidth}{$
    \CM_{21}^{\rm{SWSSB }}=\left(
\begin{array}{cccccccccccccccccccccc}
 1 & 0 & 0 & \bar{e}_B & \bar{e}_{\text{RG}} & 0 & 0 & \bar{e}_{\text{RGB}} & 0 & 0 & 0 & 0 & 0 & 0 & 0 & 0 & 0 & 0 & 0 & 0 & 0 & 0 \\
 0 & e_R \bar{e}_R & e_R \bar{e}_G & 0 & 0 & e_R \bar{e}_{\text{GB}} & e_R \bar{e}_{\text{RB}} & 0 & 0 & 0 & 0 & 0 & 0 & 0 & 0 & 0 & 0 & 0 & 0 & 0 & 0 & 0 \\
 0 & e_G \bar{e}_R & e_G \bar{e}_G & 0 & 0 & e_G \bar{e}_{\text{GB}} & e_G \bar{e}_{\text{RB}} & 0 & 0 & 0 & 0 & 0 & 0 & 0 & 0 & 0 & 0 & 0 & 0 & 0 & 0 & 0 \\
 e_B & 0 & 0 & e_B \bar{e}_B & e_B \bar{e}_{\text{RG}} & 0 & 0 & e_B \bar{e}_{\text{RGB}} & 0 & 0 & 0 & 0 & 0 & 0 & 0 & 0 & 0 & 0 & 0 & 0 & 0 & 0 \\
 e_{\text{RG}} & 0 & 0 & e_{\text{RG}} \bar{e}_B & e_{\text{RG}} \bar{e}_{\text{RG}} & 0 & 0 & e_{\text{RG}} \bar{e}_{\text{RGB}} & 0 & 0 & 0 & 0 & 0 & 0 & 0 & 0 & 0 & 0 & 0 & 0 & 0 & 0 \\
 0 & e_{\text{GB}} \bar{e}_R & e_{\text{GB}} \bar{e}_G & 0 & 0 & e_{\text{GB}} \bar{e}_{\text{GB}} & e_{\text{GB}} \bar{e}_{\text{RB}} & 0 & 0 & 0 & 0 & 0 & 0 & 0 & 0 & 0 & 0 & 0 & 0 & 0 & 0 & 0 \\
 0 & e_{\text{RB}} \bar{e}_R & e_{\text{RB}} \bar{e}_G & 0 & 0 & e_{\text{RB}} \bar{e}_{\text{GB}} & e_{\text{RB}} \bar{e}_{\text{RB}} & 0 & 0 & 0 & 0 & 0 & 0 & 0 & 0 & 0 & 0 & 0 & 0 & 0 & 0 & 0 \\
 e_{\text{RGB}} & 0 & 0 & e_{\text{RGB}} \bar{e}_B & e_{\text{RGB}} \bar{e}_{\text{RG}} & 0 & 0 & e_{\text{RGB}} \bar{e}_{\text{RGB}} & 0 & 0 & 0 & 0 & 0 & 0 & 0 & 0 & 0 & 0 & 0 & 0 & 0 & 0 \\
\end{array}
\right)
$}\ee

\item The SWSSB phase with $D_4^w$ weak symmetry, within which $\bbZ_{2,a^2}^s$ is strong, given by the Lagrangian algebra
\begin{equation}
\adjustbox{width = 0.92\textwidth}{$
    \CM_{22}^{\rm{SWSSB }}=\left(
\begin{array}{cccccccccccccccccccccc}
 1 & 0 & 0 & 0 & 0 & 0 & 0 & \bar{e}_{\text{RGB}} & 0 & 0 & 0 & 0 & 0 & 0 & 0 & 0 & 0 & 0 & 0 & 0 & 0 & 0 \\
 0 & e_R \bar{e}_R & 0 & 0 & 0 & e_R \bar{e}_{\text{GB}} & 0 & 0 & 0 & 0 & 0 & 0 & 0 & 0 & 0 & 0 & 0 & 0 & 0 & 0 & 0 & 0 \\
 0 & 0 & e_G \bar{e}_G & 0 & 0 & 0 & e_G \bar{e}_{\text{RB}} & 0 & 0 & 0 & 0 & 0 & 0 & 0 & 0 & 0 & 0 & 0 & 0 & 0 & 0 & 0 \\
 0 & 0 & 0 & e_B \bar{e}_B & e_B \bar{e}_{\text{RG}} & 0 & 0 & 0 & 0 & 0 & 0 & 0 & 0 & 0 & 0 & 0 & 0 & 0 & 0 & 0 & 0 & 0 \\
 0 & 0 & 0 & e_{\text{RG}} \bar{e}_B & e_{\text{RG}} \bar{e}_{\text{RG}} & 0 & 0 & 0 & 0 & 0 & 0 & 0 & 0 & 0 & 0 & 0 & 0 & 0 & 0 & 0 & 0 & 0 \\
 0 & e_{\text{GB}} \bar{e}_R & 0 & 0 & 0 & e_{\text{GB}} \bar{e}_{\text{GB}} & 0 & 0 & 0 & 0 & 0 & 0 & 0 & 0 & 0 & 0 & 0 & 0 & 0 & 0 & 0 & 0 \\
 0 & 0 & e_{\text{RB}} \bar{e}_G & 0 & 0 & 0 & e_{\text{RB}} \bar{e}_{\text{RB}} & 0 & 0 & 0 & 0 & 0 & 0 & 0 & 0 & 0 & 0 & 0 & 0 & 0 & 0 & 0 \\
 e_{\text{RGB}} & 0 & 0 & 0 & 0 & 0 & 0 & e_{\text{RGB}} \bar{e}_{\text{RGB}} & 0 & 0 & 0 & 0 & 0 & 0 & 0 & 0 & 0 & 0 & 0 & 0 & 0 & 0 \\
\end{array}
\right)
$}\ee

\item The SWSSB phase with $\bbZ_{2,a^2}^s\times \bbZ_{2,b}^w$, given by Lagrangian algebra
\begin{equation}
\adjustbox{width = 0.92\textwidth}{$
    \CM_{23}^{\rm{SWSSB }}=\left(
\begin{array}{cccccccccccccccccccccc}
 1 & \bar{e}_R & 0 & 0 & 0 & \bar{e}_{\text{GB}} & 0 & \bar{e}_{\text{RGB}} & 0 & 0 & 0 & 0 & 0 & 0 & 0 & 0 & 0 & 0 & 0 & 0 & 0 & 0 \\
 e_R & e_R \bar{e}_R & 0 & 0 & 0 & e_R \bar{e}_{\text{GB}} & 0 & e_R \bar{e}_{\text{RGB}} & 0 & 0 & 0 & 0 & 0 & 0 & 0 & 0 & 0 & 0 & 0 & 0 & 0 & 0 \\
 0 & 0 & e_G \bar{e}_G & e_G \bar{e}_B & e_G \bar{e}_{\text{RG}} & 0 & e_G \bar{e}_{\text{RB}} & 0 & 0 & 0 & 0 & 0 & 0 & 0 & 0 & 0 & 0 & 0 & 0 & 0 & 0 & 0 \\
 0 & 0 & e_B \bar{e}_G & e_B \bar{e}_B & e_B \bar{e}_{\text{RG}} & 0 & e_B \bar{e}_{\text{RB}} & 0 & 0 & 0 & 0 & 0 & 0 & 0 & 0 & 0 & 0 & 0 & 0 & 0 & 0 & 0 \\
 0 & 0 & e_{\text{RG}} \bar{e}_G & e_{\text{RG}} \bar{e}_B & e_{\text{RG}} \bar{e}_{\text{RG}} & 0 & e_{\text{RG}} \bar{e}_{\text{RB}} & 0 & 0 & 0 & 0 & 0 & 0 & 0 & 0 & 0 & 0 & 0 & 0 & 0 & 0 & 0 \\
 e_{\text{GB}} & e_{\text{GB}} \bar{e}_R & 0 & 0 & 0 & e_{\text{GB}} \bar{e}_{\text{GB}} & 0 & e_{\text{GB}} \bar{e}_{\text{RGB}} & 0 & 0 & 0 & 0 & 0 & 0 & 0 & 0 & 0 & 0 & 0 & 0 & 0 & 0 \\
 0 & 0 & e_{\text{RB}} \bar{e}_G & e_{\text{RB}} \bar{e}_B & e_{\text{RB}} \bar{e}_{\text{RG}} & 0 & e_{\text{RB}} \bar{e}_{\text{RB}} & 0 & 0 & 0 & 0 & 0 & 0 & 0 & 0 & 0 & 0 & 0 & 0 & 0 & 0 & 0 \\
 e_{\text{RGB}} & e_{\text{RGB}} \bar{e}_R & 0 & 0 & 0 & e_{\text{RGB}} \bar{e}_{\text{GB}} & 0 & e_{\text{RGB}} \bar{e}_{\text{RGB}} & 0 & 0 & 0 & 0 & 0 & 0 & 0 & 0 & 0 & 0 & 0 & 0 & 0 & 0 \\
\end{array}
\right)
$}\ee

\item The SWSSB phase with $\bbZ_{2,a^2}^s\times \bbZ_{2,ab}^w$, given by Lagrangian algebra
\begin{equation}
\adjustbox{width = 0.92\textwidth}{$
    \CM_{24}^{\rm{SWSSB }}=\left(
\begin{array}{cccccccccccccccccccccc}
 1 & 0 & \bar{e}_G & 0 & 0 & 0 & \bar{e}_{\text{RB}} & \bar{e}_{\text{RGB}} & 0 & 0 & 0 & 0 & 0 & 0 & 0 & 0 & 0 & 0 & 0 & 0 & 0 & 0 \\
 0 & e_R \bar{e}_R & 0 & e_R \bar{e}_B & e_R \bar{e}_{\text{RG}} & e_R \bar{e}_{\text{GB}} & 0 & 0 & 0 & 0 & 0 & 0 & 0 & 0 & 0 & 0 & 0 & 0 & 0 & 0 & 0 & 0 \\
 e_G & 0 & e_G \bar{e}_G & 0 & 0 & 0 & e_G \bar{e}_{\text{RB}} & e_G \bar{e}_{\text{RGB}} & 0 & 0 & 0 & 0 & 0 & 0 & 0 & 0 & 0 & 0 & 0 & 0 & 0 & 0 \\
 0 & e_B \bar{e}_R & 0 & e_B \bar{e}_B & e_B \bar{e}_{\text{RG}} & e_B \bar{e}_{\text{GB}} & 0 & 0 & 0 & 0 & 0 & 0 & 0 & 0 & 0 & 0 & 0 & 0 & 0 & 0 & 0 & 0 \\
 0 & e_{\text{RG}} \bar{e}_R & 0 & e_{\text{RG}} \bar{e}_B & e_{\text{RG}} \bar{e}_{\text{RG}} & e_{\text{RG}} \bar{e}_{\text{GB}} & 0 & 0 & 0 & 0 & 0 & 0 & 0 & 0 & 0 & 0 & 0 & 0 & 0 & 0 & 0 & 0 \\
 0 & e_{\text{GB}} \bar{e}_R & 0 & e_{\text{GB}} \bar{e}_B & e_{\text{GB}} \bar{e}_{\text{RG}} & e_{\text{GB}} \bar{e}_{\text{GB}} & 0 & 0 & 0 & 0 & 0 & 0 & 0 & 0 & 0 & 0 & 0 & 0 & 0 & 0 & 0 & 0 \\
 e_{\text{RB}} & 0 & e_{\text{RB}} \bar{e}_G & 0 & 0 & 0 & e_{\text{RB}} \bar{e}_{\text{RB}} & e_{\text{RB}} \bar{e}_{\text{RGB}} & 0 & 0 & 0 & 0 & 0 & 0 & 0 & 0 & 0 & 0 & 0 & 0 & 0 & 0 \\
 e_{\text{RGB}} & 0 & e_{\text{RGB}} \bar{e}_G & 0 & 0 & 0 & e_{\text{RGB}} \bar{e}_{\text{RB}} & e_{\text{RGB}} \bar{e}_{\text{RGB}} & 0 & 0 & 0 & 0 & 0 & 0 & 0 & 0 & 0 & 0 & 0 & 0 & 0 & 0 \\
\end{array}
\right)
$}\ee

\item The SWSSB phase with $\bbZ_{4,a}^w$ weak symmetry, within which $\bbZ_{2,a^2}^s$ is strong, formulating an intrinsic ASPT. This phase is given by Lagrangian algebra 
\begin{equation}
\adjustbox{width = 0.92\textwidth}{$
    \CM_{25}^{\rm{SWSSB }}=\left(
\begin{array}{cccccccccccccccccccccc}
 1 & 0 & 0 & 0 & \bar{e}_{\text{RG}} & \bar{e}_{\text{GB}} & \bar{e}_{\text{RB}} & 0 & 0 & 0 & 0 & 0 & 0 & 0 & 0 & 0 & 0 & 0 & 0 & 0 & 0 & 0 \\
 0 & e_R \bar{e}_R & e_R \bar{e}_G & e_R \bar{e}_B & 0 & 0 & 0 & e_R \bar{e}_{\text{RGB}} & 0 & 0 & 0 & 0 & 0 & 0 & 0 & 0 & 0 & 0 & 0 & 0 & 0 & 0 \\
 0 & e_G \bar{e}_R & e_G \bar{e}_G & e_G \bar{e}_B & 0 & 0 & 0 & e_G \bar{e}_{\text{RGB}} & 0 & 0 & 0 & 0 & 0 & 0 & 0 & 0 & 0 & 0 & 0 & 0 & 0 & 0 \\
 0 & e_B \bar{e}_R & e_B \bar{e}_G & e_B \bar{e}_B & 0 & 0 & 0 & e_B \bar{e}_{\text{RGB}} & 0 & 0 & 0 & 0 & 0 & 0 & 0 & 0 & 0 & 0 & 0 & 0 & 0 & 0 \\
 e_{\text{RG}} & 0 & 0 & 0 & e_{\text{RG}} \bar{e}_{\text{RG}} & e_{\text{RG}} \bar{e}_{\text{GB}} & e_{\text{RG}} \bar{e}_{\text{RB}} & 0 & 0 & 0 & 0 & 0 & 0 & 0 & 0 & 0 & 0 & 0 & 0 & 0 & 0 & 0 \\
 e_{\text{GB}} & 0 & 0 & 0 & e_{\text{GB}} \bar{e}_{\text{RG}} & e_{\text{GB}} \bar{e}_{\text{GB}} & e_{\text{GB}} \bar{e}_{\text{RB}} & 0 & 0 & 0 & 0 & 0 & 0 & 0 & 0 & 0 & 0 & 0 & 0 & 0 & 0 & 0 \\
 e_{\text{RB}} & 0 & 0 & 0 & e_{\text{RB}} \bar{e}_{\text{RG}} & e_{\text{RB}} \bar{e}_{\text{GB}} & e_{\text{RB}} \bar{e}_{\text{RB}} & 0 & 0 & 0 & 0 & 0 & 0 & 0 & 0 & 0 & 0 & 0 & 0 & 0 & 0 & 0 \\
 0 & e_{\text{RGB}} \bar{e}_R & e_{\text{RGB}} \bar{e}_G & e_{\text{RGB}} \bar{e}_B & 0 & 0 & 0 & e_{\text{RGB}} \bar{e}_{\text{RGB}} & 0 & 0 & 0 & 0 & 0 & 0 & 0 & 0 & 0 & 0 & 0 & 0 & 0 & 0 \\
\end{array}
\right)
$}\ee

\item The SWSSB phase with $\bbZ_{2,ab}^w$, given by Lagrangian algebra
\begin{equation}
\adjustbox{width = 0.92\textwidth}{$
    \CM_{26}^{\rm{SWSSB }}=\left(
\begin{array}{cccccccccccccccccccccc}
 1 & 0 & \bar{e}_G & 0 & 0 & 0 & 0 & 0 & 0 & 0 & 0 & 0 & \bar{m}_B & 0 & 0 & 0 & 0 & 0 & 0 & 0 & 0 & 0 \\
 0 & e_R \bar{e}_R & 0 & 0 & e_R \bar{e}_{\text{RG}} & 0 & 0 & 0 & 0 & 0 & 0 & 0 & e_R \bar{m}_B & 0 & 0 & 0 & 0 & 0 & 0 & 0 & 0 & 0 \\
 e_G & 0 & e_G \bar{e}_G & 0 & 0 & 0 & 0 & 0 & 0 & 0 & 0 & 0 & e_G \bar{m}_B & 0 & 0 & 0 & 0 & 0 & 0 & 0 & 0 & 0 \\
 0 & 0 & 0 & 0 & 0 & 0 & 0 & 0 & 0 & 0 & 0 & 0 & 0 & 0 & 0 & 0 & 0 & 0 & 0 & 0 & 0 & 0 \\
 0 & e_{\text{RG}} \bar{e}_R & 0 & 0 & e_{\text{RG}} \bar{e}_{\text{RG}} & 0 & 0 & 0 & 0 & 0 & 0 & 0 & e_{\text{RG}} \bar{m}_B & 0 & 0 & 0 & 0 & 0 & 0 & 0 & 0 & 0 \\
 0 & 0 & 0 & 0 & 0 & 0 & 0 & 0 & 0 & 0 & 0 & 0 & 0 & 0 & 0 & 0 & 0 & 0 & 0 & 0 & 0 & 0 \\
 0 & 0 & 0 & 0 & 0 & 0 & 0 & 0 & 0 & 0 & 0 & 0 & 0 & 0 & 0 & 0 & 0 & 0 & 0 & 0 & 0 & 0 \\
 0 & 0 & 0 & 0 & 0 & 0 & 0 & 0 & 0 & 0 & 0 & 0 & 0 & 0 & 0 & 0 & 0 & 0 & 0 & 0 & 0 & 0 \\
\end{array}
\right)
$}\ee

\item The SWSSB phase with $\bbZ_{2,b}^w$, given by Lagrangian algebra
\begin{equation}
\adjustbox{width = 0.92\textwidth}{$
    \CM_{27}^{\rm{SWSSB }}=\left(
\begin{array}{cccccccccccccccccccccc}
 1 & \bar{e}_R & 0 & 0 & 0 & 0 & 0 & 0 & 0 & 0 & 0 & 0 & \bar{m}_B & 0 & 0 & 0 & 0 & 0 & 0 & 0 & 0 & 0 \\
 e_R & e_R \bar{e}_R & 0 & 0 & 0 & 0 & 0 & 0 & 0 & 0 & 0 & 0 & e_R \bar{m}_B & 0 & 0 & 0 & 0 & 0 & 0 & 0 & 0 & 0 \\
 0 & 0 & e_G \bar{e}_G & 0 & e_G \bar{e}_{\text{RG}} & 0 & 0 & 0 & 0 & 0 & 0 & 0 & e_G \bar{m}_B & 0 & 0 & 0 & 0 & 0 & 0 & 0 & 0 & 0 \\
 0 & 0 & 0 & 0 & 0 & 0 & 0 & 0 & 0 & 0 & 0 & 0 & 0 & 0 & 0 & 0 & 0 & 0 & 0 & 0 & 0 & 0 \\
 0 & 0 & e_{\text{RG}} \bar{e}_G & 0 & e_{\text{RG}} \bar{e}_{\text{RG}} & 0 & 0 & 0 & 0 & 0 & 0 & 0 & e_{\text{RG}} \bar{m}_B & 0 & 0 & 0 & 0 & 0 & 0 & 0 & 0 & 0 \\
 0 & 0 & 0 & 0 & 0 & 0 & 0 & 0 & 0 & 0 & 0 & 0 & 0 & 0 & 0 & 0 & 0 & 0 & 0 & 0 & 0 & 0 \\
 0 & 0 & 0 & 0 & 0 & 0 & 0 & 0 & 0 & 0 & 0 & 0 & 0 & 0 & 0 & 0 & 0 & 0 & 0 & 0 & 0 & 0 \\
 0 & 0 & 0 & 0 & 0 & 0 & 0 & 0 & 0 & 0 & 0 & 0 & 0 & 0 & 0 & 0 & 0 & 0 & 0 & 0 & 0 & 0 \\
\end{array}
\right)
$}\ee

\item The SWSSB phase with $\bbZ_{2,a^2}^s\times \bbZ_{2,b}^s$ strong symmetry and additional $\bbZ_{2,ab}^w$ weak symmetry, formulating an ASPT. This phase is given by Lagrangian algebra
\begin{equation}
\adjustbox{width = 0.92\textwidth}{$
    \CM_{28}^{\rm{SWSSB }}=\left(
\begin{array}{cccccccccccccccccccccc}
 1 & 0 & 0 & \bar{e}_B & 0 & 0 & 0 & 0 & 0 & 0 & \bar{m}_G & 0 & 0 & 0 & 0 & 0 & 0 & 0 & 0 & 0 & 0 & 0 \\
 0 & e_R \bar{e}_R & 0 & 0 & 0 & 0 & e_R \bar{e}_{\text{RB}} & 0 & 0 & 0 & e_R \bar{m}_G & 0 & 0 & 0 & 0 & 0 & 0 & 0 & 0 & 0 & 0 & 0 \\
 0 & 0 & 0 & 0 & 0 & 0 & 0 & 0 & 0 & 0 & 0 & 0 & 0 & 0 & 0 & 0 & 0 & 0 & 0 & 0 & 0 & 0 \\
 e_B & 0 & 0 & e_B \bar{e}_B & 0 & 0 & 0 & 0 & 0 & 0 & e_B \bar{m}_G & 0 & 0 & 0 & 0 & 0 & 0 & 0 & 0 & 0 & 0 & 0 \\
 0 & 0 & 0 & 0 & 0 & 0 & 0 & 0 & 0 & 0 & 0 & 0 & 0 & 0 & 0 & 0 & 0 & 0 & 0 & 0 & 0 & 0 \\
 0 & 0 & 0 & 0 & 0 & 0 & 0 & 0 & 0 & 0 & 0 & 0 & 0 & 0 & 0 & 0 & 0 & 0 & 0 & 0 & 0 & 0 \\
 0 & e_{\text{RB}} \bar{e}_R & 0 & 0 & 0 & 0 & e_{\text{RB}} \bar{e}_{\text{RB}} & 0 & 0 & 0 & e_{\text{RB}} \bar{m}_G & 0 & 0 & 0 & 0 & 0 & 0 & 0 & 0 & 0 & 0 & 0 \\
 0 & 0 & 0 & 0 & 0 & 0 & 0 & 0 & 0 & 0 & 0 & 0 & 0 & 0 & 0 & 0 & 0 & 0 & 0 & 0 & 0 & 0 \\
\end{array}
\right)
$}\ee

\item The SWSSB phase with $\bbZ_{2,b}^s\times \bbZ_{2,a^2}^w$, formulating an ASPT. This phase is given by Lagrangian algebra
\begin{equation}
\adjustbox{width = 0.92\textwidth}{$
    \CM_{29}^{\rm{SWSSB }}=\left(
\begin{array}{cccccccccccccccccccccc}
 1 & \bar{e}_R & 0 & 0 & 0 & 0 & 0 & 0 & 0 & 0 & \bar{m}_G & 0 & 0 & 0 & 0 & 0 & 0 & 0 & 0 & 0 & 0 & 0 \\
 e_R & e_R \bar{e}_R & 0 & 0 & 0 & 0 & 0 & 0 & 0 & 0 & e_R \bar{m}_G & 0 & 0 & 0 & 0 & 0 & 0 & 0 & 0 & 0 & 0 & 0 \\
 0 & 0 & 0 & 0 & 0 & 0 & 0 & 0 & 0 & 0 & 0 & 0 & 0 & 0 & 0 & 0 & 0 & 0 & 0 & 0 & 0 & 0 \\
 0 & 0 & 0 & e_B \bar{e}_B & 0 & 0 & e_B \bar{e}_{\text{RB}} & 0 & 0 & 0 & e_B \bar{m}_G & 0 & 0 & 0 & 0 & 0 & 0 & 0 & 0 & 0 & 0 & 0 \\
 0 & 0 & 0 & 0 & 0 & 0 & 0 & 0 & 0 & 0 & 0 & 0 & 0 & 0 & 0 & 0 & 0 & 0 & 0 & 0 & 0 & 0 \\
 0 & 0 & 0 & 0 & 0 & 0 & 0 & 0 & 0 & 0 & 0 & 0 & 0 & 0 & 0 & 0 & 0 & 0 & 0 & 0 & 0 & 0 \\
 0 & 0 & 0 & e_{\text{RB}} \bar{e}_B & 0 & 0 & e_{\text{RB}} \bar{e}_{\text{RB}} & 0 & 0 & 0 & e_{\text{RB}} \bar{m}_G & 0 & 0 & 0 & 0 & 0 & 0 & 0 & 0 & 0 & 0 & 0 \\
 0 & 0 & 0 & 0 & 0 & 0 & 0 & 0 & 0 & 0 & 0 & 0 & 0 & 0 & 0 & 0 & 0 & 0 & 0 & 0 & 0 & 0 \\
\end{array}
\right)
$}\ee

\item The SWSSB phase with $\bbZ_{2,b}^s\times \bbZ_{2,a^2}^w$ symmetry, given by Lagrangian algebra 
\begin{equation}
\adjustbox{width = 0.92\textwidth}{$
    \CM_{30}^{\rm{SWSSB }}=\left(
\begin{array}{cccccccccccccccccccccc}
 1 & \bar{e}_R & 0 & 0 & 0 & 0 & 0 & 0 & 0 & 0 & 0 & 0 & 0 & 0 & 0 & 0 & \bar{m}_{\text{GB}} & 0 & 0 & 0 & 0 & 0 \\
 e_R & e_R \bar{e}_R & 0 & 0 & 0 & 0 & 0 & 0 & 0 & 0 & 0 & 0 & 0 & 0 & 0 & 0 & e_R \bar{m}_{\text{GB}} & 0 & 0 & 0 & 0 & 0 \\
 0 & 0 & 0 & 0 & 0 & 0 & 0 & 0 & 0 & 0 & 0 & 0 & 0 & 0 & 0 & 0 & 0 & 0 & 0 & 0 & 0 & 0 \\
 0 & 0 & 0 & 0 & 0 & 0 & 0 & 0 & 0 & 0 & 0 & 0 & 0 & 0 & 0 & 0 & 0 & 0 & 0 & 0 & 0 & 0 \\
 0 & 0 & 0 & 0 & 0 & 0 & 0 & 0 & 0 & 0 & 0 & 0 & 0 & 0 & 0 & 0 & 0 & 0 & 0 & 0 & 0 & 0 \\
 0 & 0 & 0 & 0 & 0 & e_{\text{GB}} \bar{e}_{\text{GB}} & 0 & e_{\text{GB}} \bar{e}_{\text{RGB}} & 0 & 0 & 0 & 0 & 0 & 0 & 0 & 0 & e_{\text{GB}} \bar{m}_{\text{GB}} & 0 & 0 & 0 & 0 & 0 \\
 0 & 0 & 0 & 0 & 0 & 0 & 0 & 0 & 0 & 0 & 0 & 0 & 0 & 0 & 0 & 0 & 0 & 0 & 0 & 0 & 0 & 0 \\
 0 & 0 & 0 & 0 & 0 & e_{\text{RGB}} \bar{e}_{\text{GB}} & 0 & e_{\text{RGB}} \bar{e}_{\text{RGB}} & 0 & 0 & 0 & 0 & 0 & 0 & 0 & 0 & e_{\text{RGB}} \bar{m}_{\text{GB}} & 0 & 0 & 0 & 0 & 0 \\
\end{array}
\right)
$}\ee

\item The SWSSB phase with $D_4^w$ weak symmetry, within which $\bbZ_{2,a^2}^s\times \bbZ_{2,b}^s$ is strong, given by Lagrangian algebra
\begin{equation}
\adjustbox{width = 0.92\textwidth}{$
    \CM_{31}^{\rm{SWSSB }}=\left(
\begin{array}{cccccccccccccccccccccc}
 1 & 0 & 0 & 0 & 0 & 0 & 0 & \bar{e}_{\text{RGB}} & 0 & 0 & 0 & 0 & 0 & 0 & 0 & 0 & \bar{m}_{\text{GB}} & 0 & 0 & 0 & 0 & 0 \\
 0 & e_R \bar{e}_R & 0 & 0 & 0 & e_R \bar{e}_{\text{GB}} & 0 & 0 & 0 & 0 & 0 & 0 & 0 & 0 & 0 & 0 & e_R \bar{m}_{\text{GB}} & 0 & 0 & 0 & 0 & 0 \\
 0 & 0 & 0 & 0 & 0 & 0 & 0 & 0 & 0 & 0 & 0 & 0 & 0 & 0 & 0 & 0 & 0 & 0 & 0 & 0 & 0 & 0 \\
 0 & 0 & 0 & 0 & 0 & 0 & 0 & 0 & 0 & 0 & 0 & 0 & 0 & 0 & 0 & 0 & 0 & 0 & 0 & 0 & 0 & 0 \\
 0 & 0 & 0 & 0 & 0 & 0 & 0 & 0 & 0 & 0 & 0 & 0 & 0 & 0 & 0 & 0 & 0 & 0 & 0 & 0 & 0 & 0 \\
 0 & e_{\text{GB}} \bar{e}_R & 0 & 0 & 0 & e_{\text{GB}} \bar{e}_{\text{GB}} & 0 & 0 & 0 & 0 & 0 & 0 & 0 & 0 & 0 & 0 & e_{\text{GB}} \bar{m}_{\text{GB}} & 0 & 0 & 0 & 0 & 0 \\
 0 & 0 & 0 & 0 & 0 & 0 & 0 & 0 & 0 & 0 & 0 & 0 & 0 & 0 & 0 & 0 & 0 & 0 & 0 & 0 & 0 & 0 \\
 e_{\text{RGB}} & 0 & 0 & 0 & 0 & 0 & 0 & e_{\text{RGB}} \bar{e}_{\text{RGB}} & 0 & 0 & 0 & 0 & 0 & 0 & 0 & 0 & e_{\text{RGB}} \bar{m}_{\text{GB}} & 0 & 0 & 0 & 0 & 0 \\
\end{array}
\right)
$}\ee

\item The SWSSB phase with $D_4^w$ weak symmetry, within which $\bbZ_{2,a^2}^s\times \bbZ_{2,ab}^s$ is strong, formulating an ASPT. This phase is given by the Lagrangian algebra
\begin{equation}
\adjustbox{width = 0.92\textwidth}{$
    \CM_{32}^{\rm{SWSSB }}=\left(
\begin{array}{cccccccccccccccccccccc}
 1 & 0 & 0 & \bar{e}_B & 0 & 0 & 0 & 0 & \bar{m}_R & 0 & 0 & 0 & 0 & 0 & 0 & 0 & 0 & 0 & 0 & 0 & 0 & 0 \\
 0 & 0 & 0 & 0 & 0 & 0 & 0 & 0 & 0 & 0 & 0 & 0 & 0 & 0 & 0 & 0 & 0 & 0 & 0 & 0 & 0 & 0 \\
 0 & 0 & e_G \bar{e}_G & 0 & 0 & e_G \bar{e}_{\text{GB}} & 0 & 0 & e_G \bar{m}_R & 0 & 0 & 0 & 0 & 0 & 0 & 0 & 0 & 0 & 0 & 0 & 0 & 0 \\
 e_B & 0 & 0 & e_B \bar{e}_B & 0 & 0 & 0 & 0 & e_B \bar{m}_R & 0 & 0 & 0 & 0 & 0 & 0 & 0 & 0 & 0 & 0 & 0 & 0 & 0 \\
 0 & 0 & 0 & 0 & 0 & 0 & 0 & 0 & 0 & 0 & 0 & 0 & 0 & 0 & 0 & 0 & 0 & 0 & 0 & 0 & 0 & 0 \\
 0 & 0 & e_{\text{GB}} \bar{e}_G & 0 & 0 & e_{\text{GB}} \bar{e}_{\text{GB}} & 0 & 0 & e_{\text{GB}} \bar{m}_R & 0 & 0 & 0 & 0 & 0 & 0 & 0 & 0 & 0 & 0 & 0 & 0 & 0 \\
 0 & 0 & 0 & 0 & 0 & 0 & 0 & 0 & 0 & 0 & 0 & 0 & 0 & 0 & 0 & 0 & 0 & 0 & 0 & 0 & 0 & 0 \\
 0 & 0 & 0 & 0 & 0 & 0 & 0 & 0 & 0 & 0 & 0 & 0 & 0 & 0 & 0 & 0 & 0 & 0 & 0 & 0 & 0 & 0 \\
\end{array}
\right)
$}\ee

\item The SWSSB phase with $\bbZ_{2,ab}^s\times \bbZ_{2,a^2}^w$ symmetry, formulating an ASPT. This phase is given by Lagrangian algebra
\begin{equation}
\adjustbox{width = 0.92\textwidth}{$
    \CM_{33}^{\rm{SWSSB }}=\left(
\begin{array}{cccccccccccccccccccccc}
 1 & 0 & \bar{e}_G & 0 & 0 & 0 & 0 & 0 & \bar{m}_R & 0 & 0 & 0 & 0 & 0 & 0 & 0 & 0 & 0 & 0 & 0 & 0 & 0 \\
 0 & 0 & 0 & 0 & 0 & 0 & 0 & 0 & 0 & 0 & 0 & 0 & 0 & 0 & 0 & 0 & 0 & 0 & 0 & 0 & 0 & 0 \\
 e_G & 0 & e_G \bar{e}_G & 0 & 0 & 0 & 0 & 0 & e_G \bar{m}_R & 0 & 0 & 0 & 0 & 0 & 0 & 0 & 0 & 0 & 0 & 0 & 0 & 0 \\
 0 & 0 & 0 & e_B \bar{e}_B & 0 & e_B \bar{e}_{\text{GB}} & 0 & 0 & e_B \bar{m}_R & 0 & 0 & 0 & 0 & 0 & 0 & 0 & 0 & 0 & 0 & 0 & 0 & 0 \\
 0 & 0 & 0 & 0 & 0 & 0 & 0 & 0 & 0 & 0 & 0 & 0 & 0 & 0 & 0 & 0 & 0 & 0 & 0 & 0 & 0 & 0 \\
 0 & 0 & 0 & e_{\text{GB}} \bar{e}_B & 0 & e_{\text{GB}} \bar{e}_{\text{GB}} & 0 & 0 & e_{\text{GB}} \bar{m}_R & 0 & 0 & 0 & 0 & 0 & 0 & 0 & 0 & 0 & 0 & 0 & 0 & 0 \\
 0 & 0 & 0 & 0 & 0 & 0 & 0 & 0 & 0 & 0 & 0 & 0 & 0 & 0 & 0 & 0 & 0 & 0 & 0 & 0 & 0 & 0 \\
 0 & 0 & 0 & 0 & 0 & 0 & 0 & 0 & 0 & 0 & 0 & 0 & 0 & 0 & 0 & 0 & 0 & 0 & 0 & 0 & 0 & 0 \\
\end{array}
\right)
$}\ee

\item The SWSSB phase with $\bbZ_{2,ab}^s\times \bbZ_{2,a^2}^w$ symmetry, given by Lagrangian algebra
\begin{equation}
\adjustbox{width = 0.92\textwidth}{$
    \CM_{34}^{\rm{SWSSB }}=\left(
\begin{array}{cccccccccccccccccccccc}
 1 & 0 & \bar{e}_G & 0 & 0 & 0 & 0 & 0 & 0 & 0 & 0 & 0 & 0 & 0 & 0 & 0 & 0 & 0 & \bar{m}_{\text{RB}} & 0 & 0 & 0 \\
 0 & 0 & 0 & 0 & 0 & 0 & 0 & 0 & 0 & 0 & 0 & 0 & 0 & 0 & 0 & 0 & 0 & 0 & 0 & 0 & 0 & 0 \\
 e_G & 0 & e_G \bar{e}_G & 0 & 0 & 0 & 0 & 0 & 0 & 0 & 0 & 0 & 0 & 0 & 0 & 0 & 0 & 0 & e_G \bar{m}_{\text{RB}} & 0 & 0 & 0 \\
 0 & 0 & 0 & 0 & 0 & 0 & 0 & 0 & 0 & 0 & 0 & 0 & 0 & 0 & 0 & 0 & 0 & 0 & 0 & 0 & 0 & 0 \\
 0 & 0 & 0 & 0 & 0 & 0 & 0 & 0 & 0 & 0 & 0 & 0 & 0 & 0 & 0 & 0 & 0 & 0 & 0 & 0 & 0 & 0 \\
 0 & 0 & 0 & 0 & 0 & 0 & 0 & 0 & 0 & 0 & 0 & 0 & 0 & 0 & 0 & 0 & 0 & 0 & 0 & 0 & 0 & 0 \\
 0 & 0 & 0 & 0 & 0 & 0 & e_{\text{RB}} \bar{e}_{\text{RB}} & e_{\text{RB}} \bar{e}_{\text{RGB}} & 0 & 0 & 0 & 0 & 0 & 0 & 0 & 0 & 0 & 0 & e_{\text{RB}} \bar{m}_{\text{RB}} & 0 & 0 & 0 \\
 0 & 0 & 0 & 0 & 0 & 0 & e_{\text{RGB}} \bar{e}_{\text{RB}} & e_{\text{RGB}} \bar{e}_{\text{RGB}} & 0 & 0 & 0 & 0 & 0 & 0 & 0 & 0 & 0 & 0 & e_{\text{RGB}} \bar{m}_{\text{RB}} & 0 & 0 & 0 \\
\end{array}
\right)
$}\ee

\item The SWSSB phase with $D_4^w$ weak symmetry, within which $\bbZ_{2,a^2}^s \times \bbZ_{2,ab}^s$ is strong, given by Lagrangian algebra
\begin{equation}
\adjustbox{width = 0.92\textwidth}{$
    \CM_{35}^{\rm{SWSSB }}=\left(
\begin{array}{cccccccccccccccccccccc}
 1 & 0 & 0 & 0 & 0 & 0 & 0 & \bar{e}_{\text{RGB}} & 0 & 0 & 0 & 0 & 0 & 0 & 0 & 0 & 0 & 0 & \bar{m}_{\text{RB}} & 0 & 0 & 0 \\
 0 & 0 & 0 & 0 & 0 & 0 & 0 & 0 & 0 & 0 & 0 & 0 & 0 & 0 & 0 & 0 & 0 & 0 & 0 & 0 & 0 & 0 \\
 0 & 0 & e_G \bar{e}_G & 0 & 0 & 0 & e_G \bar{e}_{\text{RB}} & 0 & 0 & 0 & 0 & 0 & 0 & 0 & 0 & 0 & 0 & 0 & e_G \bar{m}_{\text{RB}} & 0 & 0 & 0 \\
 0 & 0 & 0 & 0 & 0 & 0 & 0 & 0 & 0 & 0 & 0 & 0 & 0 & 0 & 0 & 0 & 0 & 0 & 0 & 0 & 0 & 0 \\
 0 & 0 & 0 & 0 & 0 & 0 & 0 & 0 & 0 & 0 & 0 & 0 & 0 & 0 & 0 & 0 & 0 & 0 & 0 & 0 & 0 & 0 \\
 0 & 0 & 0 & 0 & 0 & 0 & 0 & 0 & 0 & 0 & 0 & 0 & 0 & 0 & 0 & 0 & 0 & 0 & 0 & 0 & 0 & 0 \\
 0 & 0 & e_{\text{RB}} \bar{e}_G & 0 & 0 & 0 & e_{\text{RB}} \bar{e}_{\text{RB}} & 0 & 0 & 0 & 0 & 0 & 0 & 0 & 0 & 0 & 0 & 0 & e_{\text{RB}} \bar{m}_{\text{RB}} & 0 & 0 & 0 \\
 e_{\text{RGB}} & 0 & 0 & 0 & 0 & 0 & 0 & e_{\text{RGB}} \bar{e}_{\text{RGB}} & 0 & 0 & 0 & 0 & 0 & 0 & 0 & 0 & 0 & 0 & e_{\text{RGB}} \bar{m}_{\text{RB}} & 0 & 0 & 0 \\
\end{array}
\right)
$}\ee

\item The SWSSB phase with $D_4^w$ weak symmetry, within which  $\bbZ_{4,a}^s$ is strong, formulating an ASPT. This phase is given by Lagrangian algebra
\begin{equation}
\adjustbox{width = 0.92\textwidth}{$
    \CM_{36}^{\rm{SWSSB }}=\left(
\begin{array}{cccccccccccccccccccccc}
 1 & 0 & 0 & \bar{e}_B & 0 & 0 & 0 & 0 & 0 & 0 & 0 & 0 & 0 & 0 & \bar{m}_{\text{RG}} & 0 & 0 & 0 & 0 & 0 & 0 & 0 \\
 0 & 0 & 0 & 0 & 0 & 0 & 0 & 0 & 0 & 0 & 0 & 0 & 0 & 0 & 0 & 0 & 0 & 0 & 0 & 0 & 0 & 0 \\
 0 & 0 & 0 & 0 & 0 & 0 & 0 & 0 & 0 & 0 & 0 & 0 & 0 & 0 & 0 & 0 & 0 & 0 & 0 & 0 & 0 & 0 \\
 e_B & 0 & 0 & e_B \bar{e}_B & 0 & 0 & 0 & 0 & 0 & 0 & 0 & 0 & 0 & 0 & e_B \bar{m}_{\text{RG}} & 0 & 0 & 0 & 0 & 0 & 0 & 0 \\
 0 & 0 & 0 & 0 & e_{\text{RG}} \bar{e}_{\text{RG}} & 0 & 0 & e_{\text{RG}} \bar{e}_{\text{RGB}} & 0 & 0 & 0 & 0 & 0 & 0 & e_{\text{RG}} \bar{m}_{\text{RG}} & 0 & 0 & 0 & 0 & 0 & 0 & 0 \\
 0 & 0 & 0 & 0 & 0 & 0 & 0 & 0 & 0 & 0 & 0 & 0 & 0 & 0 & 0 & 0 & 0 & 0 & 0 & 0 & 0 & 0 \\
 0 & 0 & 0 & 0 & 0 & 0 & 0 & 0 & 0 & 0 & 0 & 0 & 0 & 0 & 0 & 0 & 0 & 0 & 0 & 0 & 0 & 0 \\
 0 & 0 & 0 & 0 & e_{\text{RGB}} \bar{e}_{\text{RG}} & 0 & 0 & e_{\text{RGB}} \bar{e}_{\text{RGB}} & 0 & 0 & 0 & 0 & 0 & 0 & e_{\text{RGB}} \bar{m}_{\text{RG}} & 0 & 0 & 0 & 0 & 0 & 0 & 0 \\
\end{array}
\right)
$}\ee

\item The SWSSB phase with $D_4^w$ weak symmetry, within which  $\bbZ_{4,a}^s$ is strong, given by Lagrangian algebra
\begin{equation}
\adjustbox{width = 0.92\textwidth}{$
    \CM_{37}^{\rm{SWSSB }}=\left(
\begin{array}{cccccccccccccccccccccc}
 1 & 0 & 0 & 0 & 0 & 0 & 0 & \bar{e}_{\text{RGB}} & 0 & 0 & 0 & 0 & 0 & 0 & \bar{m}_{\text{RG}} & 0 & 0 & 0 & 0 & 0 & 0 & 0 \\
 0 & 0 & 0 & 0 & 0 & 0 & 0 & 0 & 0 & 0 & 0 & 0 & 0 & 0 & 0 & 0 & 0 & 0 & 0 & 0 & 0 & 0 \\
 0 & 0 & 0 & 0 & 0 & 0 & 0 & 0 & 0 & 0 & 0 & 0 & 0 & 0 & 0 & 0 & 0 & 0 & 0 & 0 & 0 & 0 \\
 0 & 0 & 0 & e_B \bar{e}_B & e_B \bar{e}_{\text{RG}} & 0 & 0 & 0 & 0 & 0 & 0 & 0 & 0 & 0 & e_B \bar{m}_{\text{RG}} & 0 & 0 & 0 & 0 & 0 & 0 & 0 \\
 0 & 0 & 0 & e_{\text{RG}} \bar{e}_B & e_{\text{RG}} \bar{e}_{\text{RG}} & 0 & 0 & 0 & 0 & 0 & 0 & 0 & 0 & 0 & e_{\text{RG}} \bar{m}_{\text{RG}} & 0 & 0 & 0 & 0 & 0 & 0 & 0 \\
 0 & 0 & 0 & 0 & 0 & 0 & 0 & 0 & 0 & 0 & 0 & 0 & 0 & 0 & 0 & 0 & 0 & 0 & 0 & 0 & 0 & 0 \\
 0 & 0 & 0 & 0 & 0 & 0 & 0 & 0 & 0 & 0 & 0 & 0 & 0 & 0 & 0 & 0 & 0 & 0 & 0 & 0 & 0 & 0 \\
 e_{\text{RGB}} & 0 & 0 & 0 & 0 & 0 & 0 & e_{\text{RGB}} \bar{e}_{\text{RGB}} & 0 & 0 & 0 & 0 & 0 & 0 & e_{\text{RGB}} \bar{m}_{\text{RG}} & 0 & 0 & 0 & 0 & 0 & 0 & 0 \\
\end{array}
\right)
$}\ee
\end{enumerate}

\end{document}